\documentclass[11pt,twoside]{article} 
\usepackage{euler,acta-info}
\usepackage[latin2]{inputenc}
\usepackage{pdflscape}
\title{
\ \,Deciding football sequences
}

\newcommand{\vol}{\textbf{4,} 1 (2012) 130--183}   %% to be completed by the editor, do not modify it
\newcommand{\amss}{\amssubj{%
05C85, 68R10
}}   %% Source:   http://www.ams.org/msc/    
\newcommand{\CRs}{\CRsubj{%
G.2.2. 
}}   %% Source: http://oldwww.acm.org/class/1998/
\newcommand{\keyw}{\keywords{%
tournament, score sequence, football tournament, polynomial algorithm 
}}

\newcommand{\ideze}{\setbox0=\hbox{\lower1.38ex\hbox{''}}\dp0=0pt\box0}

\begin{document}
\maketitle

\centerline{\emph{Dedicated to the memory of Antal Bege (1962--2012)}}

%% SINGLE AUTHOR. If you are a single author, please, use the following command and delete 
%%                the \twoauthors command completely. 

\twoauthors{%
\href{http://compalg.inf.elte.hu/tanszek/tony/oktato.php?oktato=tony&angolul=1}{Antal IVÁNYI}}
{%
\href{http://www.elte.hu/en}{Eötvös Loránd University},
\href{http://www.inf.elte.hu/english/Lapok/default.aspx}{Faculty of Informatics}, Hungary}
{%
 \href{mailto:ivanyi.antal2@upcmail.hu}{ivanyi.antal2@upcmail.hu}
}{%
Jon E. SCHOENFIELD
}{%
 Huntsville, Alabama, USA}
{%
 \href{mailto:jonscho@hiwaay.net}{jonscho@hiwaay.net}}

 %% Short name of the authors and short title, to be included in heading.

\short{%
A. Iv\'anyi, J. E. Schoenfield
}{Deciding football sequences}

\begin{abstract}
An open problem posed by the first author \cite{Frank2011,Isaak2010,Ivanyi2001,Ivanyi2012Egres,Schoenfield2008A064626} is the 
complexity to decide whether a sequence of nonnegative integer numbers can be the final score 
of a football tournament.  In this paper we propose polynomial time approximate and exponential time exact algorithms which solve the problem.  
\end{abstract}

\hyphenation{cur-rent
}

%%%%%%%%%%%%%%%%%%%%%%%%%%%%%%%%%%%%%%%%%%%%%%%%%%%%%%%%%%%%%%%%%%%%%%%%%%%%%%%%
%%%%%%%%%%%%%%%%%%%%%%%%%%%%%%%%%%%%%%%%%%%%%%%%%%%%%%%%%%%%%%%%%%%%%%%%%%%%%%%%
%%%%%%%%%%%%%%%%%%%%%%%%%%%%%%%%%%%%%%%%%%%%%%%%%%%%%%%%%%%%%%%%%%%%%%%%%%%%%%%%
\normalsize
\section{Introduction\label{sec-Intro}}
Let $a, \ b$ and $n$ be nonnegative integers $(b \geq a \geq 0, \ n \geq 1)$,   
$\mathcal{T}(a,b,n)$ be the set of directed multigraphs $T = (V,E)$, where $|V| = n$, 
and  each pair of different vertices $u, \ v \in V$ are connected 
with at least $a$ and at most $b$ arcs \cite{Ivanyi2009,Ivanyi2010}. 
$T \in \mathcal{T}(a,b,n)$ is called $(a,b,n)$\textit{-tournament.} 
$(1,1,n)$-tournaments are the usual tournaments, and $(0,1,n)$-tournaments are also called 
\textit{oriented graphs} or \textit{simple directed graphs} \cite{GrossY2004,Pirzada2012}. The set $\mathcal{T}$ is defined by
$$
\mathcal{T} = \bigcup _{b\geq a \geq 0, \ n \geq 1} \mathcal{T}(a,b,n).
$$

The definition of (undirected) $(a,b,n)$\textit{-graphs} is similar. The $(0,1,n)$-graphs are the usual \textit{simple graphs}.

An $(a,b,n)$-tournament is called \textit{complete,} if the set of permitted results is $\{0:c, 1:c - 1, 
\ldots,c:0\}$ for all possible $c \ (a \leq c \leq b).$ If some of these results are prohibited, then the 
tournament is called \textit{incomplete} \cite{Ivanyi2002,Ivanyi2009,Ivanyi2010} . 

For example football is an incomplete $(2,3,n)$-tournament since the permitted results are $0:3$, $1:1$ and 
$3:0$, while $0:2$, $1:2$, $2:0$, and $2:1$ are prohibited.

According to this definition $\mathcal{T}$ is the set of the finite directed loopless multigraphs. We remark, that if 
$a' \leq a \leq b \leq b'$ then an $(a,b,n)$-tournament is also an $(a',b',n)$-tournament. The outdegree sequence of an $(a,b,n)$-tourna\-ment 
we call \textit{the score sequence} of the tournament \cite{GrossY2004,Pirzada2012,Reid1996}. 
 
Let $l$, $u$, and $m$ be integer numbers with $u \geq l$ and $m \geq 1$.  The sequence $s = (s_1, \ldots, s_m)$ of integer numbers  
with $l \leq s_1 \leq \cdots \leq s_m \leq u$ is called $(l,u,m)$\textit{-regular}. It is well-known that the number of $(l,u,m)$-regular 
sequences is  
\begin{equation}  
R(l,u,m) = \binom{u - l + m}{m}. \label{eq-lum}
\end{equation}

In this paper we consider only the graph theoretical aspects of the investigated problems, although they have many applications 
\cite{AnholzerBBK2011,BozokiFP2011,BozokiFR2010,Keri2011,LiljerosEASA2001,NewmanBW2006,Temesi2011}. We analyze only sequential algorithms. The Reader can find parallel 
results e.g. in \cite{ArikatiM1996,DessmarkLG1994, PecsySz2000,Siklosi2001,Soroker1990}.

The structure of the paper is as follows. After this introduction in Section \ref{sec-completetest} we deal with the filtering of  potential complete sequences, then in 
Section \ref{sec-incomplete} describe incomplete sequences. 
Section \ref{sec-filter} contains  filtering and  Section \ref{sec-recon} reconstruction algorithms of potential football sequences.  Finally in Section \ref{sec-enum} 
we deal with the enumeration of football sequences. 

%%%%%%%%%%%%%%%%%%%%%%%%%%%%%%%%%%%%%%%%%%%%%%%%%%%%%%%%%%%%%%%%%%%%%%%%%%%%%%%%%%%%%%%%
%%%%%%%%%%%%%%%%%%%%%%%%%%%%%%%%%%%%%%%%%%%%%%%%%%%%%%%%%%%%%%%%%%%%%%%%%%%%%%%%%%%%%%%%
%%%%%%%%%%%%%%%%%%%%%%%%%%%%%%%%%%%%%%%%%%%%%%%%%%%%%%%%%%%%%%%%%%%%%%%%%%%%%%%%%%%%%%%%
\section{Filtering  of potential complete sequences \label{sec-completetest}}
We are seeking football sequences. Taking into account that a score sequence of an incomplete $(a,b,n)$-tournament is at the same time a score sequence of the complete  
$(a,b,n)$-tournament, the properties of score sequences of complete tournaments allow some filtering among the regular sequences.  

In 1953 Landau \cite{Landau1953} proved the following popular theorem. 
About ten proofs are summarized by Reid \cite{Reid1996}. Further proofs are in 
\cite{BangS1979,BrauerGS1968,BrualdiK2009,BrualdiS2001,BuschCJ2010,GriggsR1999,GuiduliGyTW1998,SierksmaH1991,SzekelyCE1992,TripathiT2008}.

\begin{theorem} \emph{(Landau \cite{Landau1953})} A $(0,n-1,n)$-regular sequence $s=(s_1, \ldots, s_n)$ is 
the outdegree sequence of some $(1,1,n)$-tournament if and only if
\begin{equation}
\sum_{i = 1}^k s_i \geq \frac{k(k - 1)}{2}\mathrm{,} \quad 1 \leq k \leq n\mathrm{,} \label{eq-Landau}
\end{equation}
with equality when $k = n$.
\end{theorem} 

\begin{proof} See \cite{IvanyiP2011,Landau1953,Moon1968}.
\end{proof}

Moon \cite{Moon1963} proved the following generalization of 
Landau's theorem (we present it in reformulated form). Later Takahashi \cite{Takahashi2007} reproved the theorem. 

\begin{theorem} \emph{(Moon \cite{Moon1963})} A $(0,b(n-1),n)$-regular sequence $s = (s_1, \ldots, s_n)$  
 is the score sequence of some $(b,b,n)$-tournament if and only if
\[ \sum_{i = 1}^k s_i \geq \frac{bk(k - 1)}{2}\mathrm{,} \ 1 \leq k \leq n, \]
with equality when $k = n.$  
\end{theorem}

\begin {proof} See \cite{Moon1963}.
\end{proof}

We define a \textit{point-loss function} $P_k \ (k = 0, \ldots, n)$ by the following recursion: 
$P_0 = 0$ and if $1 \leq k \leq n$, then 
\[ P_k = \max \left( P_{k - 1}\mathrm{,} \ \frac{bk(k - 1)}{2} - \sum_{i = 1}^{k} s_i  \right ). \]

Now $P_k$ gives a lower bound for the number of lost points 
in the matches among the teams T$_1, \ \ldots, \ T_k$  (not the exact value since the teams 
$T_1, \ \ldots, \ T_k$ could win points against $T_{k+1}, \ \ldots, \ T_n$).

\begin{theorem} \emph{(Iványi \cite{Ivanyi2009})} A $(0,b(n-1),n)$-regular\label{th-P0} sequence 
$s = (s_1, \ldots, s_n)$ is the score sequence of some complete $(a,b,n)$-tournament if and only if
\[ \frac{ak(k - 1)}{2} \leq \sum_{i = 1}^k s_i \leq \frac{bn(n - 1)}{2} - 
   P_k - (n - k)s_k \ (1 \leq k \leq n). \]
\end{theorem}
\begin{proof} See \cite{Ivanyi2009}. 
\end{proof}

%%%%%%%%%%%%%%%%%%%%%%%%%%%%%%%%%%%%%%%%%%%%%%%%%%%%%%%%%%%%%
%%%%%%%%%%%%%%%%%%%%%%%%%%%%%%%%%%%%%%%%%%%%%%%%%%%%%%%%%%%%%
%%%%%%%%%%%%%%%%%%%%%%%%%%%%%%%%%%%%%%%%%%%%%%%%%%%%%%%%%%%%%
\section{Incomplete tournaments \label{sec-incomplete}}
We know only the following three results on the score sequences of incomplete tournaments. 

\textit{Semicomplete digraphs} (semicomplete tournaments) are defined as $(1,2,n)$-digraphs in which if two vertices are connected with two arcs then these arcs have different directions. 

\begin{theorem} \emph{(Reid, Zhang \cite{ReidZ1998})} A $(0,n-1,n)$-regular sequence $s = (s_1, \ldots, s_n)$ is 
the score sequence of some semicomplete tournament if and only if
\begin{equation}
\sum_{i = 1}^k s_i \geq \frac{k(k - 1)}{2} \quad  \textit{and} \quad s_k \leq n - 1, \quad 1 \leq k \leq n. 
\end{equation}
\end{theorem}

\begin{proof} See \cite{ReidZ1998}.
\end{proof}

Antal Bege asked in 1999 \cite{Bege1999} how many wins are necessary in a football tournament of $n$ teams to get a strictly monotone score sequence. 
If $n = 2$ then 1, if $n = 3$ then 1, and if $n = 4$ then 2  are sufficient and necessary. The following assertion gives the general answer.

\begin{theorem} \emph{(Iványi \cite{Ivanyi2002})} If $N(n)$ denotes the minimal number of necessary and sufficient wins for different scores in a football tournament 
of $n$ teams then
\begin{equation}
N(n) = \left (\frac{3}{2} - \sqrt{2} \right )n^2 + \Theta(n).
\end{equation}
\end{theorem}

Recently Berger \cite{Berger2011arX} published the following criterion for special incomplete $(0,2,n)$-tournaments.

\begin{theorem} \emph{(Berger \cite{Berger2011arX})} Sequence\label{th-02} $\sigma = \left (\binom{a_1}{b_1}, \ldots, \binom{a_n}{b_n} \right )$ with $a_1 \geq \cdots \geq a_n$ is the 
score sequence of special incomplete $(0,2,n)$-tournaments---in which \linebreak 
\noindent $0:0$, $0:1$, $1:0$, and $1:1$ are the permitted results---if and only if  
\begin{equation}
\sum _{i=1}^k a_i \leq \sum _{i =1}^k \min(b_i\mathrm{,} k - 1) + \sum _{i = k + 1}^n \min(b_i, k) \label{eq-02} 
\end{equation} 
\textit{ for all } $k = 1, \ldots, n,$ \textit{ with equality for }  $n$.
\end{theorem} 

\begin{proof}  See \cite{Berger2011arX}.
\end{proof} 

Earlier (weaker) results can be found in \cite{Chen1966,Fulkerson1960,Gale1957,Ryser1957}.

%%%%%%%%%%%%%%%%%%%%%%%%%%%%%%%%%%%%%%%%%%%%%%%%%%%%%%%%%%%%%%%%
%%%%%%%%%%%%%%%%%%%%%%%%%%%%%%%%%%%%%%%%%%%%%%%%%%%%%%%%%%%%%%%%
%%%%%%%%%%%%%%%%%%%%%%%%%%%%%%%%%%%%%%%%%%%%%%%%%%%%%%%%%%%%%%%%
\section{Filtering of potential football sequences \label{sec-filter}}
There are many exact results deciding whether a given sequence is the degree/outdegree sequence of a given type of undirected  
(e.g. \cite{Choudum1986,Chungphaisan1974,Eggleton1975,EggletonH1979, ErdosG1960,ErdosMT2011,ErdosR1993,FordF1962,GargGT2011,GuiduliGyTW1998,Hakimi1962,Hakimi1965,Havel1955,HellK2009,
IvanyiL2012AML,KimTMESz2009,LaMar2010,Miller2012,Moon1962,Palvolgyi2009,Weisstein2012GS,ZverovichZ1992}) or directed (e.g. \cite{BergerM20115,BergerM20116,Landau1953,Moon1963,Ivanyi2009,Ivanyi2010,PirzadaZI2011,Weisstein2012DS}) graphs. 
Several authors studied the case when the indegree and outdegree sequences are together prescribed \cite{Berger2011PhD,Berger2011arX,BergerM2012,ErdosMT2011,Hakimi1965,PatrinosH1976}. 

The score sequences of the football tournaments we call \textit{football sequences}. A $(0,3n - 3,n)$-regular sequence $s = (s_1,\ldots,s_n)$ is called \textit{good} if there exists a 
football tournament whose score sequence is $s$, and $s$ is called \textit{bad} otherwise.  We denote the football sequences by $f = (f_1, \ldots, f_n)$.

In this section we present approximate algorithms which filter only some part of the bad sequences. Since these filtering algorithms have short running time they help to reduce the expected running 
time of the exact algorithms.

The filtering algorithms are classified according to their worst running time as constant, linear, and other polynomial type ones.

%%%%%%%%%%%%%%%%%%%%%%%%%%%%%%%%%%%%%%%%%%%%%%%%%%%%%%%%%%%%%%%%%
%%%%%%%%%%%%%%%%%%%%%%%%%%%%%%%%%%%%%%%%%%%%%%%%%%%%%%%%%%%%%%%%%
\subsection{Constant time filtering algorithms \label{sec-constant}}
The expected running time can be substantially decreased if we can filter some part of the investigated sequences 
in constant time. 

Let $n \geq 2$ and $f = (f_1,\ldots,f_n)$ a football sequence.

\begin{lemma} \emph{(C1 test)} \label{lemma-C1} $f_n \neq 3n - 4.$
\end{lemma}

\begin{proof} If a team wins all matches then its score is $3n - 3.$ If not, then it loses at least two points 
making a draw, so its score is at most $3n - 5.$
\end{proof}

\begin{lemma} \emph{(C2 test)} \label{lemma-C2} If $f_n = 3n - 3$ then $f_{n-1} \leq 3n - 6$.
\end{lemma}

\begin{proof} $f_n$ can be $3n - 3$ only so, that T$_n$ wins all matches. Then the score T$_{n-1}$ is at most  
$3n - 6.$
\end{proof}

\begin{lemma} \emph{(C3 test)} \label{lemma-C3} If $f_1 = 0$ then $f_2 \geq 3.$
\end{lemma}

\begin{proof} If $f_1 = 0$ then T$_1$ lost all matches therefore T$_2$ has at least one win and so $f_2$ 
is at least $3.$
\end{proof}

\begin{lemma} \emph{(C4 test)} \label{lemma-C4} If $f_1 = f_2 = 1$ then $f_3 \geq 6.$
\end{lemma}

\begin{proof} If $f_1 = f_2 = 1$ then the match of T$_1$ and T$_2$ ended with a draw implying that T$_3$ has at least 
two wins and so at least six points.    
\end{proof}

\begin{lemma} \emph{(C5 test)} \label{lemma-C5} If $f_n = f_{n-1} = 3n - 5$, then $f_{n-2} \leq 3n - 9.$
\end{lemma}

\begin{proof} If the joint score of T$_n$ and T$_{n-1}$ is $3n - 5$ then the result of their match has to be a draw. 
In this case T$_{n-2}$ lost at least two matches and so $f_{n-2} \leq 3n - 9.$   
\end{proof}

\begin{lemma} \emph{(C6 test)} \label{lemma-C6} If $f_n = 3n - 3$ and $f_{n-1} = 3n - 6$,   
then $f_{n-2} \leq 3n - 9$.
\end{lemma}

\begin{proof} If $f_n = 3n - 3$, then T$_n$ won all matches. In this case the score of T$_{n-1}$ can be 
$3n - 6$ only then if T$_{n-1}$ loses against T$_n$ but wins all remaining matches. Then T$_{n-2}$ lost at least 
two matches and so $f_{n-2} \leq 3n - 9.$
\end{proof}

\begin{lemma} \emph{(C7 test)} \label{lemma-C7} If $f_1 = 0$ and $f_2 = 3$ then $f_3 \geq 6.$
\end{lemma}

\begin{proof}  See the proof of Lemma \ref{lemma-C6}.  
\end{proof}

\begin{lemma} \emph{(C8 test)} \label{lemma-C8} If $f_1 = 1$ and $f_2 = 2$ then  $f_3 \geq 4.$ 
\end{lemma}

\begin{proof}  Since T$_1$ and T$_2$ gathered points only with draws their match ended with a draw.  Therefore T$_3$ won against T$_1$ 
and either won against T$_2$ or they made a draw, so T$_3$ has at least 4 points. 
\end{proof}

\begin{lemma} \emph{(C9 test)} \label{lemma-C9} If $f_n = 3n - 5$ and $f_{n-1} = 3n - 7$ then 
$f_{n-2} \leq 3n - 8.$ 
\end{lemma}

\begin{proof} If $f_n = 3n - 5$ then T$_n$ has a draw and $n - 2$ wins. If $f_{n-1} = 3n - 7$ then T$_{n-1}$ 
has two draws and $n - 3$ wins, and the match between T$_n$ and $T_{n-1}$ ended with a draw. In this case T$_{n-2}$
has at least a loss and a draw implying $f_{n-2} \leq 3n - 8.$
\end{proof}

The following program \textsc{Constant} realizes the tests of the previous 9 lemmas. This and later programs 
are written using the pseudocode conventions described in \cite{CormenLRS2009}. In this and in the further pseudocodes input variables are 
$n$: the length of the investigated sequence $(n \geq 3)$; $s = (s_1,\ldots,s_n)$: a $(0,3n-3,n)$-regular sequence; output variable is  
$L$: $L = 0$ means that the investigated input is bad, $L = 1$ means that it is good while $L = 2$ shows that the given algorithm  could not decide.

\newpage
\noindent \textsc{Constant}$(n, s)$
\vspace{-2mm}
\begin{tabbing}
199 \= x\=x\=x\=x\=x\=x\=x\=x \+ \kill
\hspace{-7mm}01 $L = 0$                                                                                   \` \textbf{//} line 01: initialization of $L$   \\                                                 
\hspace{-7mm}02 \textbf{if} \= $s_n == 3n - 4$                                                 \` \textbf{//} line 02--03: C1 \\
\hspace{-7mm}03              \> \textbf{return} $L$                                                \\
\hspace{-7mm}04 \textbf{if} \= $s_n == 3n - 3$ and  $s_{n-1} \geq 3n - 5$          \` \textbf{//} line 04--05: C2 \\
\hspace{-7mm}05              \> \textbf{return} $L$  \\
\hspace{-7mm}06 \textbf{if} \= $s_1 == 0$ and $s_2 \leq 2$  \` \textbf{//} line 06--07: C3 \\
\hspace{-7mm}07              \> \textbf{return} $L$     \\
\hspace{-7mm}08 \textbf{if} \= $s_1 == 1$ and $s_2 == 1$ and $s_3 \leq 5$  \` \textbf{//} line 08--09: C4 \\
\hspace{-7mm}09              \> \textbf{return} $L$  \\
\hspace{-7mm}10 \textbf{if} \= $s_n == 3n - 5$ and $s_{n-1} = 3n - 5$ and $s_{n-2} \geq 3n - 8$ \`  \textbf{//} line 10--11: C5 \\
\hspace{-7mm}11              \> \textbf{return} $L$     \\
\hspace{-7mm}12 \textbf{if} \= $s_n  \!== \! 3n - 3$ and $s_{n-2} \!== \! 3n - 6$ and $s_{n-3} \geq 3n - 8$     \` \textbf{//} line 12--13: C6 \\
\hspace{-7mm}13              \> \textbf{return} $L$     \\
\hspace{-7mm}14 \textbf{if} \= $s_1 == 0$ and $s_2 == 3$ and $s_3 \leq 5$ \` \textbf{//} line 14--15: C7 \\
\hspace{-7mm}15              \> \textbf{return} $L$     \\
\hspace{-7mm}16 \textbf{if} \= $s_1 == 1$ and $s_2 == 2$ and $s_3 \leq 3$ \` \textbf{//} line 16--17: C8 \\
\hspace{-7mm}17               \> \textbf{return} $L$     \\
\hspace{-7mm}18 \textbf{if} \= $s_n \! == 3n - 5$ and $s_{n-1} \! ==  \!3n - 7$ and $s_{n-2} \geq 3n - 8$        \` \textbf{//} line 18--19: C9 \\
\hspace{-7mm}19              \> \textbf{return} $L$                                                                                  \\
\hspace{-7mm}20 $L = 2$                                                                                       \` \textbf{//} line 20--21: these tests can not decide     \\
\hspace{-7mm}21 \textbf{return} $2$                                                                  \\
\end{tabbing}

\vspace*{-6pt}
Tables \ref{table-C1-C3tests}, \ref{table-C4-C7tests}, and \ref{table-C8-C9tests} show the filtering results of \textsc{Constant}. 
The numbers in the tables show how many sequences are accepted from the sequences accepted by the previous 
filtering algorithm. The exact results in these tables are printed with bold font (such emphasizing will be used in the later tables too). 

The programs are written in  C  by Loránd Lucz and run on an Inter Core i7 processor (3.4 GHz) with optimization level O3. The running times are given in seconds. 

\begin{table}[!ht]
\centering
\begin{footnotesize}
\begin{tabular}{||r|r|r|r|r||} \hline \hline 
$n$ &          $R$                &         C1                  &               C2               &            C3                      \\ \hline  \hline  
1  &                    \text{1}  &                \textbf{1}  &               \textbf{1}    &                     \textbf{1}    \\   \hline 
2  &                            10 &                             7 &                               4 &                     \textbf{2}  \\    \hline 
3  &                            84 &                           63 &                             45 &                                30  \\     \hline  
4  &                          715 &                         550 &                           414 &                              311  \\   \hline  
5  &                       6 188 &                      4 823 &                        3 718 &                           2 911  \\    \hline 
6  &                     54 264 &                    42 636 &                      33 320 &                        26 650  \\    \hline 
7  &                   480 700 &                  379 753 &                    299 421 &                      242 624   \\    \hline 
8  &                4 292 145 &               3 404 115 &                 2 700 775 &                  2 207 800   \\   \hline 
9  &              38 567 375 &             30 678 375 &               24 452 220 &                20 116 030  \\   \hline 
10 &           348 330 136 &           277 722 676 &             222 146 496 &              183 629 160 \\  \hline 
11 &        3 159 461 960 &        2 523 716 572 &          2 024 386 180 &         1 679 655 640  \\   \hline  
12 &      28 760 021 745 &      23 008 017 396 &        18 498 140 232 &       15 394 304 500   \\   \hline 
13 &    262 596 783 764 &    210 345 382 913 &     169 436 070 190  &     141 355 053 635    \\    \hline 
14 & 2 403 979 904 200 & 1 927 719 734 500 &  1 555 302 958 664  &  1 300 210 775 786  \\ \hline 
15 & {\scriptsize  22 057 981 462 440} &{\scriptsize  17 704 432 489 590} &{\scriptsize  14 303 680 429 990} &{\scriptsize  11 978 596 958 384}  \\ \hline \hline  
\end{tabular} 
\end{footnotesize}
\caption{Number of $(0,3n-3,n)$-regular sequences $(R)$   
accepted by C1, C2, and C3 for $n = 1, \ \ldots, \ 15$ teams. \label{table-C1-C3tests}}
\end{table}

Table \ref{table-C4-C7tests} shows the filtering results of C4, C5, C6 and C7. 

\begin{table}[!ht]
\centering
\begin{footnotesize}
\begin{tabular}{||r|r|r|r|r||} \hline \hline 
$n$&         C4                     &               C5                &              C6              &           C7           \\ \hline 
1  &                    \textbf{1} &                    \textbf{1} &                 \textbf{1} &                     \textbf{1}                  \\ \hline
2  &                    \textbf{2} &                    \textbf{2} &                 \textbf{2} &                     \textbf{2}            \\ \hline
3  &                               26 &                             22 &                            19 &                              17             \\ \hline
4  &                             281 &                           255 &                           237&                            222             \\ \hline
5  &                         2 691 &                         2 501 &                        2 374&                         2 271        \\ \hline
6  &                       24 000 &                       23 373 &                     22 302 &                       21 596         \\ \hline
7  &                     227 770 &                    215 227  &                  207 042  &                      200 609       \\ \hline
8  &                  2 700 775 &                2 207 800   &               2 097 803  &                   1 972 783 \\ \hline
9  &                19 155 258 &              18 065 694   &             17 460 916  &                 16 989 609 \\ \hline
10 &             175 138 885 &            165 526 269  &             160 206 767  &             156 070 967 \\ \hline
11 &         1 591 808 376  &         1 518 385 621  &          1 471 133 714  &          1 434 460 309  \\ \hline 
12 &       14 605 778 836  &       13 947 629 921  &        13 524 714 862  &       13 196 925 716 \\ \hline 
13 &     134 230 657 710  &     128 305 394 396  &      124 497 616 840  &     121 549 435 860   \\ \hline 
14 &  1 235 669 598 354  &  1 181 962 750 733  &   1 147 511 569 252  &  1 208 609 923 538   \\ \hline 
15 & \scriptsize{11 391 620 617 874}  & \scriptsize{10 903 053 416 141} & 
\scriptsize{10 590 098 238 918}  &\scriptsize{10 348 178 700 655}  \\ \hline \hline   
\end{tabular} 
\end{footnotesize}
\caption{Number of $(0,3n-3,n)$-regular sequences accepted by C4, C5, C6, and C7 
for $n = 1, \ \ldots, \ 15$ teams. \label{table-C4-C7tests}}
\end{table}

Table \ref{table-C8-C9tests} shows the filtering results of algorithms C8 and C9, further the number of football sequences $(F)$ and the running time 
of \textsc{Linear}  for $n = 1, \ \ldots, \ 15$ teams. 
Column $R$ in Table \ref{table-C1-C3tests} and column $t$ in Table \ref{table-C8-C9tests} 
show that the running time is approximately proportional with the number of the regular sequences.  

\begin{table}[!ht]
\centering 
\begin{footnotesize}
\begin{tabular}{||r|r|r|r|r||} \hline \hline 
$n$&                     C8          &                C9             &        $F$   
       &            $t$                  \\ \hline 
2 &                      \textbf{2} &                  \textbf{2} &  \textbf{2}     
  &          0.000                         \\  \hline
 3 &                               15 &                            14 &  \textbf{7} 
      &          0.000                             \\  \hline
 4 &                             209 &                         203  &  \textbf{40}  
    &          0.000                           \\  \hline
 5 &                         2 175 &                      2 133   & \textbf{355}    
  &          0.000                        \\   \hline
 6 &                       20 039 &                    20 510  & \textbf{3 678}     
&          0.000                          \\  \hline
 7 &                     194 333 &                   191 707 & \textbf{37 263}     &
         0.016                         \\  \hline
 8 &                  1 795 074 &                1 772 842 & \textbf{361 058}    &  
       0.062               \\  \hline
 9 &                16 524 335 &              16 332 091  & \textbf{3 403 613}    & 
     0.499        \\ \hline
10 &             154 361 149 &            150 288 309   & \textbf{31 653 777}  &    
  4.602   \\ \hline
11 &          1 398 051 547 &         1 383 099 467 &  \textbf{292 547 199}  &    
41.771        \\  \hline
12 &        12 870 899 770 &       12 737 278 674 & \textbf{2 696 619 716} &  
380.984              \\  \hline
13 &     118 612 802 828  &     117 411 184 292 &                                &  
3 489.299                  \\ \hline 
14 &   1 094 282 911 155 &  1 083 421 567 482  &                              &   34
079.254  \\ \hline  
15 & 10 106 678 997 431 &10 008 094 941 133   &                             &  316
965.954  \\ \hline \hline 
\end{tabular}
\end{footnotesize}
 \caption{Number of $(0,3n-3,n)$-regular sequences accepted by C8
  and  C9,  the number of football sequences $(F)$, and the running time $(t)$ of
\textsc{C9} 
for $n = 1, \ \ldots, \ 15$ teams. \label{table-C8-C9tests}}
\end{table}  

For example if $n = 2$ then C1, C2 and C3 filter 80 $\%$ of the regular and 100 $\%$
of the bad 
sequences. If $n = 3$ then they filter 54 from the 84 regular sequences while C1,
\ldots, C9 filter 70 sequences which represent 90.90 $\%$ 
of the bad sequences. If $n = 15$ then the nine constant time algorithms filter
54.73 $\%$ of the bad sequences. This is surprisingly high efficiency but smaller 
than the sum of the individual asymptotic efficiency of the 9 algorithms. The reason
is simple: e. g. the sequence 
$s = (0,0,5)$ would be filtered by C1 and C3 too.

%%%%%%%%%%%%%%%%%%%%%%%%%%%%%%%%%%%%%%%%%%%%%%%%%%%%%%%%%%%%%%%%%
%%%%%%%%%%%%%%%%%%%%%%%%%%%%%%%%%%%%%%%%%%%%%%%%%%%%%%%%%%%%%%%%%
\subsection{Efficiency of the constant time testing algorithms \label{sec-constanteff}}
Using \eqref{eq-lum} we give the efficiency of the nine constant time filtering algorithms.

\begin{lemma} \emph{(efficiency of C1)} The ratio\label{lemma-C1eff} of sequences with $s_n = 3n - 4$ among $(0,3n - 3,n)$-regular 
sequences is 
\begin{equation}
\frac{\binom{4n - 5}{n - 1}}{\binom{4n - 3}{n}} = \frac{n(3n - 3)}{(4n - 4)(4n - 3)} = \frac{3}{16} 
+ \frac{9}{16(4n - 3)} = \frac{3}{16} + o(1).\label{eq-C1eff1}
\end{equation}
\end{lemma}

\begin{proof} The sequences satisfying the given condition are such $(0,3n-3,n)$-regular ones, whose lower bound 
is $l = 0$, upper bound is $u = 3n - 4$, and contain $m = n - 1$ elements. So according to \eqref{eq-lum} 
the required ratio is
\begin{equation}
\frac{R(0,3n-4,n-1)}{R(0,3n-3,n)} = \frac{n(3n - 3)}{(4n - 4)(4n - 3)} = \frac{3}{16} + o(1). \label{eq-C1eff2}
\end{equation} 
\end{proof}

\begin{lemma} \emph{(efficiency of C2)} The ratio\label{lemma-C2eff} of the sequences satisfying the conditions  
$s_n = 3n - 3$ and $s_{n-1} \geq 3n - 5$ among the $(0,3n - 3,n)$-regular sequences is 
\begin{equation}
\frac{37}{256} + o(1). \label{eq-C2eff1}
\end{equation}
\end{lemma}

\begin{proof} Since $R(0,3n-3,n-2)$ sequences satisfy the conditions $s_n = 3n - 3$ and $s_{n - 1} = 3n - 3$,  
the corresponding ratio is  
\begin{equation}
\frac{R(0,3n-3,n-2)}{R(0,3n-3,n)} = \frac{n(n - 1)}{(4n - 4)(4n - 3)} = \frac{1}{16} + o(1). \label{eq-C2eff2}  
\end{equation}
$R(0,3n-4,n-2)$ sequences satisfy $s_n = 3n - 3$ and $s_{n - 1} = 3n - 4$, so the corresponding ratio is 
\begin{equation}
\frac{R(0,3n-4,n-2)}{R(0,3n-3,n)} = \frac{n(n - 1)(3n - 3)}{(4n - 3)(4n - 4)(4n - 5)} = 
\frac{3}{64} + o(1).\label{eq-C2eff3}
\end{equation} 
$R(0,3n-5,n-2)$ sequences have the properties $s_n = 3n - 3$ and $s_{n - 1} = 3n - 5$, so the corresponding 
ratio is 
\begin{equation}
\frac{R(0,3n-5,n-2)}{R(0,3n-3,n)} = \frac{n(n - 1)(3n - 3)(3n - 4)}{(4n - 3)(4n - 4)(4n - 5)(4n-6)} 
\label{eq-C2eff4}
= \frac{9}{256} + o(
1).
\end{equation}
Summing up the right sides  \eqref{eq-C2eff2}, \eqref{eq-C2eff3}, and \eqref{eq-C2eff4} we get 
the value \eqref{eq-C2eff1}.
\end{proof}

\begin{lemma} \emph{(efficiency of C3)}  The ratio of the sequences satisfying the conditions \label{lemma-C3eff} 
$s_1 = 0$ and $s_2 \leq 2$ among the $(0,3n - 3,n)$-regular sequences is
\begin{equation}
\frac{37}{256} + o(1). \label{eq-C3eff1}
\end{equation}
\end{lemma}

\begin{proof} Similar to the proof of Lemma \ref{lemma-C2}.
\end{proof}

\begin{lemma} \emph{(efficiency of C4)} The ratio of the sequences satisfying the conditions \label{lemma-C4eff1} 
$s_1 = 1$ and $s_2 = 1$ and $s_3 \leq 5$  among the $(0,3n - 3,n)$-regular sequences is
\begin{equation}
\frac{2343}{4^8} + o(1). \label{eq-C4eff1}
\end{equation}
\end{lemma}

\begin{proof} Since $R(1,3n-3,n-3)$ sequences satisfy the conditions $s_1 = s_2 = s_3 = 1$  the corresponding ratio is  
\begin{equation}
\frac{R(1,3n-3,n-3)}{R(0,3n-3,n)} = \frac{n(n - 1)(n -2)(3n - 3)}{(4n - 3)(4n - 4)(4n - 5)(4n - 6)} = \frac{3}{4^4} + o(1). \label{eq-C4eff2}  
\end{equation}

The sequences with $s_1 = s_2  = 1$ and $s_3 = 2$, $s_1 = s_2 = 1$ and $s_3 = 3, $ $s_1 = s_2 = 1$ and $s_3 = 4$,  and 
$s_1 = s_2 = 1$ and $s_3 = 5$ have the asymptotic ratio $3/4^5$,  $3/4^6$, $3/4^7$, and $3/4^8$ resp. 

The sum of the received five ratios is 
\begin{equation}
\frac{3}{4^4} + \frac{3^2}{4^5} + \frac{3^3}{4^6} + \frac{3^4}{4^7} + \frac{3^5}{4^7} = \frac{2343}{4^8}, \label{eq-C4eff3}
\end{equation}  
implying \eqref{eq-C4eff1}.
\end{proof}

\begin{lemma} \emph{(efficiency of C5)} The ratio of the sequences satisfying the conditions of \label{lemma-C5eff} 
$s_n = s_{n - 1} = 3n -5$ and $s_{n - 3} \geq 3n - 8$ among the $(0,3n - 3,n)$-regular sequences is
\begin{equation}
\frac{1575}{4^8} + o(1). \label{eq-C5eff1}
\end{equation}
\end{lemma}

\begin{proof} We have to sum the contributions of $R(0,3n - 5,n-2)$, $R(0,3n - 6,n-2)$,  $R(0,3n - 7,n-2)$, and $R(0,3n - 8,n-2)$ sequences: 
\begin{equation}
\frac{3^2}{4^5} + \frac{3^3}{4^6} + \frac{3^5}{4^7} + \frac{3^6}{4^8} = \frac{1575}{4^8}, \label{eq-C5eff2}
\end{equation}
implying \eqref{eq-C5eff1}.
\end{proof}

\begin{lemma} \emph{(efficiency of C6)} The ratio of the sequences satisfying the conditions of \label{lemma-C6eff}
$s_n = 3n - 3$,  $s_{n -1} = 3n - 6$, and $s_{n-2} \geq 3n - 8$ among the $(0,3n - 3,n)$-regular sequences is
\begin{equation}
\frac{999}{4^8} + o(1). \label{eq-C6eff1}
\end{equation}
\end{lemma}

\begin{proof} In this case  we sum the contributions of $R(0,3n - 6,n - 3)$,  $R(0,3n - 7,n - 1)$, and 
$R(0,3n - 8,n - 1)$ sequences:
\begin{equation}
\frac{3^3}{4^6} + \frac{3^4}{4^7} + \frac{3^5}{4^8} = \frac{999}{4^8}, \label{eq-C6-eff2}  
\end{equation}
implying \eqref{eq-C6eff1}.
\end{proof}

\begin{lemma} \emph{(efficiency of C7)} The ratio of the sequences satisfying the conditions of  \label{lemma-C7eff} 
$s_1 = 0$, $s_2 = 3$, and $s_3 \leq 5$ among the $(0,3n - 3,n)$-regular sequences is
\begin{equation}
\frac{999}{4^8} + o(1). \label{eq-C7eff1}
\end{equation}
\end{lemma}

\begin{proof} Similar to the proof of Lemma \ref{lemma-C6eff}.
\end{proof}

\begin{lemma} \emph{(efficiency of C8)} The ratio of the sequences satisfying the conditions of \label{lemma-C8eff} 
$s_1 = 1$, $s_2 = 2$, and $s_3 \leq 3$  among the $(0,3n - 3,n)$-regular sequences is
\begin{equation}
\frac{63}{4^6} + o(1). \label{eq-C8eff1}
\end{equation}
\end{lemma}

\begin{proof} We sum the contributions of $R(2,3n - 3,n - 3)$ and  $R(3,3n - 3,n - 3)$ sequences:
\begin{equation}
\frac{3^2}{4^5} + \frac{3^3}{4^6} = \frac{63}{4^6}, \label{eq-C8-eff2}  
\end{equation}
implying \eqref{eq-C8eff1}.
\end{proof}

\begin{lemma} \emph{(efficiency of C9)} The ratio of the sequences satisfying the conditions of \label{lemma-C9eff} 
$s_n = 3n - 5$,  $s_{n - 1} = 3n - 7$, and $s_{n - 2} \geq 3n - 7$  among the $(0,3n - 3,n)$-regular sequences is
\begin{equation}
\frac{3^4}{4^7} + o(1). \label{eq-C9eff3}
\end{equation}
\end{lemma}

\begin{proof} Similar to the proof of Lemma \ref{lemma-C8eff}.
\end{proof}

The cumulated asymptotic efficiency of the constant time algorithms is 
\begin{equation}
\frac{3}{16} + \frac{2 \cdot 37}{4^4} + \frac{2343}{4^8} + \frac{1575}{4^8} + \frac{2 \cdot 999}{4^8} + \frac{63}{4^6} + \frac{3^4}{4^7} = \frac{38480}{4^8}. \label{eq-C9eff4}
\end{equation}  

The cumulated efficiency of the nine constant time algorithms is about 58.72~$\%.$ According to Table \ref{table-C1-C3tests} the practical joint efficiency of C1, C2 and C3 is 
64.28~$\%$ for $n = 3$ and $45.91$~$\%$ for $n = 14.$ According to Table \ref{table-C8-C9tests} the total practical efficiency of the nine constant 
time algorithms is 91.67~$\%$ for $n = 3$ and $54.93$~$\%$ for $n = 14.$ 

The practical cumulated efficiency is smaller than the theoretical one, since some part of the sequences is filtered by several algorithms: e.g. the sequence 
$s = (0,0,5)$ is filtered by C1 and C3 too.

We remark that the algorithms of \textsc{Constant} are sorted on the base of their nonincreasing asymptotic efficiency. We get the same order of the practical efficiency of these algorithms shown on the small values of $n$.

%%%%%%%%%%%%%%%%%%%%%%%%%%%%%%%%%%%%%%%%%%%%%%%%%%%%%%%%%%%%%%%%%%%%%%
%%%%%%%%%%%%%%%%%%%%%%%%%%%%%%%%%%%%%%%%%%%%%%%%%%%%%%%%%%%%%%%%%%%%%%
\subsection{Filtering algorithms with linear running time \label{subsec-linear}}
We investigate the following filtering algorithms whose worst running time is linear: \textsc{Complete} = L1, \textsc{Point-Losses} = L2, 
\textsc{Reduction0} = L3, \textsc{Reduction1} = L4, \textsc{Draw-Unique} = L5, \textsc{Balanced} = L6,    \textsc{Draw-Uniform} = L7,  \textsc{Draw-Sorted-Unique} = L8.

%%%%%%%%%%%%%%%%%%%%%%%%%%%%%%%%%%%%%%%%%%%%%%%%%%%%%%%%%%%%%%%%%%%%%%%
\subsubsection{Linear filtering algorithm L1 = \sc{Complete} \label{subsub-L1}}
The first linear time filtering algorithm L1 = \textsc{Complete} is based on the following special case of Lemma 3 in \cite{Ivanyi2009}.
\begin{corollary} \emph{((2,3,$n$)-complete test, \cite{Ivanyi2009})} If $n \geq 1$ and $ (f_1, \ldots, f_n)$ is a football sequence then\label{cor-23complete}
\begin{equation}
2 \binom{k}{2} \leq  \sum_{i=1}^k f_i \leq  3 \binom{n}{2} - (n - k)f_k \quad (k = 1, \ldots, n). \label{eq-23complete}
\end{equation}
\end{corollary}

Basic parameters of \textsc{Complete} are the usual ones, further $S$: the current sum of the first $i$ elements of $s$. 

\bigskip
\noindent \textsc{Complete}$(n, s)$
\vspace{-2mm}
\begin{tabbing}
199 \= x\=x\=x\=x\=x\=x\=x\=x \+ \kill
\hspace{-7mm}01 $S = 0$                                  \` \textbf{//}    line 01: initialization of $S$ \\
\hspace{-7mm}02 \textbf{for} \= $i = 1$ \textbf{to} $n$      \` \textbf{//}   line 02--06:  test \\
\hspace{-7mm}03                  \> $S = S + s_i$ \\
\hspace{-7mm}04                  \>  \textbf{if} \= $(S < 2 \binom{i}{2}) \vee  (S > 3 \binom{n}{2} - (n - i)s_i) = \textsc{true}$            \\
\hspace{-7mm}05                  \>                 \> $L = 0$   \\
\hspace{-7mm}06                  \>                 \> \textbf{return} $L$  \\
\hspace{-7mm}07 $L = 2$                                                       \` \textbf{//} line 07--08: $s$ is undecided \\
\hspace{-7mm}08 \textbf{return} $L$                                                                
\end{tabbing}

%%%%%%%%%%%%%%%%%%%%%%%%%%%%%%%%%%%%%%%%%%%%%%%%%%%%%%%%%%%
\subsubsection{Linear filtering algorithm L2 = \sc{Point-Losses} \label{subsub-L2}}
The second linear time filtering algorithm L2 = 
\textsc{Point-Losses} is based on the following assertion which is an extension of Lemma 3 in \cite{Ivanyi2009}. 
The basic idea is, that the small sums of the prefixes of $s$ and the mod 3 remainders of 
the elements of $s$ signalize lost points. 

\begin{lemma} If $(f_1, \ldots, f_n)$ is a football sequence then\label{lemma-point-losses}
\begin{equation}
2\binom{k}{2} \leq  \sum_ {i=1}^k f_i \leq 3\binom{n}{2} - (n - k)f_k - P_k \quad (k = 1,\ldots, n), \label{eq-losses}
\end{equation}
where $P_{0} = 0$ and 
\begin{equation}
P_k = \max \left( P_{k-1}, 3\binom{k}{2} - \sum_ {i=1}^k f_i, \left \lceil \frac{\sum_{i=1}^k  (f_i - 3 \lfloor f_i/3 \rfloor)}{2} \right \rceil \right ). \label{eq-Pdef}
\end{equation}
\end{lemma} 

\begin{proof} The sum of the $k$ smallest scores is at least $2 \binom{k}{2}$ and at most $3 \binom{n}{2}$ minus the following point-losses:
\begin{enumerate}
\item the sum of the remaining scores, which is at least $(n - k)f_k$;
\item the point-losses due to draws documented by the mod $3$ remainders;
\item the point-losses documented by differences $3 \binom{k}{2} - \sum _ {i=1}^k f_i$;  
\end{enumerate}
\end{proof}

Basic parameters of \textsc{Point-Losses} are the usual ones, further $S$: the current sum of the first $i$ elements of $s$, and $P$: the current value of the point-losses.

\bigskip
\noindent \textsc{Point-Losses}$(n, s)$
\vspace{-2mm}
\begin{tabbing}
199 \= x\=x\=x\=x\=x\=x\=x\=x \+ \kill
\hspace{-7mm}01 $S = P = L = 0$                                  \` \textbf{//}    line 01: initialization of $S$, $P$, and $L$ \\
\hspace{-7mm}02 \textbf{for} \= $k = 1$ \textbf{to} $n$      \` \textbf{//}   line 02--06:  filtering \\
\hspace{-7mm}03                  \> $S = S + s_k$ \\
\hspace{-7mm}04                  \> $P = \max  \left ( P_{k-1}, 3\binom{k}{2} - S,\left \lceil \frac{\sum _{i=1}^k  (s_i - 3 \lfloor s_i/3 \rfloor}{2} \right \rceil \right )$  \\ 
\hspace{-7mm}05                  \>  \textbf{if} \= $S >  3\binom{n}{2} - (n - k)s_k - P$            \\
\hspace{-7mm}06                  \>                 \> \textbf{return} $L$  \\
\hspace{-7mm}07 $L = 2$   \\
\hspace{-7mm}08 \textbf{return} $L$                                                                \` \textbf{//} line 08: $s$ is undecided
\end{tabbing}

%%%%%%%%%%%%%%%%%%%%%%%%%%%%%%%%%%%%%%%%%%%%%%%%%%%%%%%%%%%%%%%%%%%%
\subsubsection{Linear filtering algorithm L3 = \sc{Reduction0} \label{subsub-L3}}
The third linear test is based on the observation that if the sum of the $k$ smallest scores is minimal then all matches among the first $k$ teams 
ended by a draw and if the sum of the $k$ largest scores is maximal then the corresponding scores are multiples of 3 and further if $k<n$ then $f_{n-k} \leq 3(n-k-1)$. 
\begin{lemma} If \label{lemma-reduction0} $n \geq 2$, $1 \leq k \leq n$, and $f = (f_1, \ldots, f_n)$ is a football sequence  then 

1) if  the sum of the first $k$ scores is $k(k - 1)$ then $f _1 = \cdots =  f_k = k - 1$ and if further $k < n$ then  $f_{k+1} \geq 3k$;

2) if  the sum of the last $k$ scores is $3(n - k)k + 3\binom{k}{2}$ then $f_{n - k +1}, \ldots, f_n$ are multiples of 3 and if further $k<n$ then $f_{n-k}\leq 3(n-k-1)$.
\end{lemma}

\begin{proof} If $f_1 + \cdots + f_k = k(k - 1)$ then all matches among T$_1$, \ldots, T$_k$ ended with a draw and these teams lost all matches 
against the remaining teams implying assertions 1). 

 If $f_{k +1} + \cdots + f_n  =  3(n - k)k + 3\binom{k}{2}$ then T$_{k + 1}$, \ldots, T$_n$ won all matches against the remaining teams and have no draws 
implying assertion  2.
\end{proof} 

Parameters of \textsc{Reduction0} are the usual ones, further $S$: the current sum of the  $i$ smallest scores; $Q$: the current sum of the $i$  largest scores; $B$ 
is a logical variable characterizing the remainders mod $3$ of the $i$ largest scores. 

\bigskip
\noindent \textsc{Reduction0}$(n, s)$
\vspace{-2mm}
\begin{tabbing}
199 \= x\=x\=x\=x\=x\=x\=x\=x \+ \kill
\hspace{-7mm}01 $L = B = S = Q = 0$                                                              \` \textbf{//}   line 01: initialization of $L$, $B$, $S$, and $Q$  \\
\hspace{-7mm}02 \textbf{for} \= $i = 1$ \textbf{to} $n - 1$                     \` \textbf{//}   line 02--12:  test of the small scores  \\
\hspace{-7mm}03                  \> $S = S + s_i$                       \\
\hspace{-7mm}04                  \> \textbf{if} \= $S == i(i - 1)$ \\ 
\hspace{-7mm}05                  \>                \> \textbf{if} \= $s_1 < i - 1 \vee  s_i > i - 1$   \\
\hspace{-7mm}06                  \>                \>                \> \textbf{return} $L$   \\
\hspace{-7mm}07                  \>                \> \textbf{if} \> $s_{i+1} < 3i$          \\
\hspace{-7mm}08                  \>                \>                \> \textbf{return} $L$   \\
\hspace{-7mm}09 $S = S + s_n$                                                                         \\
\hspace{-7mm}10 \textbf{if} \= $S == n(n - 1)$ \\
\hspace{-7mm}11                \>   \textbf{if} \= $s_1 < n - 1$   \\
\hspace{-7mm}12                \>                  \>  \textbf{return} $L$                                              \\ 
\hspace{-7mm}13 \textbf{for} \= $i = n$ \textbf{downto} $2$                     \` \textbf{//}   line 13--25:  test of the large scores  \\   
\hspace{-7mm}14                  \> $Q = Q + s_i$                                                                                          \\
\hspace{-7mm}15                  \> \textbf{if} \= $s_{i - 1}  > 3(n - i - 1)$                  \\  
\hspace{-7mm}16                  \>                \> \textbf{return} $L$                                \\ 
\hspace{-7mm}17                  \> \textbf{if} \= $s_i - 3 \lfloor s_i/3 \rfloor > 0$   \\
\hspace{-7mm}18                  \>                \> $B = 1$                                                              \\
\hspace{-7mm}19                  \> \textbf{if} \= $B == 1$   \\
\hspace{-7mm}20                  \>                \> \textbf{return} $L$                                 \\  
\hspace{-7mm}21 $Q = Q + s_1$                               \\                                 
\hspace{-7mm}22 \textbf{if} \= $s_1 - \lfloor s_1/3 \rfloor > 0$     \\
\hspace{-7mm}23                \>                \> $B = 1$                 \\
\hspace{-7mm}24                \>                \> \textbf{if} \= $B == 1$   \\
\hspace{-7mm}25                \>                \>                \> \textbf{return} $L$   \\
\hspace{-7mm}26  $L = 2$                                                                                     \` \textbf{//} line 26--27: $s$ is undecided       \\
\hspace{-7mm}27  \textbf{return} $L$                                                                             
\end{tabbing}

Even this simple filtering algorithm finds a football sequence: if the condition of line 11 does not hold then the sum of all scores is minimal therefore all matches ended with draw. For the 
sake of the simplicity of the program we left this sequence undecided.

%%%%%%%%%%%%%%%%%%%%%%%%%%%%%%%%%%%%%%%%%%%%%%%%%%%%%%%%%%%%%%%%%%%%%
\subsubsection{Linear filtering algorithm L4 = \sc{Reduction1} \label{subsub-L4}}
The fourth linear test is based on the observations that if the sum of the $i$ smallest scores is $i(i -1) + 1$ then either zero or one match among the first $i$  teams ended with a win and 
if the sum of the $i$ largest scores has near the maximal $3i(n - i) + 3i(i - 1)/2$ value then among the $i$ maximal scores $i - 2$ are multiples of 3 and 2 give 1 as remainder mod $3$.  

\begin{lemma} If \label{lemma-reduction1} $n \geq 3$,  $f = (f_1, \ldots, f_n)$ is a football sequence,  $1 \leq k \leq n$ then   

1) if 

\vspace*{-12pt}\begin{equation}
\sum _{i=1}^k f_i = k(k -1) + 1
\end{equation}

\vspace*{-6pt}
\noindent then 

a) either  $f _1 = \cdots =  f_{k- 1} = k - 1$, $f_k = k$, and if $k + 1 \leq n$, and $f_{k+1} \geq 3k - 2$; 

b) or $f_1 = k - 2$, $f_2, \ldots, f_{k-1} = k - 1$, $f_k = k +1$, and $f_{k + 1} \geq 3k$;

2) if 

\vspace*{-12pt}\begin{equation} 
\sum _{i = 1}^k f_{n - i + 1} = 3k(n - k) + 3 \binom{k}{2} - 1
\end{equation} 

\vspace*{-6pt}
\noindent then 

a) $\sum _{i=1, \ f_i - 3 \lfloor f_i/3 \rfloor = 0}^k   1= k - 2$; 

b) $\sum _{i=1, \ f_i -3 \lfloor f_i/3 \rfloor = 1} ^k  1 = 2$;

c) $\sum _{i = 1, \ f_i - 3\lfloor f_i/3 \rfloor = 2} ^k 1  = 0$;

d) if $n - k > 0$ then $f_{n - k} \leq 3(n - k -1)$.
\end{lemma}

\vspace*{-6pt}
\begin{proof} 1) If $f_1 + \cdots + f_k = k(k - 1) + 1$ then either all matches among T$_1$, \ldots, T$_k$ ended with a draw and these teams lost all but one matches 
against the remaining teams and T$_k$ made a draw with one of the teams T$_k$ implying assertions a) or T$_k$ won against T$_1$, the remaining matches among T$_1$, \ldots, T$_k$ 
ended with a draw and the teams T$_{k+1}$, \ldots,  T$_n$ has no draw and won all matches against the first $n - k$ teams implying assertions b).

2) In case 2) of the lemma the teams T$_{k +1}$,  \ldots,  T$_n$ won all matches against the first $n - k$ teams, and made exactly one draw.
\end{proof}

Parameters of \textsc{Reduction1} are the usual ones, further $S$: the current sum of the first $i$ scores; 
$Q$: the current sum of the last $i$ scores; $L_1$ and $L_2$: logical variables; $B$ is the number of scores giving remainder 1 mod $3$; $C$ is the number of scores giving remainder 0 mod $3$.

\medskip
\noindent \textsc{Reduction1}$(n, s)$
\vspace*{-2mm}
\begin{tabbing}
199 \= x\=x\=x\=x\=x\=x\=x\=x \+ \kill
\hspace{-7mm}01 $L = B = C = S = Q = 0$                                                              \` \textbf{//}   line 01: initialization of $L$, $B$, $C$, $S$, and $Q$  \\
\hspace{-7mm}02 \textbf{for} \= $i = 1$ \textbf{to} $n - 1$                     \` \textbf{//}   line 02--12:  test of the small scores  \\
\hspace{-7mm}03                  \> $S = S + s_i$                       \\
\hspace{-7mm}04                  \> \textbf{if} \= $S == i(i - 1) + 1$ \\
\hspace{-7mm}05                  \>                \> $L_1 =  (s_1 == i  - 1) \wedge (s_{i-1} == i-1) \wedge (s_i == i) \wedge (s_{i+1} \geq 3i - 2)$          \\
\hspace{-7mm}06                  \>                \> $L_2 =  (s_1 == i - 2) \wedge (s_2 == i - 1) \wedge (s_{i-1} == i - 1) $   \\
                                            \>                \> \quad \;\;\;\;\;         $\wedge \;(s_{i} == i + 1) \wedge (s_{i+1} \geq 3i)$ \\ 
\hspace{-7mm}07                  \>                \> \textbf{if} \= $(L_1 == \textsc{false}) \wedge (L_2 == \textsc{false}) ==  \textsc{true}$   \\
\hspace{-7mm}08 \> \> \> \textbf{return} $L$   \` \textbf{//}    line 07--08: $s$ is not good \\
\hspace{-7mm}09 $S = S + s_n$                                           \\
\hspace{-7mm}10 \textbf{if} \= $S == n(n - 1) + 1$ \\
\hspace{-7mm}11                \> \textbf{if} \= $(s_1 < n - 2) \wedge (s_2 == n - 1) \wedge (s_{n-1} == n -1) \wedge (s_n == n + 1)$ \\
     \> \>   \;\;\;\;\;\;    $== \textsc{false}$    \\
\hspace{-7mm}12                \>                \> \textbf{return} $L$ \\                
\hspace{-7mm}13 \textbf{for} \= $i = n$ \textbf{downto} $2$                     \` \textbf{//}   line 13--35:  test of the large scores  \\   
\hspace{-7mm}14                  \> $Q = Q + s_i$                                                  \\
\hspace{-7mm}15                  \> \textbf{if} \= $s_i - 3\lfloor{s_i/3} \rfloor == 2$     \\
\hspace{-7mm}16                  \>                \> \textbf{return} $L$    \\
\hspace{-7mm}17                  \> \textbf{if} \= $s_i - 3\lfloor{s_i/3} \rfloor == 1$     \\
\hspace{-7mm}18                  \>                \> $B = B + 1$                                    \\
\hspace{-7mm}19                  \> \textbf{if} \= $s_i - 3\lfloor s_i/3 \rfloor > 0$   \\
\hspace{-7mm}20                  \>                \> $C = C + 1$                                  \\
\hspace{-7mm}21                  \> \textbf{if} \= $Q == 3(n - i)i + 3i(i - 1)/2 - 1)$                   \\
\hspace{-7mm}22                  \>                \> \textbf{if} \= $s_{n - i} > 3(n - i - 1)$                \\
\hspace{-7mm}23                  \>                \>                \> \textbf{return} $L$                        \\ 
\hspace{-7mm}24                  \>                \> \textbf{if} \= $(B == 2) \wedge (C == i - 2) == \textsc{false}$     \\
\hspace{-7mm}25                  \>                \>                \> \textbf{return} $L$                        \\ 
\hspace{-7mm}26 $Q = Q + s_1$                                           \\ 
\hspace{-7mm}27 \textbf{if} \= $s_i - 3\lfloor{s_i/3} \rfloor == 2$     \\
\hspace{-7mm}28                \> \textbf{return} $L$    \\
\hspace{-7mm}29 \textbf{if} \= $s_i - 3\lfloor{s_i/3} \rfloor == 1$     \\
\hspace{-7mm}30                \> $B = B + 1$                                    \\
\hspace{-7mm}31 \textbf{if} \= $s_i - 3\lfloor s_i/3 \rfloor > 0$   \\
\hspace{-7mm}32                \> $C = C + 1$                                 \\ 
\hspace{-7mm}33 \textbf{if} \= $Q == 3n(n - 1)/2 - 1$                   \\
\hspace{-7mm}34                \> \textbf{if} \= $(B == 2) \wedge (C == i - 2) == \textsc{false}$     \\
\hspace{-7mm}35                \>                \> \textbf{return} $L$          \\ 
\hspace{-7mm}36  $L = 2$    \` \textbf{//} line 36--37: $s$ is undecided       \\
\hspace{-7mm}37  \textbf{return} $L$                                                                             
\end{tabbing}

%%%%%%%%%%%%%%%%%%%%%%%%%%%%%%%%%%%%%%%%%%%%%%%%%%%%%%%%%%%%%%%%%%%%%%%%%%
\subsubsection{Linear filtering algorithm L5 = \sc{Draw-Unique} \label{subsub-L5}}
A \textit{draw sequence} $d(s) = (d_1,\ldots,d_n)$ belonging to a $(0,3(n-1),n)$-regular sequence $s$ accepted by L4 is defined as a sequence of nonnegative 
integers having the following properties for $i = 1,\ldots,n$:
\begin{enumerate}\addtolength{\itemsep}{-0.5\baselineskip} 
\item $0 \leq d_i \leq 2$;
\item $d_i = s_i \mod \ 3$;
\item $d_i \leq \min(s_i,n-1)$;
\item $d_i + 3(n - 1 - d_i) \geq s_i$,
\end{enumerate}  
further
\begin{equation}
\sum_{i=1}^n d_i = 2\left ( 3\binom{n}{2} - \sum_{i=1}^n s_i \right ).
\end{equation}

A draw sequence $d = (d_1,\ldots,d_n)$ is called $(0,1,n)$\textit{-graphic}  (or simply graphic or good), if there exists a $(0,1,n)$-graph whose degree sequence is $d$.  

The fifth linear filtering algorithm is based on the following assertion. 

\begin{lemma} If a $(2,3,n)$-regular sequence $s$ has only a unique draw sequence $d(s)$ which is not graphical then $s$ is not football sequence. 
\end{lemma} 
\begin{proof} Since the football sequences have at least one graphical draw sequence, the regular sequences without graphical draw sequence are not football sequences. 
\end{proof}  

Basic parameters of \textsc{Draw-Unique} are the usual ones, further $S$: ithe current sum of the elements of $s$; 
$R$: the number of obligatory draws; 
$Dn$: the number of the draws in the investigated potential tournament;  
$d = (d_1, \ldots, d_n)$: $d_i$ is the number of draws allocated to T$_i$; 
$r = (r_1,\ldots,r_n)$: $r_i$ is the remainder of $s_i$  mod 3;  
$y$: is the current number of allocated draws;    
$x$: is the current maximal number of draw packets acceptable by T$_i$.

\bigskip
\noindent \textsc{Draw-Unique}$(n, s)$
\vspace{-2mm}
\begin{tabbing}
199 \= x\=x\=x\=x\=x\=x\=x\=x \+ \kill
\hspace{-7mm}01 $S = R = L = x = y = 0$                                                               \` \textbf{//}    line 01: initialization of $S$, $R$, $L$, $x$ and $y$ \\
\hspace{-7mm}02 \textbf{for} \= $i = 1$ \textbf{to} $n$                     \` \textbf{//}   line 02--03:  computation of $S$  \\
\hspace{-7mm}03                  \> $S = S + s_i$                                  \\
\hspace{-7mm}04 $Dn =  3 \binom{n}{2} - S$                                    \` \textbf{//}   line 04:  computation of $Dn$  \\                                   
\hspace{-7mm}05 \textbf{for} \= $i = 1$ \textbf{to} $n$                     \` \textbf{//}   line 05--17:  allocation of draws   \\
\hspace{-7mm}06                  \> $r_i = s_i - 3 \lfloor\frac{s_i}{3} \rfloor$     \\ 
\hspace{-7mm}07                  \> $R = R + r_i$                                              \\
\hspace{-7mm}08                  \> $x = \min \left ( \frac{s_i - r_i}{3}, \lfloor \frac{n - 1 - r_i}{3} \rfloor, 
               \lfloor \frac{3(n - 1) - 2r_i - s_i}{6} \rfloor \right )$        \\
\hspace{-7mm}09                  \> $d_i = r_i + 3x$                                                     \\
\hspace{-7mm}10                  \> $y = y + d_i$                                                       \\
\hspace{-7mm}11 \textbf{if} \= $R > 2Dn$                                                    \\
\hspace{-7mm}12                \> \textbf{return} $L, \ d$                           \\
\hspace{-7mm}13 \textbf{if} \= $y < 2Dn$                                              \\ 
\hspace{-7mm}14             \> \textbf{return} $L, d$                             \\
\hspace{-7mm}15 \textbf{if} \> $y \geq 2Dn$                                                       \\
\hspace{-7mm}16                \> $L = 2$                                                \\
\hspace{-7mm}17                \> \textbf{return} $L, \ d$                                          \\
\hspace{-7mm}18 sort $d$ in decreasing order by \textsc{Counting-Sort} resulting $d'$           \\ 
\hspace{-7mm}19 HHL$(d') $                                                      \\
\hspace{-7mm}20  \textbf{return} $L, \ d$                       \` \textbf{//} line 20: $s$ is undecided
\end{tabbing}

Procedure HHL (\textsc{Havel-Hakimi-Linear}) is described in \cite{Ivanyi2012Comp}. We remark that the original Havel-Hakimi algorithm requires in worst 
case $\Theta(n^2)$ time. Recently Király \cite{Kiraly2012} published a version which uses the data structure proposed by van Emde Boas \cite{Kiraly2012DS,Emde1975} and 
requires $O(n \log \log n)$ time. Our algorithm is linear and works also for some multigraphs.  

A natural requirement is $d_i \leq n - 1$ but $d_i > n - 1$ can occur only in the cases $s = (0,2)$ and $s = (1,2)$ which are filtered by the constant time algorithms.

We get a stronger filtering algorithm \textsc{Draw-Sorted-Unique} using the definition of the uniqueness of the sorted draw sequence. For example in the case of the sequence $s = (3,3,3,5)$  
we have three possibilities to allocate two draw packets but only the teams having 3 points can accept a packet therefore we get in each case 
the bad draw sequence $(3,3)$. 

We remark that the problem of unicity of graphs determined in a unique way by their degree sequences was studied for some graph classes (see e.g. the papers of Tetali \cite{Tetali1998},  Tyskevich \cite{Tyskevich2000}, and Barrus \cite{Barrus2012}).

%%%%%%%%%%%%%%%%%%%%%%%%%%%%%%%%%%%%%%%%%%%%%%%%%%%%%%%%%%%%%%%%%%%%%%%%%%
\subsubsection{Linear filtering algorithm L6 = \sc{Balanced-Lin} \label{subsub-L6}}
The sixth linear filtering algorithm L6 = \textsc{Balanced-Lin} is based on the observation that if the draw sequence is unique, then the victory sequence $w = (w_1,\ldots,w_n)$ and 
the loss sequence $l = (l_1,\ldots,l_n)$ are also unique. The following assertion gives a necessary condition for the reconstructability of the sequence 
pair $(w,l)$. 

\begin{lemma} \emph{(Lucz \cite{Lucz2012})} If $n \geq 2$,  $w = (w_1, \ldots, w_n)$ is the win sequence and $l = (l_1, \ldots, l_n)$ is the loss sequence 
of a football sequence $f = (f_1, \ldots, f_n)$ with $\sum _{i=1}^n f_i > n(n - 1)$ then let  
\begin{equation}
w_i = \max _{1 \leq j \leq n} w_j   \quad   \textit{and} \quad     l_j = \max _ {1 \leq i \leq n} l_i.
\end{equation}

In this case
\begin{equation}
w_i \leq \sum _{j=1}^{i-1} \left \lceil \frac{l_j}{n - 1} \right \rceil + \sum _{j=i+1}^n \left \lceil \frac{l_j}{n - 1} \right \rceil
\end{equation}  
and
\begin{equation}
l_j \leq \sum _{i=1}^{j-1} \left \lceil  \frac{w_i}{n - 1} \right \rceil  + \sum _{i=j+1}^n \left \lceil  \frac{w_i}{n - 1} \right \rceil 
\end{equation}
for $n = 1, \ldots, n$.
\end{lemma}

\begin{proof}  The wins (losses) of the team T$_i$ (T$_j$) having the maximal number of wins (losses) can be paired with losses (wins) only if there are at least $w_i$ ($l_j$) teams 
having at least one loss (win).   
\end{proof}

%%%%%%%%%%%%%%%%%%%%%%%%%%%%%%%%%%%%%%%%%%%%%%%%%%%%%%%%%%%%%%%%%%%%%%%%%%
\subsubsection{Linear filtering algorithm L7 = \sc{Sport-Uniform} \label{subsub-L7}}
The seventh linear filtering algorithm L7 = \textsc{Sport-Uniform} is connected with a popular concept called  in the world of sport \textit{sport matrix}. 
It is an $n \times 5$ sized matrix containing the basic data of the teams  
of a tournament. We use the following formal definition of \textit{sport matrix} for $n$ teams.

\begin{definition} Let\label{def-sportmat} $n \geq 1$ and $s = (s_1, \ldots, s_n)$ be a $(0,3(n - 1),n)$-regular sequence.  
Then the sport matrices $S(s)$ corresponding to $s$  are defined by the following properties:
\begin{enumerate}
\item the size of the matrix is $n \times 5$, its elements are nonnegative integers;
\item $w_i + d_i + l_i = n - 1$ for $i =1, \ldots, n$;
\item $3w_i + d_i = s_i$  for $i =1, \ldots, n$;
\item $\sum _{i=1}^n w_i = \sum _{i=1}^n l_i = \sum _{i = 1}^n s_i - n(n - 1)$;
\item $\sum _{i=1}^n d_i = 2 \left ( 3 \binom{n}{2} - \sum _{i=1}^n s_i \right )$. \hfill 
\end{enumerate}
\end{definition} 

We remark that the $i$th row of the sport matrices contains data of T$_i$  for $i =1, \ldots, n$: index $i$, number of wins $w_i$, number of draws $d_i$, number of losses $l_i$ 
and number of points $s_i$ ($w_i$, $d_i$ and $l_i$ are estimated values). These formal requirements are only \textit{necessary} for $\mathcal{S}$ to contain the basic characteristics of some football tournament.

A sequence $s = (s_1,\ldots,s_n)$ is called \textit{sport sequence} if there exists at least one sport matrix corresponding to $s$.

Another useful concept is the \textit{obligatory sport matrix} belonging to given regular sequence $s$. 

\begin{definition} Let $n \geq 1$ be\label{def-oblsport} and $s = (s_1, \ldots, s_n)$ be a $(0,3(n - 1),n)$-regular sequence.  Then the obligatory sport matrix $\mathcal{O}(s)$ corresponding 
to $s$  is defined by the following properties:
\begin{enumerate}
\item  the size of the matrix is $n \times 5$, its elements are nonnegative integers;
\item $wo_i = \max \left (0, \lceil \frac{s_i -  (n - 1)}{2} \rceil \right )$ for  $i =1, \ldots, n$;
\item $do_i = s_i - 3 \lfloor \frac{s_i}{3} \rfloor$  for $i =1, \ldots, n$;
\item $lo_i = \max  (0, n - 1 - s_i)$  for $i =1, \ldots, n$.    \hfill $\square$
\end{enumerate}
\end{definition}

 The $i$-th row of the matrix contains the (partially estimated) data of T$_i$  for $i =1, \ldots, n$: index $i$, number of obligatory wins $wo_i$, number of obligatory draws $do_i$, 
number of obligatory losses $lo_i$ and number of points $s_i$ (the obligatory values are lower bounds for the correct values, the index and the number of points are exact values).

\begin{definition} We say that the obligatory sport matrix $\mathcal{O}(s)$ of $s$ is \textit{extendable to a sport matrix} $\mathcal{S}(s)$ corresponding to $s$ if 
\begin{enumerate}
\item $\mathcal{O}(s)$ is a sport matrix belonging to $s$ or
\item we can increase some $w_i$, $d_i$ and $l_i$ values so that the result will be a sport matrix $\mathcal{S}(s)$.
\end{enumerate} 
\end{definition}

According to the following assertion we get a linear filtering algorithm using the obligatory sport matrix. 

\begin{lemma} The obligatory sport matrix $\mathcal{O}(s)$ belonging to a  $(0,3(n-1),n)$-regular sequence $s$ is unique. If $\mathcal{O}(s)$ is not extendable 
to a sport matrix $\mathcal{S}(s)$ then $s$ is not a football sequence. 
\end{lemma} 
\begin{proof} The obligatory sport matrix is defined by unique formulas therefore it is unique. If $s$ is a football sequence then its obligatory sport matrix $\mathcal{O}(s)$ contains lower bounds for  
$w_i$, $d_i$, and $l_i$ of any sport matrix $\mathcal{S}(s)$ therefore any sport matrix $\mathcal{S}(s)$ can be constructed by the extension of $\mathcal{O}(s)$. 
\end{proof} 

The following \textsc{Sport-Uniform} is a \textit{draw-based} algorithm which at first constructs the obligatory sport matrix belonging to $s$  then tries 
to extend it to a sport matrix so that it allocates the draw packets in a greedy way as uniformly as possible. If the so received draw sequence is not graphic then 
the investigated sequence is not good. 

The base of the uniform allocation of the draws  is the following assertion.

\begin{lemma}  If \label{lemma-sportuni} $n \geq 1$,  $d = (d_1, \ldots, d_n)$ is graphical and $d_i < d_j$ then the sequence $d'$---received  increasing $d_i$ by $1$ and decreasing $d_j$ 
by $1$---is also graphical. 
\end{lemma} \begin{proof} Let $G$ be a $(0,1,n)$-graph on vertices $V_1, \ \ldots, \ V_n$ having the degree sequence $d = (d_1, \ldots,d_n)$ in which $d_i < d_j$. Then there exists a vertex 
$V_k$ which is connected with $V_j$ and not connected with $V_i$. In $G$ delete the edge between $V_j$ and $V_k$ and add the edge between $V_i$ and $V_k$. Then the received 
new graph is graphical with the required degree sequence.
\end{proof} 

This lemma has a useful corollary.

\begin{corollary} If $n \geq 1$, $s = (s_1, \ldots, s_n)$ is a $(2,3,n)$-regular sequence, and its uniform draw sequence $u(s) = (u_1, \ldots, u_n)$ is not graphical, then 
$s$ is not a football sequence.
\end{corollary} 
\begin{proof} By the recursive application of Lemma \ref{lemma-sportuni} we get that if $s$ has a graphical draw sequence then its uniform draw sequence is also graphical.  
\end{proof}

We remark that the problem of the pairing of the draws  has a reach bibliography as the problem of degree sequences of simple graphs 
\cite{Choudum1986,ErdosG1960,Hakimi1962,Havel1955,HellK2009,IvanyiLMS2011Acta,MeierlingV2009,Ozkan2011,Takahashi2007,TripathiT2008,TripathiV2003,TripathiVW2010}.

Basic parameters of \textsc{Sport-Uniform} are the usual ones further 
$S$: the sum of the elements of $s$; 
$S_0$: auxiliary variable; 
$wo = (wo_1,\ldots,wo_n)$: $wo_i$ is the number of obligatory wins of T$_i$; 
$do = (do_1,\ldots,do_n)$: $do_i$ is the number of obligatory draws of T$_i$;   
$lo = (lo_1,\ldots,lo_n)$: $lo_i$ is the number of obligatory losses of T$_i$; 
$WO = (WO_0,\ldots,WO_n)$: $WO_i$ is the total number of wins of the first $i$ teams;   
$DO = (DO_0,\ldots,DO_n)$. $DO_i$ is the total number of draws of the first $i$ teams;
$LO = (LO_0,\ldots,LO_n)$. $LO_i$ is the total number of the first $i$ teams;   
$wm = (wm_1,\ldots,wm_n)$: $wm_i$ is the maximal  number of wins of T$_i$; 
$dm = (dm_1,\ldots,dm_n)$: $dm_i$ is the maximal number of draw packets of T$_i$; 
$lm = (lm_1,\ldots,lm_n)$: $lm_i$ is the maximal  number of losses of T$_i$; 
$Wn$: the number of the wins in the tournament;   
$Ln$: the number of the losses in the tournaments;   
$Dn$: the number of draws in the tournament;
$D$: the current number of yet not allocated draws;   
$da = (da_1, \ldots, da_n)$: $da_i$ is the number of allocated to T$_i$ draw packets;
$wa = (wa_1,\ldots,wa_n)$: $wa_i$ is the number of allocated wins of T$_i$; 
$la = (la_1,\ldots,la_n)$: $la_i$ is the number of allocated losses of T$_i$; 
$R = (R_0,R_1,R_2)$: $R_i$ is the number of elements of $s$ giving remainder $i$ mod $3$;   
$c$: average number of draw packets to allocate for a team.  

\bigskip
\noindent \textsc{Sport-Uniform}$(n, s)$  
\vspace{-2mm}
\begin{tabbing}
199 \= x\=x\=x\=x\=x\=x\=x\=x \+ \kill
\hspace{-7mm}01 $S_0 = WO_0 = DO_0 = LO_0 = R_0 = R_1 = R_2 = L = 0$                                                               \` \textbf{//}    line 01: initialization \\
\hspace{-7mm}02 \textbf{for} \= $i = 1$ \textbf{to} $n$                     \` \textbf{//}   line 02--03:  computation of the parameters  \\
\hspace{-7mm}03                  \> $S = S + s_i$                                  \\
\hspace{-7mm}04                  \> $wo_i =  \max \left (0, \lceil \frac{s_i -  (n - 1)}{2} \rceil \right )$    \\
\hspace{-7mm}05                  \> $WO_i = WO_{i-1} + wo_i$   \\
\hspace{-7mm}06                  \> $do_i =  s_i - 3 \lfloor \frac{s_i}{3} \rfloor$    \\ 
\hspace{-7mm}07                  \> $DO_i = DO_{i-1} + do_i$    \\
\hspace{-7mm}08                  \> $R_{do_i} = R_{do_i} + 1$   \\
\hspace{-7mm}09                  \> $lo_i =  \max (n - 1 - s_i,0)$  \\
\hspace{-7mm}10                  \> $LO_i = LO_{i-1} + lo_i$    \\
\hspace{-7mm}11                  \> $dm_i = \min \left (\frac{s_i - do_i}{3},n - 1 - do_i,\lfloor \frac{3(n - 1) -2do_i - s_i}{6} \rfloor \right )$   \\
\hspace{-7mm}12                  \> $wm_i = \frac{s_i - do_i}{3}$ \\
\hspace{-7mm}13                  \> $lm_i = \left \lfloor \frac{3(n - 1) - s_i}{3} \right \rfloor$   \\  
\hspace{-7mm}14 $Wn = Ln = S_n - n(n - 1)$                                               \` \textbf{//}   line 14:  computation of $Wn$, $Ln$  \\                \hspace{-7mm}15 $Dn = D = 3n(n -1)/2 - S$                                                 \` \textbf{//}   line 15:  computation of $Dn$, $D$   \\
\hspace{-7mm}16 \textbf{if} \= $\frac{D - DO_n}{3} > \left \lfloor \frac{D - DO_n}{3}  \right \rfloor$         \` \textbf{//}   line 16--43:  allocation of draw packets  \\                                           
\hspace{-7mm}17                \> \textbf{return} $L$   \\ 
\hspace{-7mm}18 \textbf{while} \= $D > 0$   \\
\hspace{-7mm}19                      \> $c = \left  \lfloor \frac{D}{R_0 + R_1 + R_2}  \right \rfloor$                \\
\hspace{-7mm}20                      \> \textbf{while} \= $c \geq 1$   \\
\hspace{-7mm}21                      \>                      \> $R_0 = R_1 = R_2 = 0$   \\
\hspace{-7mm}22                      \>                      \> \textbf{for} \= $i = 1$ \textbf{to} $n$   \\ 
\hspace{-7mm}23                      \>                      \>      \> $da_i = \min(\frac{s_i - d_i}{3}, c)$   \\
\hspace{-7mm}24                      \>                      \>       \> $d_i = do_i + 3da_i$   \\
\hspace{-7mm}25                      \>                      \>  \> $D = D - 3da_i$   \\
\hspace{-7mm}26                      \>                      \>                  \> \textbf{if} \= $d_i < dm_i$  \\ 
\hspace{-7mm}27                      \>                      \>                  \>                \> $R_{do_i} = R_{do_i} + 1$   \\
\hspace{-7mm}28                      \>                      \> $c = \left  \lfloor \frac{D}{R_0 + R_1 + R_2}  \right \rfloor$   \\
\hspace{-7mm}29                          \> \textbf{if} \= $0 < \frac{D}{3} \leq R_0$  \\ 
\hspace{-7mm}30                      \>                \> \textbf{for} \= $i = 1$ \textbf{to} $n$   \\ 
\hspace{-7mm}31                      \>                \>                  \> \textbf{if} \= ($D > 0 \wedge do_i == 0$) == \textsc{true}         \\
\hspace{-7mm}32                      \>                \>                  \>                \> $d_i = d_i + 3$   \\
\hspace{-7mm}33                      \>                \>                  \>                \> $D = D - 3$     \\                
\hspace{-7mm}34                      \> \textbf{if} \= $R_0 < \frac{D}{3} \leq R_0 + R_1$  \\ 
\hspace{-7mm}35                      \>                \> \textbf{for} \= $i = 1$ \textbf{to} $n$   \\ 
\hspace{-7mm}36                      \>                \>                  \> \textbf{if} \= ($D > 0 \wedge do_i == 0 \vee do_i == 1$) == \textsc{true}   \\
\hspace{-7mm}37                      \>                \>                  \>                \> $d_i = d_i + 3$   \\
\hspace{-7mm}38                      \>                \>                  \>                \> $D = D - 3$     \\ 
\hspace{-7mm}39                      \> \textbf{if} \= $R_0 + R_1 > \frac{D}{3}$  \\ 
\hspace{-7mm}40                      \>                \> \textbf{for} \= $i = 1$ \textbf{to} $n$   \\ 
\hspace{-7mm}41                      \>                \>                  \> \textbf{if} \= $D > 0$   \\
\hspace{-7mm}42                      \>                \>                  \>                \> $d_i = d_i + 3$   \\
\hspace{-7mm}43                      \>                \>                  \>                \> $D = D - 3$     \\              
\hspace{-7mm}44 sort $d$ in decreasing order resulting $d'$      \\ 
\hspace{-7mm}45 HHL$(d')$     \` \textbf{//} line 44--45: sorting of the draw sequence                \\
\hspace{-7mm}46 \textbf{return} $L$, $d$                                              \` \textbf{//} line 46: $s$ is undecided (if $L = 2$) or bad (if $L = 0$)
\end{tabbing}

%%%%%%%%%%%%%%%%%%%%%%%%%%%%%%%%%%%%%%%%%%%%%%%%%%%%%%%%%%%%%%%%%%%%%%%%%%
%%%%%%%%%%%%%%%%%%%%%%%%%%%%%%%%%%%%%%%%%%%%%%%%%%%%%%%%%%%%%%%%%%%%%%%%%%
\subsubsection{Linear filtering algorithm L8 = \sc{Draw-Sorted-Unique} \label{subsub-L8}}
The fifth linear filtering algorithm \textsc{Draw-Unique} exploits the fact that some football sequences have unique sport matrix implying the uniqueness of the draw sequence. 
The eighth linear algorithm L8 = \textsc{Draw-Sorted-Unique} exploits that the uniqueness of the sport matrix is not necessary to have a unique sorted draw sequence. 

\textit{Sorted version} of a sport matrix $\mathcal{S}(s)$ is denoted by $\overline{\mathcal{S}}(s)$ and is defined by the following property: if $1 \leq i < j \leq n$ then either $d'_i < d'_j$ or $d'_i = d'_j$ and 
$w'_i < w'_j$ or $d'_i = d'_j$ and $w'_i = w'_j$ and $i' < j'$ ($d'_i$ is the draw value in the $i$-th row of the sorted matrix and $i'$ is the original index belonging to $d'_i$). 

\textsc{Draw-Sorted-Unique} is based on the following assertion.

\begin{lemma} If $n \geq 1$, $s = (s_1,\ldots,s_n)$ is a $(2,3,n)$-regular sequence, the sorted versions of the sport matrices $\mathcal{S}(s)$ are identical and their joint draw sequence is not graphical, then $s$ is not a football sequence.    
\end{lemma} 
%\begin{proof} 
%\end{proof}

Basic parameters of \textsc{Draw-Sorted-Unique} are the usual ones, further $S$: the current sum of the elements of $s$; 
$D$: the number of the draws in the investigated potential tournament;  
$d = (d_1, \ldots, d_n)$: $d_i$ is the number of draws allocated to T$_i$; 
$do = (do_1,\ldots,do_n)$: $do_i$ is the number of obligatory draws of T$_i$; 
DO: the number of the obligatory draws in the tournament; 
$lo = (lo_1,\ldots,lo_n)$: $lo_i$ is the number of obligatory losses of T$_i$;
$LO$ is the number of obligatory losses in the tournament; 
$wo = (wo_1,\ldots,wo_n)$: $wo_i$ is the number of obligatory wins of T$_i$; 
$WO$ is the number of obligatory wins in the tournament;  
$dm = (dm_1,\ldots,dm_n)$: $dm_i$ is the maximal number of draw packets which can be accepted by $T_i$;
$DM$: the sum of the $dm_i$'s; 
$lm = (lm_1,\ldots,lm_n)$: $lm_i$ is of the maximal  number of losses of T$_i$; 
$LM$: the sum of the $lm_i$'s; 
$wm = (wm_1, \ldots,wm_n)$: $wm_i$ is the maximal number of wins of T$_i$; 
$WM$: the sum of the $wm_i$'s; 
$Wn$: the number of the wins in the tournament;   
$Ln$: the number of the losses in the tournaments;   
$Dn$: the number of draws in the tournament; 
$D$: the number of yet not allocated draws;   
$da = (da_1, \ldots, da_n)$: $da_i$ is the number of allocated to T$_i$ draw packets;
$wa = (wa_1,\ldots,wa_n)$: $w_i$ is the number of allocated to T$_i$ wins; 
$la = (la_1,\ldots,la_n)$: $la_i$ is the number of allocated to T$_i$ losses;    
$h$: the maximal number of draw packets assigned to a team;
$R = R_{i,j}$: a $3 \times h$ sized matrix, where $R_{i,j}$ gives the number of teams which are able at most $i$ draw packets and 
having score of form $3k + j$;
$A = (A_0,A_1,A_2)$: $A_j$ is the number of  scores giving $i$ mod $(3$; 
$B = (B_0,\ldots,B_h)$: $B_i$ is the number of teams which are able to accept at most $i$ draw packets; 
$z$: number of draw pockets which the program tries to allocate to all teams;
$fs$: first score among  the scores receiving maximal number of draw pockets; 
$Rm$:   critical value of the remainder (mod 3) of the scores.

\bigskip
\noindent \textsc{Draw-Sorted-Unique}$(n, s)$
\vspace{-2mm}
\begin{tabbing}
199 \= x\=x\=x\=x\=x\=x\=x\=x \+ \kill
\hspace{-7mm}01  $S = WO = DO = LO = A_0 = A_1 = A_2 = L = 0$                                                               \` \textbf{//}    line 01: initialization \\
\hspace{-7mm}02  \textbf{for} \= $i = 1$ \textbf{to} $n$                                         \` \textbf{//}   line 02--29: test of the obligatory sport matrix             \\            
\hspace{-7mm}03                   \> $S = S + s_i$                                                       \\
\hspace{-7mm}04                   \> $do_i = s_i - 3 \lfloor s_i/3 \rfloor$                     \\
\hspace{-7mm}05                   \> $DO = DO + do_i$                                                       \\
\hspace{-7mm}06                   \> $wo_i =  \max(0,\lceil \frac{s_i - (n - 1)}{2} \rceil$                    \\ 
\hspace{-7mm}07                   \> $WO = WO + wo_i$                                                         \\
\hspace{-7mm}08                   \> $lo_i = \max(0, n - 1 -  s_i)$                                                       \\ 
\hspace{-7mm}09                   \> $LO = LO + lo_i $                                                \\
\hspace{-7mm}10                   \> $dm_i = \min \left (\frac{s_i - do_i}{3},\frac{n - 1 - do_i}{3},\frac{3(n - 1) - 2d_i - s_i}{6} \right )$           \\
\hspace{-7mm}11                   \> $DM = DM + dm_i$                                                                  \\
\hspace{-7mm}12                   \> $wm_i = \min \left ( \frac{s_i - do_i}{3},(N - 1) - DO_i  \right )$                         \\
\hspace{-7mm}13                   \> $WM = WM + wm_i$                                           \\
\hspace{-7mm}14                   \> $lm_i = \min \left (\lfloor\frac{3(n - 1) - s_i}{3} \rfloor, n - 1 - do_i \right )$                    \\
\hspace{-7mm}15                  \> $LM = LM + lm_i$                                           \\
\hspace{-7mm}16 $Dn = 3 \binom{n}{2} - S$                                                      \\
\hspace{-7mm}17 $Wn = Ln = S - 2\binom{n}{2}$                                     \\
\hspace{-7mm}18 \textbf{if} \= $DO > 2Dn$                                    \\
\hspace{-7mm}19                \> \textbf{return} $L$                         \\ 
\hspace{-7mm}20 \textbf{if} \= $3DM < 2Dn$                                    \\
\hspace{-7mm}21                \> \textbf{return} $L$                         \\
\hspace{-7mm}22 \textbf{if} \= $WO > Wn$                                    \\
\hspace{-7mm}23                \> \textbf{return} $L$                         \\
\hspace{-7mm}24 \textbf{if} \= $WM < Wn$                                    \\
\hspace{-7mm}25                \> \textbf{return} $L$                         \\
\hspace{-7mm}26 \textbf{if} \= $LO > Ln$                                    \\
\hspace{-7mm}27                \> \textbf{return} $L$                         \\
\hspace{-7mm}28 \textbf{if} \= $LM < Ln$                                    \\
\hspace{-7mm}29                \> \textbf{return} $L$                         \\
\hspace{-7mm}30 $h = \lfloor (n - 1)/3 \rfloor$                                      \` \textbf{//}    line 30--45: preparation of the allocation                         \\ 
\hspace{-7mm}31 \textbf{for} \= $i = 0$ \textbf{to} $h$                                                  \\ 
\hspace{-7mm}32                  \> $B_i = 0$                                       \\
\hspace{-7mm}33                  \> \textbf{for} \= $j = 0$ \textbf{to} $2$                                                  \\ 
\hspace{-7mm}34                  \>                  \> $R_{j,i} = 0$                                  \\
\hspace{-7mm}35 \textbf{for} \= $i = 1$ \textbf{to} $n$                                                   \\
\hspace{-7mm}36                  \> $R_{do_i,dm_i} = R_{do_i,dm_i} + 1$                   \\ 
\hspace{-7mm}37 \textbf{for} \= $i = 1$ \textbf{to} $h$                                                  \\ 
\hspace{-7mm}38                  \> \textbf{for} \= $j = 0$ \textbf{to} $2$                                                  \\ 
\hspace{-7mm}39                  \>                  \> $A_j = A_j + R_{i,j}$                                  \\
\hspace{-7mm}40                  \>                  \> $B_i = B_i + R_{i,j}$                                                      \\
\hspace{-7mm}41 $q = 0$                                                                                                         \\
\hspace{-7mm}42 $A = A_0 + A_1 + A_2$                                                                                  \\ 
\hspace{-7mm}43 $D = 2Dn - DO$                                                      \\
\hspace{-7mm}44 $c =  \lfloor \frac{D}{A} \rfloor $                                       \\ 
\hspace{-7mm}45 $q = q + c$                                                            \\
\hspace{-7mm}46 \textbf{while} \=  $c \geq 1$                         \` \textbf{//}       line 46--78: allocation of the draws                          \\ 
\hspace{-7mm}47                      \> $z = 0$                                                        \\ 
\hspace{-7mm}48                      \> \textbf{for} \= $i = q - c + 1$ \textbf{to} $q$                \\
\hspace{-7mm}49                      \>               \> $z = z + iB_i$                         \\
\hspace{-7mm}50                      \>               \> $D = D - 3z$                                                 \\ 
\hspace{-7mm}51                      \>               \> \textbf{for} \= $i = 0$ \textbf{to} $2$                \\
\hspace{-7mm}52                      \>               \>                   \> \textbf{for} \= $j = q - c + 1$ \textbf{to} $q$                \\
\hspace{-7mm}53                      \>               \>                   \>                  \>$A_i = A_i - R_{i,j}$                \\
\hspace{-7mm}54                      \>               \>  $A = A_0 + A_1 + A_2$       \\
\hspace{-7mm}55                      \>               \> $c =  \lfloor \frac{D}{A} \rfloor $                                       \\                    
\hspace{-7mm}56 \textbf{if} \= $q > 0$                                                                                       \\                      
\hspace{-7mm}57                \> \textbf{for} \=  $i = 1$ \textbf{to} $n$                                     \\
\hspace{-7mm}58                \>                  \> $d_i = d_i - 3 \min(z,dm_i)$                              \\
\hspace{-7mm}59 \textbf{if} \= $D == 0$                                      \\
\hspace{-7mm}60                \> \textbf{go to} 79                  \\
\hspace{-7mm}61 $fs = -1$                                                \\
\hspace{-7mm}62 $Rm = 2$                                                \\
\hspace{-7mm}63 \textbf{if} \= $D \leq A_1 + A2$                      \\ 
\hspace{-7mm}64                \> $Rm = 1$                             \\ 
\hspace{-7mm}65 \textbf{if} \= $D \leq A_1$                      \\ 
\hspace{-7mm}66                \> $Rm = 0$                             \\ 
\hspace{-7mm}67   \textbf{for} \= $i = 1$ \textbf{to} $n$               \\
\hspace{-7mm}68                    \> \textbf{if} \= $(dm_i > q) \wedge (do_i \leq Rm)$ == \textsc{true}                \\
\hspace{-7mm}69                    \>                 \> \textbf{if} \= $fs == -1$      \\ 
\hspace{-7mm}70                    \>                 \>                \> $fs = s_i$                        \\
\hspace{-7mm}71                    \>                 \> \textbf{else} \textbf{if} \=  $s_i \neq  fs$                 \\
\hspace{-7mm}72                    \>                 \>                                  \> \textbf{return} $L$, $d$    \\
\hspace{-7mm}73                    \>                 \> \textbf{if} \= $do_i < Rm$     \\
\hspace{-7mm}74                    \>                 \>                \> $d_i = d_i + 3$                  \\ 
\hspace{-7mm}75                    \>                 \>                \> $D = D - 3$                  \\
\hspace{-7mm}76                    \>                 \> \textbf{if} \> $(do_i == Rm) \wedge (D > 0) == \textsc{true}$                   \\
\hspace{-7mm}77                    \>                 \>                \> $d_i = d_i + 3$                  \\ 
\hspace{-7mm}78                    \>                 \>                \> $D = D - 3$                  \\ 
\hspace{-7mm}79 sort $d$ in nonincreasing order resulting $d'$                   \` \textbf{//}    line 79--80: sorting of $d$                                               \\ 
\hspace{-7mm}80 HHL$(d')$                                                                  \\
\hspace{-7mm}81 \textbf{return} $L, \ d$                                                \` \textbf{//}    line 81: return the result of HHL
\end{tabbing}

\vspace*{-4pt}
Procedure HHL (\textsc{Havel-Hakimi-Linear}) is described in \cite{Ivanyi2012Comp}. We remark that the original Havel-Hakimi algorithm requires in worst 
case $\Theta(n^2)$ time. Recently Király \cite{Kiraly2012} published a quicker algorithm which uses the data structure proposed by van Emde Boas \cite{CormenLRS2009,Kiraly2012DS,Emde1975} and 
requires only  $O(n \log \log n)$ time. Our algorithm is linear and works also for some multigraphs.  

A natural requirement is $d_i \leq n - 1$ but $d_i > n - 1$ can occur only in the cases $s = (0,2)$ and $s = (1,2)$ which are filtered by the constant time algorithms.

%%%%%%%%%%%%%%%%%%%%%%%%%%%%%%%%%%%%%%%%%%%%%%%%%%%%%%%%%%%%%%%%%%%%
\subsubsection{Efficiency of linear time filtering algorithms \label{subsub-lineareff}}
\textsc{Linear} is the union of the described linear time algorithm.

\medskip
\noindent \textsc{Linear}$(n, s)$
\vspace{-2mm}
\begin{tabbing}
199 \= x\=x\=x\=x\=x\=x\=x\=x \+ \kill
\hspace{-7mm}01 $L = 0$                                                                                   \` \textbf{//} line 01: initialization of $L$   \\                                                 
\hspace{-7mm}02 \textsc{L1}$(n,s)$                                                                    \` \textbf{//} line 02--04: filtering by \textsc{Complete}   \\
\hspace{-7mm}03 \textbf{if} \= $L = 0$                                                                \\
\hspace{-7mm}04                \> \textbf{return} $L$                                                \\
\hspace{-7mm}05 \textsc{L2}$(n,s)$                                                                    \` \textbf{//} line 05--07: filtering by \textsc{Losses}   \\
\hspace{-7mm}06 \textbf{if} \= $L = 0$                                                                \\
\hspace{-7mm}07                \> \textbf{return} $L$                                                \\
\hspace{-7mm}08 \textsc{L3}$(n,s)$                                                                    \` \textbf{//} line 08--10: filtering by \textsc{Reduction0}   \\
\hspace{-7mm}09 \textbf{if} \= $L = 0$                                                                \\
\hspace{-7mm}10                \> \textbf{return} $L$                                                \\
\hspace{-7mm}11 \textsc{L4}$(n,s)$                                                                     \` \textbf{//} line 11--13: filtering by \textsc{Reduction1}   \\
\hspace{-7mm}12 \textbf{if} \= $L = 0$                                                                \\
\hspace{-7mm}13                \> \textbf{return} $L$                                                \\
\hspace{-7mm}14 \textsc{L5}$(n,s)$                                                                    \` \textbf{//} line 14--16: filtering by \textsc{Draw-Unique}   \\
\hspace{-7mm}05 \textbf{if} \= $L = 0$                                                                \\
\hspace{-7mm}16                \> \textbf{return} $L$                                                \\
\hspace{-7mm}17 \textsc{L6}$(n,s)$                                                                    \` \textbf{//} line 17--19: filtering by \textsc{Balanced}   \\
\hspace{-7mm}18 \textbf{if} \= $L = 0$                                                                \\ 
\hspace{-7mm}19                \> \textbf{return} $L$                                                \\
\hspace{-7mm}20 \textsc{L7}$(n,s)$                                                                    \` \textbf{//} line 20--22: filtering by \textsc{Draw-Uniform}   \\
\hspace{-7mm}21 \textbf{if} \= $L = 0$                                                                \\
\hspace{-7mm}22                \> \textbf{return} $L$                                                \\
\hspace{-7mm}23 \textsc{L8}$(n,s)$                                                                    \` \textbf{//} line 23--25: filtering by \textsc{Draw-Sorted-Unique}   \\
\hspace{-7mm}24 \textbf{if} \= $L = 0$                                                                \\
\hspace{-7mm}25                \> \textbf{return} $L$                                                \\
\hspace{-7mm}26 $L = 1$                                                                                     \` \textbf{//} line 26--27: the linear time algorithms can not decide  \\
\hspace{-7mm}27 \textbf{return} $L$                                                                    \\
\end{tabbing}

Since all included algorithms have linear worst case running time, the total running time of \textsc{Linear} is also $O(n)$. Since the best running time of L1 is $O(1)$, therefore the best running time of \textsc{Linear} is also
$O(1)$.

Tables \ref{tab-linfiltering1} and \ref{tab-linfiltering2} show the concrete filtering results of the linear time filtering algorithms. Table \ref{tab-linfiltering1} contains the number of regular sequences $(R)$, the number of sequences, accepted by C9, L1 = \textsc{Complete-Test}, 
L2 = \textsc{Losses} and L3 = \textsc{Reduction0}.  

\begin{table}[!ht]
\centering
$\begin{array}{||r||r|r|r|r||} \hline \hline 
n &        C9                  &    L1                    &           L2           &    L3 + L4           \\  \hline
1 &  \textbf{1}              &\textbf{1}             &       \textbf{1}     &        \textbf{1}        \\ \hline
2 &  \textbf{2}              &\textbf{2}             &       \textbf{2}     &        \textbf{2}        \\ \hline 
3 &                     14     &   12                     & 	            10       &                 10     \\ \hline 
4 &                   203     &	134                   &	            94       &                 87     \\ \hline 
5 &                2 133      &        1 230           &                901       &	        814        \\ \hline
6 &              20 518	  &       10 947           &              8 348      &	     7 526        \\ \hline
7 &            191 707	  &       97 427           &            76 526      &           69 349            \\ \hline
8 &         1 772 842      &     872 234	         &          699 344      &         637 735              \\ \hline 
9 &       16 332 091      &     7 851 193           &      6 387 443      &      5 859 125               \\ \hline 
10&    150 288 309       &   71 001 641          &     58 367 243      &     53 817 029            \\ \hline
11&        1 383 099 467 &      644 668 154     &   538 591 486     &    494 427 384            \\ \hline
12&      12 737 278 674 &   5 873 396 400	  & 4 888 701 306    &  4 544 762 304                \\ \hline
13&    117 411 184 292  &        53 669 099 755  & 44 823 480 671   & 41 804 695 971       
          \\ \hline
14& 1 083 421 567 402 &  491 669 304 392 &411 496 549 436  &384 847 810 936         
        \\ \hline \hline 
\end{array}$
\caption{Results of filtering by linear tests tests L1, L2, and L3 + L4 for $n=1, \ \ldots, \ 14$ teams. \label{tab-linfiltering1}}
\end{table}

Table \ref{tab-linfiltering2} contains the number of sequences accepted by L4 =
\textsc{Reduction1}, L5 = \textsc{Draw-Unique}, L6 = \textsc{Balanced}, 
L7 = \textsc{Sport-Uniform} and L8 = \textsc{Inner-Draw}, further the number of the
football sequences $(F)$ and the cumulated running time and  
(the exact values of L7 are bold).

\begin{table}[!ht]
\centering
$\begin{array}{||r||r|r|r|r|r|r|r|r|r||} \hline \hline 
n &         L5 + L6            &           L7 + L8         &                $F$     
       &             $t$              \\  \hline
1 &              \textbf{1}   &           \textbf{1}     &                \textbf{1}
  &         0.000                \\ \hline
2 &              \textbf{2}   &           \textbf{2}     &                \textbf{2}
  &         0.000                \\ \hline 
3 &              \textbf{7}  &            \textbf{7}     &                \textbf{7}
  &         0.000         \\ \hline 
4 &                              46  &            \textbf{40}    &                     \textbf{40}
 &          0.000  \\ \hline 
5 &                       475 &                         365    &                        
355  &          0.000   \\ \hline
6 &                    4 459  &                        4 086    &                       3
678  &          0.015    \\ \hline
7 &                  47 867   &                44 657    &                     37
263   &          0.047    \\ \hline
8 &                460 153   &              451 213    &                         361 058   
&         0.437     \\ \hline 
9 &             4 371 783   &           4 348 655    &                      3 403 613     &
       4.196     \\ \hline 
10&          41 261 057    &        41 166 157     &                    31 653 777      &  
   40.217     \\ \hline
11&        387 821 927     &     387 416 935     &                  292 547 199      &    
393.280     \\ \hline
12&     3 635 039 265     &  3 633 749 149     &               2 696 619 716      &  3
828.002  \\ \hline
13&   34 011 137 972    & 33 821 636 274           &                                    &
37 611.185 \\ \hline\hline
14& 317 827 900 632    &316 291 028 902            &                                   &364
978.049   \\ \hline\hline
\end{array}$
\caption{Results of filtering by linear tests L5 + L6 and L7 + L8, further the
number of football sequences $(F)$ and  the running time of L8  
$(t)$  for $1, \ \ldots, \ 14$ teams. \label{tab-linfiltering2}}
\end{table}

%%%%%%%%%%%%%%%%%%%%%%%%%%%%%%%%%%%%%%%%%%%%%%%%%%%%%%%%%%%%%%%%%%%%%%%%%%%%%%%%%%%%%%%
%%%%%%%%%%%%%%%%%%%%%%%%%%%%%%%%%%%%%%%%%%%%%%%%%%%%%%%%%%%%%%%%%%%%%%%%%%%%%%%%%%%%%%%
\subsection{Quadratic filtering algorithms \label{subsecsec-quadfilter}}
In this section the quadratic recursive filtering algorithms Q1 = \textsc{Balanced-Quad}, Q2 = \textsc{Reduction-Rec-Small}, and Q3 = \textsc{Reduction-Rec-Large} are described.

%%%%%%%%%%%%%%%%%%%%%%%%%%%%%%%%%%%%%%%%%%%%%%%%%%%%%%%%%%%%%%%%%%%%%%%%%%%%%%%%%%%%%%%
\subsubsection{Quadratic filtering algorithm Q1 = \sc{Balanced-Quad} \label{subsubsec-Q1}}
The filtering algorithm Q1 = \textsc{Balanced-Quad} is based on the observation that if the draw sequence is unique, then the victory sequence $w = (w_1,\ldots,w_n)$ and 
the corresponding loss sequence $l = (l_1,\ldots,l_n)$ are also unique, further that the wins (losses) of any subset of teams have to be paired with inner and outer losses (wins).
The following assertion gives a necessary condition for the reconstructability of the sequence pair $(v,l)$.

\begin{lemma} \label{lemma-winloss} If $(a_1,\ldots, a_n)$ is the monotone nonincreaing win sequence and  $(b_1,\ldots, b_n)$  is the corresponding loss sequence of a football tournament then 
\begin{equation}
\sum _{i=1}^k a_i \leq \sum _{i =1}^k \min(b_i, k - 1) + \sum _{i = k + 1}^n \min(b_i, k) \label{eq-winloss} 
\end{equation} 
\textit{ for all } $k = 1, \ldots, n,$ \textit{ with equality for}  $n$.
\end{lemma} 

\begin{proof}  The wins included in the sum of the left side of \eqref{eq-winloss} have to be paired with the "inner losses`` (losses among T$_1$, \ldots, T$_k$) and 
"outer losses`` (losses of T$_1$, \ldots, T$_k$ in the matches against the remaining teams).
\end{proof}

We remark that this lemma is a consequence of Theorem 3 of the recent
paper due to Berger \cite{Berger2011arX} containing a necessary and
sufficient condition for some
incomplete $(0,2,n)$-tournaments. As the sequence $(1,1,8,9,9)$ satisfying
 \ref{eq-winloss} shows, in our case \eqref{eq-winloss}
 is only a necessary condition, since $s$ has a unique sport matrix shown
 in Table \ref{tab-sport11899}  which is not reconstructable.
 
 \begin{table}[!ht]
 \centering
 \begin{small}
 $\begin{array}{||r||r|r|r|r||} \hline \hline
 i  &  w_i        &   d_i          &     l_i    &    s_i        \\ \hline
 1 &   0          &   1            &       3     &      1     \\ \hline
 2 &   0          &   1            &       3    &       1     \\ \hline
 3 &   2         &    2            &       0   &        8     \\ \hline
 4 &   3         &   0            &       1    &       9     \\ \hline
  5 &   3        &    0            &       1   &        9     \\ \hline 
 \hline
 \end{array}$
 \end{small}
 \caption{Unique sport matrix belonging to the sequence $s = (1,1,8,9,9)$.
 \label{tab-sport11899}}
 \end{table}

The paper \cite{ErdosMT2011} contains an algorithm for our problem but the algorithm does not terminate for some inputs.

The following natural implementation \textsc{Balanced-Quad}  of Lemma \ref{lemma-winloss} requires quadratic time. 

Parameters of \textsc{Balanced-Quad} are the usual ones, further $w = (w_1,\ldots,$ $w_n)$: $w_i$ is the number of wins allocated to T$_i \ (0 \leq w_i \leq n - 1)$; 
$l = (l_1,\ldots,l_n)$: $l_i$ is the number of losses allocated to T$_i \ (0 \leq l_i \leq n - 1)$; 
$Sw$: the current number of the necessary wins; 
$Ss$: the maximal number of pairable losses of teams having small indices; 
$Sl$: the maximal number of pairable losses of the teams having large indices. 

\bigskip
\noindent \textsc{Balanced-Quad}$(n, w, l)$
\vspace{-2mm}
\begin{tabbing}
199 \= x\=x\=x\=x\=x\=x\=x\=x \+ \kill
\hspace{-7mm}01 $Sw = L = 0$                                                      \` \textbf{//}   line 01: initialization of $Sw$ and $L$  \\
\hspace{-7mm}02 sort $(w,l)$ nonincreasingly in $w$ using \textsc{Counting-Sort}                                                        \\
\hspace{-7mm}03 \textbf{for} \= $i = 1$ \textbf{to} $n$                     \` \textbf{//}   line 03--13:  counting of wins and losses \\
 \hspace{-7mm}04                  \> $Ss = Sl = 0$       \\
 \hspace{-7mm}05                  \> $Sw = Sw + w_i$       \\
 \hspace{-7mm}06                  \>  \textbf{for} \=  $j = 1$ \textbf{to}
 $i$                     \` \textbf{//}   line 06--07: small indices   \\
 \hspace{-7mm}07                  \>                   \> $Ss = Ss + \min
 (w_j,i - 1)$                                  \\
 \hspace{-7mm}08                  \>  \textbf{for} \= $j = i + 1$
 \textbf{to} $n$                     \` \textbf{//}   line 08--09: large
 indices   \\
 \hspace{-7mm}09                  \>                   \> $Sl = Sl + \min
 (w_j,i )$                                  \\
 \hspace{-7mm}10                  \> \textbf{if} \= $Sw > Ss + Sl$         
                                                   \` \textbf{//}    line
 10--13: $(w,l)$ is not pairable  \\
 \hspace{-7mm}11                  \>                \> \textbf{return} $L$ 
                                             \\
 \hspace{-7mm}12  \textbf{if} \= $Sw < Ss + Sl$                            
               \\
 \hspace{-7mm}13                 \> \textbf{return} $L$                    
                   \\
 \hspace{-7mm}14 \textbf{else} \= $L = 2$                                  
       \` \textbf{//} line 14--15: $s$ is undecided     \\
 \hspace{-7mm}15                    \> \textbf{return} $L$
 \end{tabbing}
 
We yet did not implemented \textsc{Balanced-Quad}.

%%%%%%%%%%%%%%%%%%%%%%%%%%%%%%%%%%%%%%%%%%%%%%%%%%%%%%%%%%%%%%%%%%%%%%
\subsubsection{Quadratic filtering algorithm Q2 = \sc{Reduction-Rec-Small} \label{subsubsec-quad2}}
Algorithm Q2 = \textsc{Reduction-Rec-Small} is based on the recursive application of \textsc{Recursive0} and \textsc{Recursive1}. Using Q2 and the next Q3  we 
shorten the input sequences  and often can filter them.

Parameters are the usual ones, further $e = (e_1,\ldots,e_N)$: work version of the investigated sequence; $n_l$: smallest index of 
not deleted elements of $s$.

\bigskip
\noindent \textsc{Reduction-Rec-Small}$(n, s)$
\vspace{-2mm}
\begin{tabbing}
199 \= x\=x\=x\=x\=x\=x\=x\=x \+ \kill
\hspace{-7mm}01 $L =  S = 0$                                                                                                    \` \textbf{//}   line 01--04: initialization of $L$, $S$, $n_l$, and $e$    \\
\hspace{-7mm}02 \textbf{for} \= $i = 1$ \textbf{to} $n$                                      \\
\hspace{-7mm}03                  \>  $e_i = s_i$                                                \\
\hspace{-7mm}04 $n_l = 1$                                                                        \\
\hspace{-7mm}05 \textbf{while} \= $n_l \leq n$                                                         \\
\hspace{-7mm}06                      \> $S = 0$                                                    \\
\hspace{-7mm}07                      \>\textbf{for} \= $i = n_l$ \textbf{to} $n$    \\
\hspace{-7mm}08                      \>                 \> $S = S + e_i$                       \\ 
\hspace{-7mm}09                      \>                 \> \textbf{if} \= $S == i(i - 1)$                   \` \textbf{//}   line 09--21:  $S$ is minimal       \\ 
\hspace{-7mm}10                      \>                 \>                \> \textbf{if} \= $i < n$                   \\
\hspace{-7mm}11                      \>                 \>                \>                \> \textbf{if} \= $(e_{n_l} \neq i - 1) \vee (e_{n_l + i - 1} \neq i - 1)$   == \textsc{true}  \\
\hspace{-7mm}12                      \>                 \>                \>                \>                \> \textbf{return} $L$                            \\
\hspace{-7mm}13                      \>                 \>                \>                \> \textbf{if} \> $e_{n_l + i} < 3i$                                                       \\
\hspace{-7mm}14                      \>                 \>                \>                \>                \> \textbf{return} $L$                                                \\
\hspace{-7mm}15                      \>                 \>                \> \textbf{if} \= $i == n$                \\
\hspace{-7mm}16                      \>                 \>                \>                \> \textbf{if} \=  $(e_{n_l} \neq i - 1) \vee (e_n \neq i - 1)$  == \textsc{true}    \\
\hspace{-7mm}17                      \>                 \>                \>                \>                \> \textbf{return} $L$                            \\
\hspace{-7mm}18                      \>                 \>                \>                \> \textbf{else} \= $L = 1$                                          \\
\hspace{-7mm}19                      \>                 \>                \>                \>                    \>\textbf{return} $L$                           \\    
\hspace{-7mm}20                      \>                 \>                \> $n_l = n_l + i$                                                             \\
\hspace{-7mm}21                      \>                 \>                \> \textbf{for} \= $j = n_l$ \textbf{to} $n$    \\
\hspace{-7mm}22                      \>                 \>                \>                   \> $e_j = e_j - 3i$                                                \\ 
\hspace{-7mm}23                      \>                 \>                \> \textbf{go to} 05                                                       \\
\hspace{-7mm}24                      \>                 \> \textbf{if} \= $S == i(i - 1) + 1$         \` \textbf{//}   line 22--35:  $S$ is minimum  plus one   \\ 
\hspace{-7mm}25                      \>                 \>                \> \textbf{if} \= $i < n$                   \\
\hspace{-7mm}26                      \>                 \>                \>                \> $L_1 =  (e_{n_l} = i - 1) \wedge (e_{n_l + i - 2} = i - 1) \wedge (e_{n_l + i -1} = i )$          \\
\>                 \>                \>                \>                           \qquad  $\;\; \wedge  (e_{n_l + i} \geq 3i - 2)$ \\
\hspace{-7mm}27                      \>                 \>                \>                \> $L_2 =  (e_{n_l} = i - 2) \wedge (e_{n_l + i - 2} = i - 1) \wedge (e_{n_l + i - 1} = i + 1) $   \\
\hspace{-7mm}28                      \>                 \>                \>                \> \textbf{if} \= $(L_1 == \textsc{false}) \wedge (L_2 == \textsc{false})$   == \textsc{true}\\    
\hspace{-7mm}29                      \>                 \>                \>                \>                \> \textbf{return} $L$  \` \textbf{//}    line 28--29: $s$ is not football sequence                        \\     
\hspace{-7mm}30                      \>                 \>                \> \textbf{if} \= $i == n$    \\
\hspace{-7mm}31                      \>                 \>                \>                \> $L_2 =  (e_{n_l} = n - 2) \wedge (e_{n_l + 1} = n - 1) \wedge (e_{n-1} = n + 1)$   \\
\hspace{-7mm}32                      \>                 \>                \>                \> \textbf{if} \= $L_2 == \textsc{False}$                                  \` \textbf{//}    line 32--33: $s$ is not football sequence  \\                                 
\hspace{-7mm}33                      \>                 \>                \>                \>                \> \textbf{return} $L$      \\
\hspace{-7mm}34                      \>                 \>                \> $n_l = n_l + i$                             \\
\hspace{-7mm}35                      \>                 \>                \> \textbf{for} \= $j = n_l$ \textbf{to} $n$    \\
\hspace{-7mm}36                      \>                 \>                \>                   \> $e_j = e_j - 3i$                                                \\
\hspace{-7mm}37                      \>                 \>                \> \textbf{go to} 05                                                                     \\                     
\hspace{-7mm}38                      \>                 \>  \textsc{Reduction-Rec-Large}$(n - n_l + 1,e)$                \\
\hspace{-7mm}39                      \>                 \>  \textbf{if} \= $L == 0$                      \\
\hspace{-7mm}40                      \>                 \>                 \> \textbf{return} $L$                      \\
\hspace{-7mm}41 \textbf{if} \= $n_u > 0$                                                                                       \\ 
\hspace{-7mm}42                \> \textsc{Filter}$(n_u,e)$                                                            \\ 
\hspace{-7mm}43 \textbf{if} \= $L == 0$                                                               \\
\hspace{-7mm}44                \> \textbf{return} $L$                                              \\ 
\hspace{-7mm}45  $L = 2$                                                                                     \` \textbf{//} line 45--46: $s$ is undecided          \\
\hspace{-7mm}46  \textbf{return} $L$                                                                                            
\end{tabbing}

\textsc{Reduction-Rec-Small} calls \textsc{Filter} which is a union of the constant and linear time filtering algorithms and \textsc{Reduction-Rec-Large} which is the next quadratic filtering algorithm.

\bigskip
\noindent \textsc{Filter}$(n, e)$
\vspace{-2mm}
\begin{tabbing}
199 \= x\=x\=x\=x\=x\=x\=x\=x \+ \kill
\hspace{-7mm}01 \textsc{Constant}$(n,e)$                                                      \` \textbf{//}   line 01--03: filtering by the constant time algorithms \\
\hspace{-7mm}02 \textbf{if} \= $L == 0$                                                               \\
\hspace{-7mm}03                \> \textbf{return} $L$                                                 \\
\hspace{-7mm}04 \textsc{Linear}$(n,e)$                                                      \` \textbf{//}   line 04--06: filtering by the linear time algorithms   \\
\hspace{-7mm}05 \textbf{if} \= $L == 0$                                                               \\
\hspace{-7mm}06                \> \textbf{return} $L$                                                       \\
\hspace{-7mm}07 $L = 2$                                                                              \` \textbf{//} line 07--08: $s$ is undecided \\
\hspace{-7mm}08  \textbf{return} $L$ 
\end{tabbing}

%%%%%%%%%%%%%%%%%%%%%%%%%%%%%%%%%%%%%%%%%%%%%%%%%%%%%%%%%%%%%%%%%%%%%%%%%%%%%%%
\subsubsection{Quadratic filtering algorithm Q3 \label{subsubsec-quad3}}
Algorithm Q3 = \textsc{Reduction-Rec-Large} is based on the recursive application of \textsc{Recursive0} and \textsc{Recursive1}.

Parameters are the usual ones, further $e = (e_1,\ldots,e_N)$: work version of the investigated sequence; $n_u$: smallest index of the 
not deleted elements of $s$; $Q$: the sum of the $i$ largest scores; $B$: the number of investigated scores giving $0$ remainder mod $3$; $C$: the number of investigated scores 
giving remainder $1$ mod $3$; $D$: the number of investigated scores giving remainder $2$ mod $3$.

\bigskip
\noindent \textsc{Reduction-Rec-Large}$(n, e)$
\vspace{-2mm}
\begin{tabbing}
199 \= x\=x\=x\=x\=x\=x\=x\=x \+ \kill
\hspace{-7mm}01 $L = Q = B = C = D = 0$                                                                                                    \` \textbf{//}   line 01--02: initialization of $L$, $Q$, $B$, $C$, $D$, $n_u$    \\
\hspace{-7mm}02 $n_u = n$                                                                        \\
\hspace{-7mm}03 \textbf{while} \= $n_u \leq 1$                                                 \` \textbf{//}   line 03--25: recursive reduction          \\
\hspace{-7mm}04                      \>\textbf{for} \= $i = n_u$ \textbf{downto} $1$             \` \textbf{//}   line 04--09: preparing of the filtering    \\
\hspace{-7mm}05                      \>                 \> $Q = Q + e_i$                       \\
\hspace{-7mm}06                      \>                 \> \textbf{if} \= $e_i - 3 \lfloor e_i/3 \rfloor == 0$       \\
\hspace{-7mm}07                      \>                 \>                \> $B = B + 1$        \\
\hspace{-7mm}08                      \>                 \> \textbf{if} \> $e_i - 3 \lfloor e_i/3 \rfloor == 1$       \\
\hspace{-7mm}09                      \>                 \>                \> $C = C + 1$        \\
\hspace{-7mm}10                      \>                 \> \textbf{if} \> $e_i - 3 \lfloor e_i/3 \rfloor == 2$       \\
\hspace{-7mm}11                      \>                 \>                \> $D = D + 1$        \\
\hspace{-7mm}12                      \>                 \> \textbf{if} \= $Q == 3i(n_u - i) + 3i(i - 1)/2$                   \` \textbf{//}   line 12--17:  $Q$ is maximal       \\ 
\hspace{-7mm}13                      \>                 \>                \> \textbf{if}  \= $B \neq i$       \\
\hspace{-7mm}14                      \>                 \>                \>                 \> \textbf{return} $L,n_u$       \\
\hspace{-7mm}15                      \>                 \>                \> \textbf{if}  \> $i > 1$                 \\
\hspace{-7mm}16                      \>                 \>                \>                 \> \textbf{if}  \= $e_{n_u - i} > 3(n_u - i - 1)$     \\
\hspace{-7mm}17                      \>                 \>                \>                                      \> \textbf{return} $L,n_u$       \\
\hspace{-7mm}18                      \>                 \>                \> $n_u = n_u - i$                                                \\ 
\hspace{-7mm}19                      \>                 \>                \> \textbf{go to} 03                                             \\
\hspace{-7mm}20                      \>                 \> \textbf{if} \= $Q == 3i(n_u - i) + 3i(i - 1)/2 - 1$                 \\
\hspace{-7mm}21                      \>                 \>                \> \textbf{if}  \= $(B == i - 2) \wedge 
(C == 2) == \textsc{false}$         \\
\hspace{-7mm}22                      \>                 \>                \>                 \>  \textbf{return} $L,n_u$                            \\
\hspace{-7mm}23                      \>                 \>                \> \textbf{if} \= $i > 1$      \` \textbf{//}   line 20--25:  $Q$ is  maximum minus 1 \\
\hspace{-7mm}24                      \>                 \>                \>                 \> \textbf{if}  \= $e_{n_u - i} > 3(n_u - i - 1)$     \\
\hspace{-7mm}25                      \>                 \>                \>                 \>                 \> \textbf{return} $L,n_u$       \\
\hspace{-7mm}26 $L = 2$                                                                                                                              \` \textbf{//}   line 26--27:  $s$ is not decided   \\
\hspace{-7mm}27 \textbf{return} $L,n_u$                \\                                                               
\end{tabbing}

The following Table \ref{tab-Q12} contains the results of quadratic filtering algorithms.
 
\begin{table}[!ht]
\centering
$\begin{array}{||r||r|r|r|r||} \hline \hline 
n &               \textsc{Linear}           &           Q2 + Q3          &         $F$                   &             $t$              \\  \hline
1 &           \textbf{1}     &                \textbf{1}   &            \textbf{1}        &       0.000                \\ \hline
2 &           \textbf{2}     &                \textbf{2}   &      \textbf{2}              &            0.000                \\ \hline 
3 &           \textbf{7}     &                \textbf{7}   &      \textbf{7}           &            0.000          \\ \hline 
4 &            \textbf{40}    &	             \textbf{40}  &      \textbf{40}          &            0.000  \\ \hline 
5 &      	           365   &             \textbf{355}   &     \textbf{355}         &            0.000   \\ \hline                                           
6 &	                4 086    &                      3 760   &     \textbf{3 678}      &            0.015    \\ \hline                                                                      
7 &                44 657    &                      39 417  &   \textbf{37 273}      &             0.109    \\ \hline                                                                                                                8 &              451 213    &	                 393 072   &   \textbf{361 058}          &           1.264    \\ \hline 
9 &           4 348 655    &	              3 804 485    &   \textbf{3 403 613}       &         15.226      \\ \hline 
10&        41 166 157     &	            36 302 148    &   \textbf{31 653 777}     &       179.249     \\ \hline
11&     387 416 935     &	          344 012 885    &   \textbf{292 547 199}    &    2 066.323     \\ \hline
12&  3 633 749 149     &	       3 246 651 763    & \textbf{2 696 619 716}  &   23 429.877 \\ \hline  
13&33 821 636 274	  &         30 405 902 165    &                                    &   \\ \hline \hline
\end{array}$
\caption{Results of filtering by \textsc{Linear} and quadratic algorithms Q2 + Q3, further the number of football sequences $(F)$ and  the running time of Q3  
$(t)$  for $n=1, \ \ldots, \ 13$ teams. \label{tab-Q12}}
\end{table}

%%%%%%%%%%%%%%%%%%%%%%%%%%%%%%%%%%%%%%%%%%%%%%%%%%%%%%%%%%%%%%%%%%%%%%%%%%%%%%%%%%%%%%%%
%%%%%%%%%%%%%%%%%%%%%%%%%%%%%%%%%%%%%%%%%%%%%%%%%%%%%%%%%%%%%%%%%%%%%%%%%%%%%%%%%%%%%%%%
%%%%%%%%%%%%%%%%%%%%%%%%%%%%%%%%%%%%%%%%%%%%%%%%%%%%%%%%%%%%%%%%%%%%%%%%%%%%%%%%%%%%%%%%
\section{Reconstruction of potential football sequences \label{sec-recon}}
In this part we investigate polynomial reconstruction algorithms, as R1 = \textsc{Reduction},  R2 = \textsc{Draw-Uniform-Rec}, and R3 = \textsc{Draw-Inner-Rec}.

%%%%%%%%%%%%%%%%%%%%%%%%%%%%%%%%%%%%%%%%%%%%%%%%%%%%%%%%%%%%%%%%%%%%%%%%%%%%
%%%%%%%%%%%%%%%%%%%%%%%%%%%%%%%%%%%%%%%%%%%%%%%%%%%%%%%%%%%%%%%%%%%%%%%%%%%%
\subsection{Reconstruction algorithm R1 = \sc{Reduction} \label{subsub-R1}}
R1 = \textsc{Reduction} is based on filtering algorithms \textsc{Reduction0} and \textsc{Reduction1}. 

%%%%%%%%%%%%%%%%%%%%%%%%%%%%%%%%%%%%%%%%%%%%%%%%%%%%%%%%%%%%%%%%%%%%%%%%%%%%
%%%%%%%%%%%%%%%%%%%%%%%%%%%%%%%%%%%%%%%%%%%%%%%%%%%%%%%%%%%%%%%%%%%%%%%%%%%%
\subsection{Reconstruction algorithm R2 = \sc{Draw-Uniform-Rec}\label{subsub-R2}}
R2 = \textsc{Draw-Uniform-Rec} is based on filtering algorithms: it tries---using the degree sequence $d$ produced by \textsc{Sport-Uniform} or \textsc{Draw-Sorted-Unique} and 
using a greedy pairing algorithm ``largest wins with largest losses"---to pair the wins and losses.

Parameters of R2 are the usual ones further $S$: sport matrix computed using the output
draw sequence $d$ of \textsc{Sport-Uniform} or $\textsc{Draw-Sorted-Unique}$ and sorted its rows so that either $w_i >w_{i+1}$ or $w_i = w_{i+1}$ and $l_i \leq l_{i+1}$; 
 $d = (d_1,\ldots,d_n)$: draw sequence of $S$; 
$\mathcal{M}_{n \times n}$ (result matrix): $\mathcal{M}_{i,j}$ is the number of points received by T$_i$ in the  match against T$_j$;  
$w = (w_1,\ldots,w_n)$:  $w_i$ is the number of wins of T$_i$;   
$l = (l_1,\ldots,l_n)$: $l_i$ is the number of losses of T$_i$.

\bigskip
\noindent \textsc{Draw-Uniform-Rec}$(n,s,d)$
\vspace{-2mm}
\begin{tabbing}
199 \= x\=x\=x\=x\=x\=x\=x\=x \+ \kill
\hspace{-7mm}01 \textbf{for} \= $i = 1$ \textbf{to} $n $                                                       \` \textbf{//}   line 01--03: initialization of $\mathcal{M}$   \\
\hspace{-7mm}02                  \>  \textbf{for} \= $j = 1$ \textbf{to} $n$                                                \\ 
\hspace{-7mm}03                  \>                   \> $\mathcal{M}_{i,j} = 0$                                   \\
\hspace{-7mm}04  \textsc{Havel-Hakimi-Draws}$(n,s,d,\mathcal{M})$                                    \` \textbf{//}   line 04: HHD allocates the draws   \\
\hspace{-7mm}05  \textbf{for} \= $i = 1$ \textbf{to} $n $                                                       \` \textbf{//}   line 05--07: computation of $w$ and $l$ \\
\hspace{-7mm}06                   \> $w_i = (s_i - d_i)/3$                                                             \\
\hspace{-7mm}07                   \> $l_i = n - 1 - d_i - w_i$                                                         \\                
\hspace{-7mm}08  \textbf{for} \= $i = n$ \textbf{downto} $1$                                                \` \textbf{//}   line 08--24: allocation of wins and losses \\                         
\hspace{-7mm}09                    \> $j = n$                                                                          \\ 
\hspace{-7mm}10                    \> \textbf{while} \= $(w_i > 0) \vee (M_{ij} \neq 1) \vee (j > 0) \vee (i \neq j) \vee (l_j > 0)==\textsc{true}$      \\
\hspace{-7mm}11                    \>                      \> $\mathcal{M}_{ij} = 3$   \\
\hspace{-7mm}12                    \>                      \> $w_i = w_ i - 1$   \\
\hspace{-7mm}13                    \>                      \> $l_j = l_j - 1$   \\
\hspace{-7mm}14                    \>                      \> $j = j - 1$ \\
\hspace{-7mm}15                    \>  \textbf{if} \= $w_i > 0$                                                      \` \textbf{//}   line 15--17: $s$ is undecided \\
\hspace{-7mm}16                    \>                 \> $L = 2$   \\
\hspace{-7mm}17                    \>                 \> \textbf{return} $L$, $\mathcal{M}$                                     \\
\hspace{-7mm}18  $L = 1$                                                                                                     \` \textbf{//} line 18--19: $s$ is a football sequence   \\
\hspace{-7mm}19  \textbf{return} $L$, $\mathcal{M}$                                                                                      
\end{tabbing}

R2 uses a special version of Havel-Hakimi algorithm called \textsc{Havel-Hakimi-Draws} (or shortly HHD). While for 
the classical Havel-Hakimi algorithm the equal scores are equivalent, in this application we have to distinguish them.

Additional parameters are $d = (d_1, \ldots, d_n)$: a draw sequence produced by \textsc{Draw-Rec};  
$\mathcal{M}$: $n \times n$ sized  matrix where $\mathcal{M}_{ij}$ is the number of points received by T$_i$ in the match with T$_j$;
$\mathcal{E} = (E_1,\ldots,E_n) =  ((e_1,h_1),\ldots,(e_n,h_n))$: current extended and sorted version of $d$; 
$H = (h_1,\ldots,h_n)$: $h_i$ is the index of $e_i$ in $d$; 
$n_l$: lower index of the essential part of $\mathcal{E}$;   
$n_u$: upper index of the essential part of $\mathcal{E}$;   
$c = (c_0,\ldots,c_n)$: $c_i$ is the number of $i$'s among $e_{n_l},\ldots,e_{n_u}$;
$C = (C_0, \ldots, C_n)$: $C_i$ is the cumulated number of $i$'s among $e_{n_l},\ldots,e_{n_u}$.

\bigskip
\noindent \textsc{Havel-Hakimi-Draws}$(n,d,\mathcal{M})$
\vspace{-2mm}
\begin{tabbing}
199 \= xxx\=xxx\=xxx\=xxx\=xxx\=xxx\=xxx\=xxx \+ \kill
\hspace{-7mm}01 $n_l = 1$                    \` \textbf{//} line 01--05: initialization of $n_l$, $n_u$, and $\mathcal{E}$;  \\
\hspace{-7mm}02 $n_u = n$ \\
\hspace{-7mm}03 \textbf{for} \= $i = n_l$ \textbf{to} $n_u$                                          \` \textbf{//} line 03--07: initialization of $G$ and $n_u$;  \\   
\hspace{-7mm}04                  \> $e_i = d_i$                                                                    \\ 
\hspace{-7mm}05                  \> $h_i = i$                                                               \\
\hspace{-7mm}07 $n_u = n$   \\
\hspace{-7mm}08  \textbf{for} \= $i = 1$ \textbf{to} $n$                                          \` \textbf{//} line 08--15: pairing of the draws;  \\  
\hspace{-7mm}09                   \> \textsc{Counting-Sort-Draws}$(n,i,n_u,\mathcal{E})$                   \` \textbf{//}   line 09: sorting   \\ 
\hspace{-7mm}10        \>           \textbf{if} \= $e_i = 0$                                        \\   
\hspace{-7mm}11        \>                          \> \textbf{return} $\mathcal{M}$   \\
\hspace{-7mm}12 \> \textbf{for} \= $k = 1$ \textbf{to} $e_i$  \\
\hspace{-7mm}13  \>                 \> $\mathcal{M}_{h_i,h_i + k} = \mathcal{M}_{h_i + k,h_i} = 1$   \` \textbf{//}   line 13: a draw is fixed   \\                    
\hspace{-7mm}14    \>               \> $e_{i + k} = e_{i + k} + 1$              \\ 
\hspace{-7mm}15  \>\textbf{while} \= $n_u == 0$   \\
\hspace{-7mm}16   \>                    \> $n_u = n_u - h_i$ \\     
\hspace{-7mm}17  \textbf{return}  $\mathcal{M}$   \`   \textbf{//}  line 17:  return the matrix containing the  paired draws
\end{tabbing} 

\textsc{Counting-Sort-Draws} is a modified version of the well-known linear time sorting algorithm \textsc{Counting-Sort} \cite{CormenLRS2009}.

Additional parameters are  $d = (d_1, \ldots, d_n)$: a draw sequence produced by \textsc{Draw-Uniform-Rec};   
$n_l$: lower index of the essential part of $\mathcal{E}$;   
$n_u$: upper index of the essential part of $\mathcal{E}$; 
$\mathcal{M}$: $n \times n$ sized  matrix where $\mathcal{M}_{ij}$ is the number of points received by T$_i$ in the match with T$_j$;
$\mathcal{E} = (E_1,\ldots,E_n) =   ((g_{11},g_{12}), \ldots , (g_{1n},g_{2n})$: current extended and sorted version of $d$ with the corresponding indices;
$\mathcal{G}$: the working version of $\mathcal{E}$; $n_l$  : lower index of the essential part of $\mathcal{E}$;  
$n_u$: upper index of the essential part of $\mathcal{E}$;
$c = (c_0,\ldots,c_{n-1})$: $c_i$ is the number of $i$'s among $g_{1,n_l},\ldots, g_{1,n_u}$; 
$C_n = 0$ working variable; $C  = (C_0,\ldots,C_{n-1})$:  $C_i$ is the number of investigated scores larger or equal with $i$.

\bigskip
\noindent \textsc{Counting-Sort-Draws}$(n,d,n_l,n_u,\mathcal{E})$
\vspace{-2mm}
\begin{tabbing}
199 \= xxx\=xxx\=xxx\=xxx\=xxx\=xxx\=xxx\=xxx \+ \kill 
\hspace{-7mm}01 \textbf{for} \= $i = n_l$ \textbf{to} $n_u$                                          \` \textbf{//} line 01--05: initialization of $G$ and $c$;  \\   
\hspace{-7mm}02                  \> $g_{1,i} = e_{1,i}$                                                                    \\ 
\hspace{-7mm}03                  \> $g_{2,i} = e_{2,i}$                                                                    \\ 
\hspace{-7mm}04 \textbf{for} \= $i = 0$ \textbf{to} $n - 1$                                           \\   
\hspace{-7mm}05                  \> $c_i = 0$                                                                  \\ 
\hspace{-7mm}06 \textbf{for} \= $i = n_l$ \textbf{to} $n_u$                                        \` \textbf{//}   line 06-10: computation of the counters    \\
\hspace{-7mm}07                  \> $c_{g_{1i}} = c_{g_{1i}} +1$                                              \\      \hspace{-7mm}08 $C_n = 1$                                                                                         \\
\hspace{-7mm}09  \textbf{for} \= $n - 1$ \textbf{downto} $0$                                         \\
\hspace{-7mm}10                   \> $C_i =  C_{i + 1} + c_i$                                                 \\
\hspace{-7mm}11 \textbf{for} \= $i = n_l$ \textbf{to} $n_u$                                        \` \textbf{//}   line 11-16: computation of the new $\mathcal{E}$    \\ 
\hspace{-7mm}12                  \>  $x = C_{g_{1,i}} +1$                                                 \\
\hspace{-7mm}13                  \> $e_{1,x} = g_{1,i}$                                                      \\                                                 
\hspace{-7mm}14                  \> $e_{2,x} = g_{2,i}$                                                        \\
\hspace{-7mm}15                  \> $C_x = C_x + 1$                                                                  \\
\hspace{-7mm}16  \textbf{return} $\mathcal{E}$
\end{tabbing} 

The running time of  \textsc{Counting-Sort-Draw} is $\Theta(n)$, of \textsc{Havel-Hakimi-Draw} is $O(n^2)$ and the one of \textsc{Draw-Uniform-Rec} 
is also  $O(n^2)$.

As an example let $s = (1,1,7,7)$. Then $s$ has a unique draw sequence $(1,1,1,1)$ and unique sport matrix shown in Table \ref{tab-sport1177}. 

\begin{table}[!ht]
\centering
\begin{small}
$\begin{array}{||r||r|r|r|r||} \hline \hline
i  &  w_i        &   d_i          &     l_i    &    s_i        \\ \hline
1 &   0        &   1             &       2     &      1     \\ \hline
2 &   0         &   1            &       2    &       1     \\ \hline 
3 &   2        &    1            &       0   &        7     \\ \hline 
4 &   2	  &    1            &       0   &	      7 \\ \hline  \hline
\end{array}$
\end{small}  
\caption{Unique sport matrix belonging to the sequence $s = (1,1,7,7)$.\label{tab-sport1177}}
\end{table}

According to the relatively quick version \textsc{Havel-Hakimi-Shifting} \cite{IvanyiLMS2011Acta}  T$_1$ plays a draw with T$_4$ and T$_2$ with T$_3$  
resulting the partial result matrix  shown in Table \ref{tab-res1177}. 

\begin{table}[!ht]
\centering
\begin{small}
$\begin{array}{||c||c|c|c|c|c||} \hline \hline
i  &  T_1        &   T_2        &     T_3    &    T_4   &  s_i       \\ \hline
1 &   -           &   ?            &       ?     &      1      & 1     \\ \hline
2 &   ?          &   -             &       1     &       ?     &  1 \\ \hline 
3 &   ?          &    1           &       -     &         ?    &   7\\ \hline 
4 &   1	   &    ?            &       ?    &	      -       &  7 \\ \hline  \hline
\end{array}$
\end{small}  
\caption{Partial result matrix  belonging to the draws of $s = (1,1,7,7)$.\label{tab-res1177}}
\end{table}

The partial result matrix containing the draws 
  in Table \ref{tab-res1177} is not reconstructible since no acceptable result for the match between T$_1$ and T$_2$.

If we use the classical Havel-Hakimi algorithm then the draws are between T$_1$ and T$_2$, further between T$_3$ and T$_4$ and our greedy algorithm  \textsc{Draw-Uniform-Rec} reconstructs the received partial result matrix.

Another example let $s = (1,1,8,8,10,13)$. Then $s$ has a unique draw sequence $(1,1,2,2,1,1)$ and a unique sport matrix shown in Table \ref{tab-sport1188}.

\begin{table}[!ht]
\centering
\begin{small}
$\begin{array}{||r||r|r|r|r||} \hline \hline
i  &  w_i        &   d_i          &     l_i    &    s_i        \\ \hline
1 &   0          &   1            &       4     &      1     \\ \hline
2 &   0          &   1            &       4    &       1     \\ \hline 
3 &   2        &    2            &       1   &        8     \\ \hline 
4 &   2         &   2            &       1    &       8     \\ \hline 
5 &   3        &    1            &       1   &        10     \\ \hline 
6 &   4	  &    1            &       0   &	      13 \\ \hline  \hline
\end{array}$
\end{small}  
\caption{Unique sport matrix belonging to the sequence $s = (1,1,8,8,10,13)$. \label{tab-sport1188}}
\end{table} 

In this case at first $\mathcal{E} = ((2,3),(2,4),(1,1),(1,2),(1,5),(1,6))$. The draws
allocated by HHD are shown in Table \ref{tab-res1188}. 

\begin{table}[!ht]
\centering
\begin{small}
$\begin{array}{||c||c|c|c|c|c|c|c||} \hline \hline
i  &  T_1        &   T_2        &     T_3    &    T_4   &   T_5   &   T_6   &   s_i 
     \\ \hline
1 &   -           &   ?            &       1     &      ?      &     ?     &     ?  
  &    1         \\ \hline
2 &   ?          &   -             &       ?     &      1     &      ?     &     ?  
  &    1    \\ \hline 
3 &   1          &    ?           &       -     &      1      &     ?      &    ?   
  &    8     \\ \hline 
4 &   ?            &    1           &       1    &       -     &     ?      &    ?      &  
  8      \\ \hline 
5 &   ?          &    ?           &       ?     &      ?      &     -      &    1   
  &   10     \\ \hline 
6 &   ?            &    ?           &       ?    &       ?     &     1      &    -      &  
  13      \\ \hline  \hline
\end{array}$
\end{small}  
\caption{Partial result matrix  belonging to the draws of $s =
(1,1,8,8,10,13)$.\label{tab-res1188}}
\end{table}

The partial result matrix in Table \ref{tab-res1188} is not reconstructible since no
acceptable result for the match between T$_1$ and T$_2$.

%%%%%%%%%%%%%%%%%%%%%%%%%%%%%%%%%%%%%%%%%%%%%%%%%%%%%%%%%%%%%%%%%%%%%%%%%%%%
%%%%%%%%%%%%%%%%%%%%%%%%%%%%%%%%%%%%%%%%%%%%%%%%%%%%%%%%%%%%%%%%%%%%%%%%%%%%
\subsection{Reconstruction algorithm R3 = \textsc{Draw-Inner-Rec} \label{subsub-R3}}
Reconstruction algorithm R3 = \textsc{Draw-Inner-Rec} is an improved version of R2: it takes into account the obligatory inner draws. 

The base of \textsc{Inner-Draws} is the following lemma.

\begin{lemma} If\label{lemma-innerdraws} $n \geq 1$, $f = (f_1, \ldots, f_n)$ is a football sequence,   
$1 \leq k \leq n$ and \
\begin{equation}
\sum _{i = 1} ^k f_i < 3 \binom{k}{2}, \label{eq-innerdraws1} 
\end{equation}
then among the teams $T_1, \ \ldots, \ T_k$ there are at least 
\begin{equation}
\left \lceil \left ( 3 \binom{k}{2} - \sum _{i = 1} ^k f_i \right )/2 \right \rceil   \label{eq-innerdraws2}
\end{equation}
draws.
\end{lemma}

\begin{proof} If 
\begin{equation}
\left \lceil 2 \left  (3 \binom{k}{2} - \sum _{i = 1} ^k f_i \right )/2 \right \rceil = q > 0,
\end{equation}
then the first $k$ teams lost at least $q$ points due to inner draws (or even more, if they gathered points 
in the matches against the remaining teams).
\end{proof} 

Trying to reconstruct the sequence $s = (1,1,8,8,10,13)$ which was the last example
of the previous Section \ref{subsub-R2}  \textsc{Draw-Inner-Rec} (see Table
\ref{tab-sport1188} and \ref{tab-res1188})  
recognizes that $s_1 + s_2 = 2$ therefore according to Lemma \ref{lemma-innerdraws}
the obligatory result between T$_1$ and T$_2$ is a draw. Then
\textsc{Draw-Inner-Rec} 
finishes the allocation of the draws as it is shown in Table \ref{tab-draws1188}. 

\begin{table}[!ht]
\centering
\begin{small}
$\begin{array}{||c||c|c|c|c|c|c|c||} \hline \hline
1 &   -           &   1            &       ?     &      ?      &     ?     &     ?  
  &    1         \\ \hline
2 &   1          &   -             &       ?     &      ?     &      ?     &     ?  
  &    1    \\ \hline 
3 &   ?          &    ?           &       -     &      1      &     1      &    ?   
  &    8     \\ \hline 
4 &   ?            &    ?           &       1    &       -     &     ?      &    1      &  
  8      \\ \hline 
5 &   ?          &    ?           &       1     &      ?      &     -      &    ?   
  &   10     \\ \hline 
6 &   ?            &    ?           &       ?    &       1     &     1?     &    -      &  
 13      \\ \hline  \hline
\end{array}$
\end{small}  
\caption{Partial result matrix  belonging to the draws of $s = (1,1,8,8,10,13)$
allocated by \textsc{Draw-Inner-Rec}. \label{tab-draws1188}}
\end{table}
 
Using the matrix of the allocated draws shown in Table \ref{tab-draws1188}
\textsc{Draw-Uniform-Rec} produces the complete result matrix shown in Table
\ref{tab-full1188}. 
proving that $s = (1,1,8,8,10,13)$ is a football sequence. 

\begin{table}[!ht]
\centering
\begin{small}
$\begin{array}{||c||c|c|c|c|c|c|c||} \hline \hline
i  &  T_1        &   T_2        &     T_3    &    T_4   &   T_5   &   T_6   &   s_i 
     \\ \hline
i  &  T_1        &   T_2        &     T_3    &    T_4   &   T_5   &   T_6   &   s_i 
     \\ \hline
1 &   -           &   1            &       0     &      0      &     0     &     0  
  &    1         \\ \hline
2 &   1          &   -             &       0     &      0     &      0     &     0  
  &    1    \\ \hline 
3 &   3          &    3           &       -     &      1      &     1      &    0   
  &    8     \\ \hline 
4 &   3            &    3           &       1    &       -     &     0      &    1      &  
  8      \\ \hline 
5 &   3          &    3           &       1     &      3      &     -      &    0   
  &   10     \\ \hline 
6 &   3            &    3           &       3    &       1     &     3     &    -      &   
13      \\ \hline  \hline
\end{array}$
\end{small}  
\caption{Partial result matrix  belonging to the draws of $s = (1,1,8,8,10,13)$
allocated by \textsc{Draw-Inner-Rec}. \label{tab-full1188}}
\end{table}

The algorithm based on this lemma yet is is not implemented.

%%%%%%%%%%%%%%%%%%%%%%%%%%%%%%%%%%%%%%%%%%%%%%%%%%%%%%%%%%%%%%%%%%%%%%%%%%%%%%%%%%%%%%%%
%%%%%%%%%%%%%%%%%%%%%%%%%%%%%%%%%%%%%%%%%%%%%%%%%%%%%%%%%%%%%%%%%%%%%%%%%%%%%%%%%%%%%%%%
%%%%%%%%%%%%%%%%%%%%%%%%%%%%%%%%%%%%%%%%%%%%%%%%%%%%%%%%%%%%%%%%%%%%%%%%%%%%%%%%%%%%%%%%
\section{Enumeration of football sequences \label{sec-enum}}
There are many publications connected with the generation \cite{BarnesS1997,Hemasinha2003,Ivanyi2011Kyoto,Schoenfield2008A064626} and enumeration of degree sequences of graphs, e.g. 
\cite{BarnesS1995,BarnesS1997,BeregI2008,Burns2007,CooperL2011,HarborthK1982,IvanyiLMS2011Acta,
IvanyiLMS2011A004251,KemnitzD1997,KleitmanW1981,LuczI2012MaCS,McKayW1996,MetropolisS1980,NarayanaB1964,PecsySz2000,RodsethST2009,RuskeyCES1994,Stanley1991,WinstonK1983}.
The problems connected with directed graphs sometimes are considered as problems of orientation of undirected graphs \cite{Frank1980,Frank2011,FrankGy1978,FrankKK2003}.

The enumeration of degree \cite{BarnesS1995,Burns2007,FrankSS2002,IvanyiLMS2011Acta,IvanyiLMS2011A004251} and score \cite{HarborthK1982,Isaak2010} sequences also has a reach literature. 

The first published enumeration results connected with football score sequences belong to 
G\'abor Kov\'acs, Norbert Pataki, Zoltán Hernyák and Tamás Hegyessy \cite{KovacsP2002} who
computed $F(n)$ for $n =  1, \ \ldots, \ 8$ in 2002. N. J. A. Sloane in May 2007 determined $F(9)$, then in June 
2008 Min Li computed $F(10).$ The newest results were received by J. E. Schoenfield who computed 
$F(11)$ in September of 2008 and $F(12)$ in December of 2008  \cite{Schoenfield2008A064626}.

Connected problems are  the listing of all degree sequences and sampling of degree sequences 
\cite{BergerM2010,BlitzsteinD2011,delGenioKTB2010,KannanTV1999,KayibiKPI2012,MiklosES2010}. 

Our basic method is similar as we enumerated the degree sequences of simple graphs 
\cite{IvanyiLMS2011Acta,Sloane2011A004251}.

From one side we try to test the elements of the possible smallest set, and from the other side we try to use 
quick as possible testing and reconstruction algorithms.

A natural idea is to investigate only the nonincreasing sequences of integers having $0$ as lower bound and 
$3(n - 1)$ as upper bound. Paul Erdős and Tibor Gallai called such sequences \textit{regular} \cite{ErdosG1960}.
The number of such sequences is given by \eqref{eq-lum}.

%%%%%%%%%%%%%%%%%%%%%%%%%%%%%%%%%%%%%%%%%%%%%%%%%%%%%%%%%%%%%%%%%%%%%%%%%%%%%%%%%%%%%%%
%%%%%%%%%%%%%%%%%%%%%%%%%%%%%%%%%%%%%%%%%%%%%%%%%%%%%%%%%%%%%%%%%%%%%%%%%%%%%%%%%%%%%%%
\subsection{Decreasing of the number of the investigated sequences \label{subsec-decrease}}
A useful tool of the enumeration of the number of football sequences is the decreasing of the number of the considered sequences.  

In Section \ref{sec-filter} we proposed and analyzed filtering of regular sequences with constant, linear and quadratic time algorithms. For $14$ teams we excluded more then the half of the regular sequences 
by the constant time algorithms. For $13$ teams the linear and quadratic algorithms left less then 10.58 percent of the regular sequences. In Section  \ref{sec-recon} the polynomial  reconstruction algorithms decreased the 
fraction of the undecided regular sequences to  $4.68$ percent of the regular sequences.

%%%%%%%%%%%%%%%%%%%%%%%%%%%%%%%%%%%%%%%%%%%%%%%%%%%%%%%%%%%%%%%%%%%%%%%%%%%%%%%%%%%%%%%
%%%%%%%%%%%%%%%%%%%%%%%%%%%%%%%%%%%%%%%%%%%%%%%%%%%%%%%%%%%%%%%%%%%%%%%%%%%%%%%%%%%%%%%
\subsection{Backtrack filtering and accepting test \label{subsec-backtrack}}
This method is due to Antal Iványi \cite{Ivanyi2001,KovacsP2002}.

The results of the filtering algorithms are summarized in Table \ref{tab-filter}.

\begin{table}[!ht]
\centering
\begin{small}
\begin{tabular}{||r|r|r|r|r||} \hline \hline 
$n$ &        \textsc{Constant}        &    \textsc{Linear}        &    \textsc{Quad}         & \textsc{Backtrack} = $F$              \\ \hline  \hline  
1    &                          \textbf{1} &                   \textbf{1} &               \textbf{1}    &        \textbf{1}          \\   \hline 
2   &                        \textbf{2}    &                   \textbf{2} &               \textbf{2}    &        \textbf{2}      \\    \hline 
3   &                                     14 &                   \textbf{7} &               \textbf{7}    &        \textbf{7}                \\     \hline  
4   &                                   203 &                  \textbf{40} &             \textbf{40}    &      \textbf{40}              \\   \hline  
5   &                                2 133 &                            365 &           \textbf{355}    &     \textbf{355}             \\    \hline
6   &                              20 518 &                          4 086 &             3 760           &   \textbf{3 678}  \\    \hline 
7   &                            191 707 &                        44 657 &            39 417         &    27 263 \\    \hline 
8   &                         1 772 442 &                      451 213 &          393 072          & 361 058 \\   \hline 
9   &                       16 332 091 &                    4 348 655 &         3 804 485        & 3 403 613 \\   \hline 
10 &                      150 288 309 &                  41 166 157 &        36 302 148       & 31 653 777 \\  \hline 
11 &                    1 383 099467 &                  387 416 935 &     344 012 885       & 292 547 199  \\   \hline 
12 &                  12 737 278 674 &               3 633 749 149 &  3 246 651 763      &  2 696 619 716  \\   \hline 
13 &                117 411 154 292 &             33 821 636 274 &30 405 902 165      &   \\    \hline 
14 &             1 083 421 567 482 &                                    &                            &  \\ \hline \hline  
\end{tabular} 
\end{small}
\caption{Numbers of sequences accepted by constant, linear and quadratic time and \textsc{Backtrack} filtering algorithms 
for $n = 1, \ \ldots, \ 14$ teams. \label{tab-filter}}
\end{table} 

The running time of the filtering algorithms are presented in Table \ref{tab-filtimes}. The times are cumulated and contain the time necessary for the generation 
of the sequences too. 

\begin{table}[!ht]
\centering
\begin{small}
\begin{tabular}{||r|r|r|r|r|r||} \hline \hline 
$n$&     \textsc{Constant}  &   \textsc{Linear}         &    \textsc{Quad}  & \textsc{Backtrack} = $F$              \\ \hline  \hline  
1   &                       0.000 &                        0.000 &                     0.000 &     0.000                                    \\   \hline 
2   &                       0.000 &                        0.000 &                     0.000 &     0.000       \\    \hline 
3   &                       0.000 &                        0.000 &                     0.000 &     0.000                       \\     \hline  
4  &                        0.000 &                        0.000 &                     0.000 &     0.000                     \\   \hline  
5  &                        0.000 &                       0.000 &                      0.000 &     0.000      \\    \hline
6  &                        0.000 &                       0.000 &                      0.000 &     0.015 \\    \hline 
7  &                        0.016 &                       0.031 &                      0.042 &     0.172 \\    \hline 
8  &                        0.046 &                       0.375 &                      0.577 &   52.603 \\   \hline 
9  &                        0.468 &                       3.572 &                      5.772 &            \\   \hline 
10 &                       4.134 &                     34.632 &                    54.741 &                   \\  \hline 
11 &                     37.612 &                    329.816 &                  525.752 &              \\   \hline  
12 &                   343.575 &                 3 145.494 &                4 998.831 &             \\   \hline 
13 &                3 142.469 &               30 541.260 &               49 035.625 &           \\    \hline 
14 &              29 438.094 &                                &                                &              \\ \hline \hline   
\end{tabular} 
\end{small}
\caption{Running times of constant, linear and quadratic time filtering algorithms for $n = 1, \ \ldots, \ 14$ teams. \label{tab-filtimes}}
\end{table}

The individual results of the reconstruction algorithms are summarized in Table \ref{tab-recon}.

\begin{table}[!ht]
\centering 
\begin{tabular}{||r|r|r|r|r||} \hline \hline 
$n$&                R1           &          R2 + R3              &  \textsc{Backtrack}       &        $F$                  \\ \hline 
 1 &                    1          &                      0               &               0                   &                      1                   \\ \hline
 2 &                    2          &                     0               &               0                   &                      2                         \\  \hline
 3 &                    6          &                     1               &               0                   &                      7                             \\  \hline
 4 &                  18          &                    22               &               0                   &                    40                           \\  \hline
 5 &                  50          &                  305                &              0                   &                   355                        \\   \hline
 6 &                 137         &               3 460               &              81                  &                3 678                          \\  \hline
 7 &                 375         &             33 993               &          2 895                 &               37 263                         \\  \hline
 8 &              1 023         &           304 349            &            56 909                 &             361 058               \\  \hline
 9 &              2 776         &        2 576 124            &                                       & 3 403 613                \\ \hline
10 &             7 498         &      21 453 751             &                                      &   31 653 777              \\ \hline 
11 &           20 177         &    177 819 555             &                                      &   292 547 199               \\  \hline
12 &           54 127         & 1 476 661 425            &                                       &  2 696 619 716              \\  \hline
13 &         144 708        &12 300 060 430             &                                      &                     \\ \hline  \hline
\end{tabular} 
\caption{Number of $(0,3n-3,n)$-regular sequences reconstructed by reconstruction algorithms R1, R2 + R3 and \textsc{Backtrack} 
 for $n = 1, \ \ldots, \ 14$ teams. \label{tab-recon}}
\end{table} 

The running times of the  reconstruction algorithms are shown in Table \ref{tab-recontimes}.

\begin{table}[!ht]
\centering 
\begin{tabular}{||r|r|r|r|r||} \hline \hline 
$n$&             R1             &               R3            &   \textsc{Backtrack}            \\ \hline 
 2 &                  0.000     &             0.000           &         0.000                         \\  \hline
 3 &                  0.000    &              0.000          &          0.000                             \\  \hline
 4 &                  0.000     &             0.000          &          0.000                           \\  \hline
 5 &                  0.000     &             0.000         &           0.000                        \\   \hline
 6 &                  0.000     &             0.015         &           0.015                      \\  \hline
 7 &                  0.063     &             0.109         &           0.172                         \\  \hline
 8 &                  0.546     &             1.264         &         52.603                   \\  \hline
 9 &                  5.491     &           15.226         &                             \\ \hline
10 &               53.880    &          179.249         &                       \\ \hline
11 &             522.386      &     2 066.323         &                     \\  \hline
12 &          4 998.831       &   23 429.877        &                    \\  \hline
13 &        49 035.625       & 261 904.750                &                    \\ \hline \hline
\end{tabular} 
\caption{Running times of the R1, R3 and \textsc{Backtrack} reconstructing algorithms for $n = 1, \ \ldots, \ 13$ teams. \label{tab-recontimes}}
\end{table} 
 
%%%%%%%%%%%%%%%%%%%%%%%%%%%%%%%%%%%%%%%%%%%%%%%%%%%%%%%%%%%%%%%%%%%%%%%%%%%%%%%%%%%%%%%
%%%%%%%%%%%%%%%%%%%%%%%%%%%%%%%%%%%%%%%%%%%%%%%%%%%%%%%%%%%%%%%%%%%%%%%%%%%%%%%%%%%%%%%
\subsection{Recursive accepting test \label{subsec-recursive}}
This method is due to Schoenfield \cite{Schoenfield2008A064626}. According to this method we compare 
the sequences of length $n$ passed through the filtering and accepting tests with the good sequences of length 
$n-1$ whether they can be derive from them.

Since if we omit a team with its results from a football matrix of size $n \times n$, then we get a football matrix 
of size $(n - 1) \times (n - 1)$, therefore we regularly delete the \textit{first} elements of the investigated 
$n$-length sequences. 

Let $n \geq 2.$ We suppose that when we enumerate the $n$-length good sequences then we know 
the $F(n - 1) \times (n - 1)$ sized 
matrix $M$ containing the $(n - 1)$-length good sequences in lexicographically increasing order, and also know 
the vector $(P_0, \ldots, P_k)$, where $k = \lfloor3(n - 1)/2 \rfloor$ and $P_i$ gives the number 
of $(n - 1)$-length good sequences starting with $i$.   

Let start the recursion with $n = 2.$ Matrix $M_1$ contains only one row $(0)$ and $P$ contains one element 
$P(1) = 1.$

The constant time filtering algorithms accept only the sequences $(0,3)$ and $(1,1).$ At first we omit 0 
from the first sequence and state that the remaining 
sequence (3) can be derived from (0) only if the team having zero points in the shorter sequence wins against 
the omitted player. So the omitted player has to have zero points. Since the omitted score is exactly zero, 
(0,3) \textit{is a good sequence.}  

Then we delete the first element from the sequence (1,1) and state that the player having zero points has 
to play a draw with the omitted team. Since it has exactly one point, therefore (1,1) is also 
\textit{a good sequence} and so $F(2) = 2.$

Now let $n = 3.$ Then $M_2$ contains two rows: (0,3) and (1,1). In this case the filtering algorithms accept only 
the seven good sequences: (0,3,6), (0,4,4), (1,1,4), (1,2,4), (1,3,4), (2,2,2) and (3,3,3).

At first we delete $0$ from $(0,3,6)$ and compare the remaining $(3,6)$ with the known good sequences. There are 
thee possibilities: the first team of the good sequence received 3, 1 or 0 points against the omitted one.    
If 3, then the good sequence has to start with 0. There is only one sequence $(0,3)$ requiring 
two losses for the omitted team. Since the omitted element is exactly zero, $(0,3,6)$ \textit{is a good sequence}.  

The second accepted sequence is $(0,4,4).$ Omitting $0$ and comparing $(4,4)$ with the good sequences we get, that 
$(1,1)$ is the only potential ancestor requiring zero points for the deleted team. Since it has exactly zero points, 
$(0,4,4)$ is also a \textit{good sequence}.  

In a similar way we can prove that the remaining five accepted sequences are also good.

When $n = 4$ then $M$ contains seven elements and $P = (1,3,6,7).$  

\textsc{Reconstruct} executes this recursive step. Its additional parameters are 
$F(n - 1)$: the number of $(n - 1)$-length good sequences; 
$\mathcal{M}_{F(n - 1) \times (n - 1)}$: matrix of good sequences of length $n - 1$ (this matrix consists of submatrices 
containing the good sequences having identical first element; 
$P = (P_0,\ldots,P_k)$, where $k = k = \lfloor3(n - 1)/2 \rfloor$and $P_i$ is the number of $n - 1$ length 
football sequences starting with $i$; 
$\mathcal{N}_{F(n) \times n}$: matrix of good sequences of length $n$;
$m = (m_1, \ldots, m_{n-1})$: the current reduced version of $s$;
$d$: the current score of the deleted team.

\bigskip
\noindent \textsc{Reconstruct}$(n,s,F,M,P)$ 
\vspace{-2mm}
\begin{tabbing}
199 \= x\=x\=x\=x\=x\=x\=x\=x \+ \kill
\hspace{-7mm}01 $L = 1$                                   line 01--02: initialization of $L$ and $u$  \\
\hspace{-7mm}02 $u = \lfloor3(n - 1)/2 \rfloor$                                                      \\
\hspace{-7mm}03 \textbf{if} \= $s_2 \leq u$      \` \textbf{//} line 03-21: omitted element starts with a loss \\
\hspace{-7mm}04             \> \> $j \leftarrow P_{s_2}$ \\
\hspace{-7mm}05             \> \textbf{while} \= $\mathcal{M}_{j,1} == s_2$         \\
\hspace{-7mm}06             \>                \> $d \leftarrow 0$ \\
\hspace{-7mm}07             \>                \> $k \leftarrow 2$ \\
\hspace{-7mm}08             \>                \> \textbf{while} \= $k \leq n$ \\
\hspace{-7mm}09             \>                \>                \> \textbf{if} \= $s_k - \mathcal{M}_{j,k} == 3$     \\
\hspace{-7mm}10             \>                \>                \>             \> $d = d + 0$ \\
\hspace{-7mm}11             \>                \>                \>             \> \textbf{go to} 19               \\
\hspace{-7mm}12             \>                \>                \> \textbf{if} \= $s_k - \mathcal{M}_{j,k} == 1$ \\
\hspace{-7mm}13             \>                \>                \>             \> $d = d + 1$ \\ 
\hspace{-7mm}14             \>                \>                \>             \> \textbf{go to} 19          \\   
\hspace{-7mm}15             \>                \>                \> \textbf{if} \= $s_k - \mathcal{M}_{j,k} == 0$    \\
\hspace{-7mm}16             \>                \>                \>             \> $d = d + 3$  \\
\hspace{-7mm}17             \>                \>                \>             \> \textbf{go to} 19   \\
\hspace{-7mm}18             \>                \>                \> \textbf{go to} 22 \\
\hspace{-7mm}19             \>                \>                \> $k \leftarrow k + 1$ \\
\hspace{-7mm}20             \> \textbf{if} \= $d == s_1$ \\
\hspace{-7mm}21             \>             \> \textbf{return} $L$ \\
\hspace{-7mm}22 \textbf{if} \= $0 \leq s_2 - 1$    \` \textbf{//} line 22-40: omitted element starts with a draw \\
\hspace{-7mm}23             \> $j \leftarrow P_{s_2 - 1}$ \\
\hspace{-7mm}24             \> \textbf{while} \= $\mathcal{M}_{j,1} == s_2 - 1$         \\
\hspace{-7mm}25             \>                \> $d \leftarrow 1$ \\
\hspace{-7mm}26             \>                \> $k \leftarrow 2$ \\
\hspace{-7mm}27             \>                \> \textbf{while} \= $k \leq n$ \\
\hspace{-7mm}28             \>                \>                \> \textbf{if} \= $s_k - \mathcal{M}_{j,k} == 1$     \\
\hspace{-7mm}29             \>                \>                \>             \>$d = d + 1$ \\
\hspace{-7mm}30             \>                \>                \>             \> \textbf{go to} 38 \`    \\
\hspace{-7mm}31             \>                \>                \> \textbf{if} \= $s_k - \mathcal{M}_{j,k} == 1$ \\
\hspace{-7mm}32             \>                \>                \>             \> $d = d + 1$ \\ 
\hspace{-7mm}33             \>                \>                \>             \> \textbf{go to} 38          \\   
\hspace{-7mm}34             \>                \>                \> \textbf{if} \> $s_k - \mathcal{M}_{j,k} == 1$    \\
\hspace{-7mm}35             \>                \>                \>             \> $d = d + 1$  \\
\hspace{-7mm}36             \>                \>                \>             \> \textbf{go to} 38   \\
\hspace{-7mm}37             \>                \>                \> \textbf{go to} 39 \\
\hspace{-7mm}38             \>                \>                \> $k = k + 1$ \\
\hspace{-7mm}39             \> \textbf{if} \= $d == s_1$ \\
\hspace{-7mm}40             \>             \> \textbf{return} $L$ \\
\hspace{-7mm}41 \textbf{if} \= $0 \leq s_2 - 3$    \` \textbf{//} line 41-59: omitted element starts with a win \\
\hspace{-7mm}42             \> \> $j \leftarrow P[s_2 - 3]$ \\
\hspace{-7mm}43             \> \textbf{while} \= $\mathcal{M}_{j,1} == s_2 - 3$         \\
\hspace{-7mm}44             \>                \> $d \leftarrow 3$ \\
\hspace{-7mm}45             \>                \> $k \leftarrow 2$ \\
\hspace{-7mm}46             \>                \> \textbf{while} \= $k \leq n$ \\
\hspace{-7mm}47             \>                \>                      \> \textbf{if} \= $s_k - \mathcal{M}_{j,k} == 3$     \\
\hspace{-7mm}48             \>                \>                      \>                \>$d = d + 3$ \\
\hspace{-7mm}49             \>                \>                      \>                \> \textbf{go to} 57 \`    \\
\hspace{-7mm}50             \>                \>                      \> \textbf{if} \= $s_k - \mathcal{M}_{j,k} == 1$ \\
\hspace{-7mm}51             \>                \>                      \>                \> $d = d + 1$ \\ 
\hspace{-7mm}52             \>                \>                      \>                \> \textbf{go to} 57          \\   
\hspace{-7mm}53             \>                \>                      \> \textbf{if} \= $s_k - \mathcal{M}_{j,k} == 1$    \\
\hspace{-7mm}54             \>                \>                      \>                \> $d = d + 1$  \\
\hspace{-7mm}55             \>                \>                      \>                \> \textbf{go to} 57   \\
\hspace{-7mm}56             \>                \>                      \> \textbf{go to} 58 \\
\hspace{-7mm}57             \>                \>                      \> $k \leftarrow k + 1$ \\
\hspace{-7mm}58             \> \textbf{if} \= $d == s_1$ \\
\hspace{-7mm}59             \>                \> \textbf{return} $L$ \\
\hspace{-7mm}60 $L =0$     \\
\hspace{-7mm}61 \textbf{return} $L$
\end{tabbing}

\begin{table}[!ht]
\centering
\begin{small}
$\begin{array}{||r||r|r|r|r|r||} \hline \hline
n &  R(n)      &   \frac{R(n+1)}{R(n)}   &       F(n) & \frac{F(n+1)}{F(n)} & \frac{F(n)}{R(n)} \\ \hline
1 &   1        &   10.000             &        1   &        2.000        &     1.0000 \\ \hline
2 &  1         &    8.400             &        2    &        3.500        &     0.2000 \\ \hline 
3 &  84        &    8.512             &        7   &        5.714        &     0.0833 \\ \hline 
4 & 715	       &    8.655             &       40   &	    8.875        &     0.0559 \\ \hline 
5 & 6188       &    8.769             &      355   &       10.361        &     0.0574 \\ \hline
6 &54264       &    8.859             &     3678   &       10.131        &     0.0678 \\ \hline
7 &480700      &    8.929             &    37263   &        9.689        &     0.0775 \\ \hline
8 &4292145     &    8.986             &   361058   &        9.427        &     0.0841 \\ \hline 
9 &38567100    &    9.032             &  3403613   &        9.300        &     0.0883 \\ \hline 
10&348330136   &    9.070             & 31653777   &        9.242        &     0.0909 \\ \hline
11&3159461968  &    9.103             &292547199   &        9.217        &     0.0926 \\ \hline
12&28760021745 &    9.131             &2696619716  &                     &     0.0938  \\ \hline
13&262596783864&  9.155               &	                 &                     &            \\ \hline 
14&2 403 979 904 20 &	               &                     &                  &            \\ \hline \hline
\end{array}$
\end{small}  
\caption{Number of regular and football sequences and the ratio of these numbers for neighboring numbers of teams 
\label{table-ratios}}
\end{table}

Table \ref{table-ratios} shows the number of regular sequences $(R(n)$, the number of football sequences $(F(n)$, 
the ratio $(R(n+1)/R(n))$, the ratio $F(n+1)/F(n)$, and the ratio $(F(n)/R(n)$ for $n = 1, \ \ldots, \ 12.$ 
In this table if $n \geq 2$ then $R(n)$ is decreasing. 

\begin{lemma} If $n$ tends\label{lemma-RperR} to infinity then $R(n + 1)/R(n)$ tends to $256/27.$
\end{lemma} 

\begin{proof} According to \eqref{eq-lum} 
\begin{equation} 
\frac{R(n + 1)}{R(n)} = \frac{(4n + 1)(4n)(4n - 1)(4n - 2)}{(n+1)(3n)(3n-1)(3n-2)} = \frac{256}{27} + o(1),
\end{equation}
implying the required limit.
\end{proof}

If $n \geq 1$ then in Table \ref{table-ratios} $F(n + 1)/F(n)$ is nondecreasing. We suppose that it tends to $1.$ 

If $5 \leq n \leq 12$ then $F(n)/R(n)$ is increasing. It is easy to see that 
\begin{equation}
\lim _{n \rightarrow \infty} \frac{F(n + 1)}{F(n)} \leq \frac{R(n + 1)}{R(n)}. \label{eq-FFRR}
\end{equation}

The behavior of $F(n)/R(n)$ is a bit surprising since the similar relative density of tournaments 
score sequences tends to zero (see \cite{Burns2007}). We suppose that $F(n)/R(n)$ also tends to zero but the convergence is slow.  

\paragraph{Acknowledgements.} The authors thank Péter Burcsi  (Eötvös Loránd University)  for many useful comments, András Pluhár (University of Szeged) and Zoltán Király  (Eötvös Loránd University) 
for the recommended references,  Loránd Lucz (Eötvös Loránd University)  and Tamás Iványi (Economical Politechnicum) for the computer experiments. The European Union 
and the European Social Fund have provided financial support to the project 
under the grant agreement no. TÁMOP 4.2.1/B-09/1/KMR-2010-0003.

\small{

}

\bigskip
\rightline{\emph{Received: March 5, 2012 {\tiny  \raisebox{2pt}{$\bullet$\!}} Revised: June 7, 2012}} %% to be completed by the editor 

\begin{thebibliography}{99}
\bibitem{AnholzerBBK2011} M. \href{http://kbo.ue.poznan.pl/anholcer/}{Anholcer}, V. Babiy, 
S. \href{http://www.oplab.sztaki.hu/cv_bs_hu.htm}{Bozóki}, 
W. W. \href{http://www.cs.laurentian.ca/wkoczkodaj/info.html}{Koczkodaj}, 
A simplified implementation of the least squares solution for pairwise comparisons matrices. 
\href{http://www.springerlink.com/content/1435-246x/}{\textit{CEJOR}} \textit{Cent. Eur. J. Oper. Res.}
\textbf{19,} 4 (2011) 439--444. 

\bibitem{ArikatiM1996} S. R. Arikati, A. Maheshwari, Realizing degree sequences in parallel, 
\textit{SIAM J. Discrete Math.} \textbf{9,} 2 (1996) 317--338. 

\bibitem{BangS1979} Ch. M. Bang, H. Sharp, Jr., Score vectors of tournaments.  
 \textit{J.} \href{http://www.sciencedirect.com/science/journal/00958956}{ \textit{Combin.}} 
\textit{Theory Ser. B} \ \textbf{26,} 1 (1979) 81--84.

\bibitem{BarnesS1995} T. M. \href{http://coitweb.uncc.edu/~tbarnes2/}{Barnes}, C. D. \href{http://www4.ncsu.edu/~savage/}{Savage}, 
A recurrence for counting graphical partitions,  
\href{http://www.combinatorics.org/Volume\_2/volume2.html\#R11}{\textit{Electron.}} \textit{J. Combin.}  
\textbf{2} (1995), Research Paper 11, 10 pages (electronic). 

\bibitem{BarnesS1997}  T. M. \href{http://coitweb.uncc.edu/~tbarnes2/}{Barnes}, C. D. \href{http://www4.ncsu.edu/~savage/}{Savage}, Efficient 
generation of graphical partitions, \href{http://www.sciencedirect.com/science/journal/0166218X}{\textit{Discrete}} \textit{Appl. Math.} 
\textbf{78,} 1--3 (1997) 17--26. 

\bibitem{Barrus2012} M. D. \href{http://www.bhsu.edu/Default.aspx?alias=www.bhsu.edu/michaelbarrus}{Barrus}, 
Havel-Hakimi residues of unigraphs, \textit{Inf. Proc.} \href{http://www.sciencedirect.com/science/journal/00200190}{\textit{Letters}} \textbf{112} (2012) 44--48.

\bibitem{Bege1999} A. \href{http://math.ubbcluj.ro/~bege/pages/cv_hu.html}{Bege}, Personal communication, Visegrád, 1999. 

\bibitem{BeregI2008} S. Bereg, H. Ito, Transforming graphs with the same 
\href{http://www.sciencedirect.com/science/article/pii/S0895717709000077}{degree sequence}, in:  \textit{Kyoto Int. Conf. on Computational Geometry and Graph Theory,}  (ed. H. Ito et al.) LNCS \textbf{4535} Springer-Verlag, Berlin, Heidelberg. 2008. pp. 25--32.

\bibitem{Berger2011PhD} A. \href{http://www.informatik.uni-halle.de/arbeitsgruppen/datenstrukturen/mitarbeiter/annabell_berger/}{Berger}, 
\href{http://wcms.uzi.uni-halle.de/download.php?down=22851\&elem=2544689}{\textit{Directed degree sequences}}, PhD Dissertation,  Martin-Luther-Universität Halle-Wittenberg, 2011. 
 

\bibitem{Berger2011arX} A. \href{http://www.informatik.uni-halle.de/arbeitsgruppen/datenstrukturen/mitarbeiter/annabell_berger/}{Berger}, 
A note on the characterization of digraph sequences, \textit{arXiv}, arXiv:1112.1215v1 [math.CO] (6 December 2011) 

\bibitem{BergerM2010} A. \href{http://www.informatik.uni-halle.de/arbeitsgruppen/datenstrukturen/mitarbeiter/annabell_berger/}{Berger}, 
M. \href{http://www.informatik.uni-halle.de/arbeitsgruppen/datenstrukturen/mitarbeiter/muellerh/}{Müller-Hannemann}, 
Uniform sampling of digraphs with a fixed degree sequence, in (ed. D. M. Thilikos)  \textit{36th Int. Workshop on Graph Theoretic  
Concepts in Computer Science} (June 28 - 30, 2010,  Zarós, Crete, Greece), LNCS \textbf{6410} (2010) 220--231.

\bibitem{BergerM20115} A. \href{http://www.informatik.uni-halle.de/arbeitsgruppen/datenstrukturen/mitarbeiter/annabell_berger/}{Berger}, 
M. \href{http://www.informatik.uni-halle.de/arbeitsgruppen/datenstrukturen/mitarbeiter/muellerh/}{Müller-Hannemann}, Dag realizations of directed degree sequences, in 
(ed. O. M. Steffen, J. A. Telle) \textit{Proc. 18th FCT} LNCS \textbf{6914} (2011) 264--275. Full version with proofs: Technical Report 2011/5 of University 
Halle-Wittenberg, Institute of Computer Science.  

\bibitem{BergerM20116} A. \href{http://www.informatik.uni-halle.de/arbeitsgruppen/datenstrukturen/mitarbeiter/annabell_berger/}{Berger}, 
M. \href{http://www.informatik.uni-halle.de/arbeitsgruppen/datenstrukturen/mitarbeiter/muellerh/}{Müller-Hannemann}, Dag characterizations of directed degree sequences, i
Technical Report 2011/6 of University Halle-Wittenberg, Institute of Computer Science.

\bibitem{BergerM2012} A. \href{http://www.informatik.uni-halle.de/arbeitsgruppen/datenstrukturen/mitarbeiter/annabell_berger/}{Berger}, 
M. \href{http://www.informatik.uni-halle.de/arbeitsgruppen/datenstrukturen/mitarbeiter/muellerh/}{Műller-Hannemann}, How to attack the NP-complete dag realization problems in practice. 
\newline \textit{arXiv}, arXiv:1203.36v1, 2012. http://arxiv.org/abs/1203.3636  

\newpage\bibitem{BlitzsteinD2011} J. K. \href{http://www.people.fas.harvard.edu/~blitz}{Blitzstein}, 
P. \href{http://www-stat.stanford.edu/~cgates/PERSI}{Diaconis}, A sequential importance sampling algorithm
for generating random graphs with prescribed degrees. \textit{Internet Mathematics}  \textbf{6,} 4 (2011) 489--522.

\bibitem{BozokiFP2011} S. \href{http://www.oplab.sztaki.hu/cv_bs_hu.htm}{Bozóki},  J. 
\href{http://www.oplab.sztaki.hu/cv_fj_hu.htm}{Fülöp},  
A. \href{http://portal.uni-corvinus.hu/index.php?id=801}{Poesz}, 
On pairwise \href{http://www.springerlink.com/content/u5564j3413400n67/fulltext.pdf}{comparison matrices} 
that can be made consistent by the modification of a few elements. 
\href{http://www.springerlink.com/content/1435-246x/}{\textit{CEJOR}} 
\textit{Cent. Eur. J. Oper. Res.} \textbf{19} (2011) 157--175.

\bibitem{BozokiFR2010} \href{http://www.oplab.sztaki.hu/cv_bs_hu.htm}{Bozóki} S., 
J. \href{http://www.oplab.sztaki.hu/cv_fj_hu.htm}{Fülöp}, L. \href{http://www.sztaki.hu/~ronyai/}{Rónyai},   
On optimal completion of incomplete pairwise comparison matrices,    
\textit{Math. Comput. Modelling} \textbf{52} (2010) 318--333.

\bibitem{BrauerGS1968} A. Brauer, I. C. Gentry, K. Shaw, A new proof of a theorem by H. G. Landau on tournament matrices. 
\textit{J. Comb. Theory} \textbf{5} (1968) 289--292. 

\bibitem{BrualdiK2009} A. R. \href{http://www.math.wisc.edu/~brualdi/}{Brualdi},   
K. Kiernan, Landau's and {R}ado's theorems and partial tournaments,   
\href{http://www.combinatorics.org/Volume_16/PDF/v16i1n2.pdf}{\textit{Electron.}}   
\textit{J. Combin.} \textbf{16,} (\#N2) (2009) 6 pages.

\bibitem{BrualdiS2001} A. R.  \href{http://www.math.wisc.edu/~brualdi/}{Brualdi}, J. Shen, 
Landau's inequalities for tournament scores and a short proof of a 
theorem on transitive sub-tournaments, \textit{J. Graph Theory} \textbf{38,} 4 (2001) 244--254. 

\bibitem{Burns2007} J. M. Burns: \textit{The number of 
\href{http://dspace.mit.edu/bitstream/handle/1721.1/38882/166267576.pdf?sequence=1}{degree sequences} of graphs} 
PhD Dissertation, \href{http://web.mit.edu/}{MIT}, 2007.

\bibitem{BuschCJ2010} A. N. Busch, G. Chen, M. S. Jacobson,  
Transitive partitions in realizations of tournament score sequences,
\textit{J.} \href{http://onlinelibrary.wiley.com/journal/10.1002/(ISSN)1097-0118/issues}{\textit{Graph}} 
\textit{Theory} \textbf{64,} 1 (2010), 52--62. 

\bibitem{Chen1966} W. Chen, On the realization of a (p,s)-digraph with prescribed degrees, \textit{J. Franklin Institute} 
\textbf{281,} (5) 406--422.

\bibitem{Choudum1986} S. A. Choudum, A simple proof of the Erdős-Gallai theorem on graph sequences, 
\textit{Bull.} \href{http://journals.cambridge.org/action/displayJournal?jid=BAZ}{\textit{Austral.}} 
\textit{Math. Soc.} \textbf{33 } (1986) 67--70. 

\bibitem{Chungphaisan1974} V. Chungphaisan, Conditions for sequences to be $r$-graphic. 
\href{http://www.sciencedirect.com/science/journal/0012365X}{\textit{Discrete Math.}} \textbf{7} (1974) 31--39.

\bibitem{CooperL2011} J. Cooper, L. Lu, Graphs with asymptotically invariant degree sequences under restriction, 
\href{http://www.tandfonline.com/toc/uinm20/current}{\textit{Internet}} \textit{Mathematics} \textbf{7,} 1 (2011)  67--80.

\bibitem{CormenLRS2009} T. H. \href{http://www.cs.dartmouth.edu/~thc/}{Cormen},   
Ch. E. \href{http://people.csail.mit.edu/cel/}{Leiserson},  
R. L. \href{http://people.csail.mit.edu/rivest/}{Rivest},  
C. \href{http://www.columbia.edu/~cs2035/}{Stein},     
\textit{Introduction to Algorithms} Third edition, The
\href{http://mitpress.mit.edu/main/home/default.asp}{MIT} 
Press/\href{http://www.mhprofessional.com/category/?cat=1012}{McGraw} Hill, 
Cambridge/New York, 2009.

\bibitem{delGenioKTB2010} C. I. \href{http://www.biond.org/user/22}{Del Genio},  
H. \href{http://icensa.nd.edu/kim.shtml}{Kim}, Z. \href{http://obelix.phys.nd.edu/~toro/}{Toroczkai},     
K. E. \href{http://phys.uh.edu/people/faculty/index.php?155622-961-5=kbassler}{Bassler},   
Efficient and exact sampling of simple graphs with given arbitrary 
\href{http://www.plosone.org/article/info$\%$3Adoi$\%$2F10.1371$\%$2Fjournal.pone.0010012}{degree sequence}, 
\href{http://www.plosone.org/home.action}{\textit{PLoS ONE}} \textbf{5,} 4  e10012 (2010).

\bibitem{DessmarkLG1994} A. Dessmark, A. Lingas, O. Garrido, On parallel complexity of maximum $f$-matching and 
the degree sequence problem. \textit{Mathematical Foundations of Computer Science 1994} (Ko\u{s}ice, 1994), 
LNCS \textbf{841,} \href{http://www.springer.com/?SGWID=1-102-0-0-0}{Springer}, Berlin, 1994, 316--325. 

\bibitem{Eggleton1975} R. B. Eggleton, Graphic sequences and graphic polynomials: a report, in \textit{Colloq. Math. Soc. J. Bolyai} \textbf{10,} North Holland, Amsterdam, 1975, 385--392.

\bibitem{EggletonH1979} R. B. Eggleton, D. A. Holton, Graphic sequences. \textit{Lecture Notes in Mathematics} \textbf{10}, Springer Verlag, Berlin, 1979, 1--10.  
 
\bibitem{ErdosG1960} P. \href{http://www-history.mcs.st-and.ac.uk/Mathematicians/Erdos.html}{Erdős},     
T. \href{http://hu.wikipedia.org/wiki/Gallai_Tibor}{Gallai}, Graphs with vertices having 
\href{http://www.renyi.hu/~p_erdos/1961-05.pdf}{prescribed degrees} (Hungarian), \textit{Mat. Lapok} 
\textbf{11} (1960) 264--274.

\newpage\bibitem{ErdosMT2011} P. L. Erdős, I. \href{http://www.renyi.hu/~miklosi/}{Mikl\'os}, Z. \href{http://obelix.phys.nd.edu/~toro/}{Toroczkai}, A simple Havel-Hakimi type
algorithm to realize graphical degree sequences of directed graphs. 
\href{http://www.combinatorics.org/ojs/index.php/eljc/index}{\textit{Electronic J. Combin.}} \textbf{17,} 1 R66 (2011).

\bibitem{ErdosR1993} P. \href{http://www-history.mcs.st-and.ac.uk/Mathematicians/Erdos.html}{Erdős}, 
L. B. Richmond, On graphical partitions, \href{http://www.combinatorica.hu/kezdolap.html}{\textit{Combinatorica}} 
\textbf{13,} 1 (1993) 57--63.

\bibitem{FordF1962} L. R. Ford, D. R. Fulkerson, \textit{Flows in Networks.} Princeton University, Press, 
Princeton, 1962.

\bibitem{Frank2011} A. \href{http://www.cs.elte.hu/~frank/}{Frank}, 
\textit{Connections in Combinatorial Optimization,}   
\href{http://oup.com/}{Oxford} University Press, Oxford, 2011.

\bibitem{Frank1980} A. \href{http://www.cs.elte.hu/~frank/}{Frank}, On the orientation of graphs. 
 \textit{J.} \href{http://www.sciencedirect.com/science/journal/00958956}{\textit{Combin.}}  
\textit{Theory Ser. B}  \textbf{28,} 3 (1980) 251--261. 

\bibitem{FrankGy1978} A. \href{http://www.cs.elte.hu/~frank/}{Frank}, A. \href{http://www.sztaki.hu/people/008001049/}{Gy\'arf\'as}, 
How to orient the edges of a graph? In \textit{Combinatorics. Vol. 1} 
(ed. A. Hajnal and V. T. S\'os), North-Holland, Amsterdam-New York, 1978. pp. 353--364. 

\bibitem{FrankKK2003} A. \href{http://www.cs.elte.hu/~frank/}{Frank}, T. 
\href{http://www.cs.elte.hu/egres/www/mp_tkiraly.html}{Király}, Z. \href{http://www.cs.elte.hu/~kiraly/}{Király}, 
On the orientation of graphs and hypergraphs. 
\href{http://www.sciencedirect.com/science/journal/0166218X}{\textit{Discrete}} \textit{Appl. Math.}, \textbf{131,} 2 (2003) 385--400.

\bibitem{FrankSS2002} D. A. Frank, C. D. \href{http://www4.ncsu.edu/~savage/}{Savage}, J. A. Sellers, 
On the number of graphical forest partitions, \textit{Ars Combin.} \textbf{65} (2002) 33--37. 

\bibitem{Fulkerson1960} D. R. Fulkerson, Zero-one matrices with zero trace, \href{http://msp.berkeley.edu/pjm/about/journal/cover.html}{\textit{Pacific J. Math}} \textbf{10} (1960) 831--836.

\bibitem{Gale1957} D. Gale, A theorem on flows in networks, \href{http://msp.berkeley.edu/pjm/about/journal/cover.html}{\textit{Pacific J. Math.}} \textbf{7} (1957) 1073--1082. 

\bibitem{GargGT2011} A. Garg, A. Goel, A.  \href{http://maths.iitd.ac.in/people/faculty/amitabh_tripathi.php}{Tripathi}, Constructive extensions of two results on graphs 
sequences. \href{http://www.sciencedirect.com/science/journal/0166218X}{\textit{Discrete}} \textit{Appl. Math.}  \textbf{159,} 17 (2011) 2170--2174. 

\bibitem{GriggsR1999} J. \href{http://www.math.sc.edu/~griggs/}{Griggs}, K. B. Reid, Landau's theorem revisited, 
\href{http://ajc.maths.uq.edu.au/}{\textit{Australas. J. Comb.}} \textbf{20} (1999), 19--24.

\bibitem{GrossY2004} J. L. Gross, J. Yellen, 
\textit{Handbook of Graph Theory,} CRC Press, Boca Raton, 2004. 

\bibitem{GuiduliGyTW1998} B. Guiduli, A.  \href{http://www.sztaki.hu/munkatars/008001049}{Gyárfás}, S. Thomassé, P. Weidl,   
$2$-partition-transitive tournaments. \textit{J.} \href{http://www.sciencedirect.com/science/journal/00958956}{ \textit{Combin.}} 
\textit{Theory Ser. B} \textbf{72}, 2 (1998) 181--196. 

\bibitem{Hakimi1962} S. L. \href{http://en.wikipedia.org/wiki/S.\_L.\_Hakimi}{Hakimi},    
On the realizability of a set of integers as degrees of the vertices of a simple graph. \textit{J.} 
\href{http://www.jstor.org/action/showPublication?journalCode=jsociinduapplmat}{\textit{SIAM}} 
\textit{Appl. Math.} \textbf{10} (1962) 496--506. 

\bibitem{Hakimi1965} S. L. \href{http://en.wikipedia.org/wiki/S.\_L.\_Hakimi}{Hakimi},  
On the degrees of the vertices of a graph, \textit{F. Franklin Institute,} \textbf{279,} 4 (1965) 290--308.

\bibitem{HarborthK1982} H. \href{http://www.mathematik.tu-bs.de/harborth/}{Harborth}, A. Kemnitz, Eine Anzahl der Fussballtabelle. \textit{Math. \href{http://www.springer.com/mathematics/journal/591}{Semester.}} 
 \textbf{29} (1982) 258--263. 

\bibitem{Havel1955} V. Havel, A remark on the existence of finite graphs (Czech), \textit{\u{C}asopis P\u{e}st. Mat.}  
\textbf{80} (1955) 477--480.

\bibitem{HellK2009} P. \href{http://www.cs.sfu.ca/~pavol/}{Hell}, D. \href{https://www.cs.ubc.ca/people/david-kirkpatrick}{Kirkpatrick}, 
Linear-time certifying algorithms for near-graphical sequences. 
\href{http://www.sciencedirect.com/science/journal/0012365X}{\textit{Discrete}} \textit{Math.} \textbf{309,} 18 (2009) 5703--5713. 

\bibitem{Hemasinha2003} R. Hemasinha, An algorithm to generate tournament score sequences, 
\textit{Math.  Comp.  Modelling} \textbf{37,} 3--4 (2003) 377--382.  

\newpage\bibitem{Isaak2010} G. \href{http://www.lehigh.edu/~gi02/gi02.html}{Isaak}, Tournaments and score sequences, in ed. by D. B. 
\href{http://www.math.uiuc.edu/~west/}{West} \textit{REGS in Combinatorics}, 2010, No. 7, \newline  
http://www.math.uiuc.edu/~west/regs/fifa.html

\bibitem{Ivanyi2001} A. \href{http://compalg.inf.elte.hu/tanszek/tony/oktato.php?oktato=tony}{Iványi}, Testing of football score sequences (Hungarian), 
in: \textit{Abstracts of 25th Hungarian Conf. on Operation Research} (Debrecen, October 17--20, 2001), MOT, Budapest, 2001, 53--53.  

\bibitem{Ivanyi2002} A.  \href{http://compalg.inf.elte.hu/tanszek/tony/oktato.php?oktato=tony}{Iványi}, 
Maximal tournaments.   \href{http://www.bke.hu/Pure/}{\textit{Pure}} \textit{Math. Appl.} \textbf{13,} 1--2 (2002) 171--183.


\bibitem{Ivanyi2009} A. \href{http://compalg.inf.elte.hu/tanszek/tony/oktato.php?oktato=tony}{Iványi},   
Reconstruction of complete interval tournaments, \textit{Acta Univ. Sapientiae,} 
\href{http://www.acta.sapientia.ro/acta-info/informatica-main.htm}{\textit{Inform.}}, 
\textbf{1,} 1 (2009) 71--88.

\bibitem{Ivanyi2010} A. \href{http://compalg.inf.elte.hu/tanszek/tony/oktato.php?oktato=tony}{Iványi},     
Reconstruction of complete interval tournaments. II, \textit{Acta Univ. Sapientiae,}  
\href{http://www.acta.sapientia.ro/acta-math/matematica-main.htm}{\textit{Math.}} 
\textbf{2,} 1 (2010) 47--71.

\bibitem{Ivanyi2011Kyoto} A. \href{http://compalg.inf.elte.hu/tanszek/tony/oktato.php?oktato=tony}{Iványi},   
Directed graphs with prescribed score sequences (ed by S. Iwata), \textit{The 7th Hungarian-Japanese Symposium 
on Discrete Mathematics and Applications} (Kyoto, May 31 - June 3, 2011, ed by S. Iwata), 114--123.

\bibitem{Ivanyi2012Egres} A. \href{http://compalg.inf.elte.hu/tanszek/tony/oktato.php?oktato=tony}{Iványi},    
Deciding the validity of the score sequence of a soccer tournament, in 
(ed. by A. \href{http://www.cs.elte.hu/~frank/}{Frank}): \textit{Open problems of the    
 Egerváry Research Group,} Budapest, 2012. \href{http://lemon.cs.elte.hu/egres/open/}{http://lemon.cs.elte.hu/egres/open/} 

\bibitem{Ivanyi2012Comp} A. \href{http://compalg.inf.elte.hu/tanszek/tony/oktato.php?oktato=tony}{Iványi}, 
Degree sequences of multigraphs. \textit{Annales Univ. Sci. Budapest., } \href{http://ac.inf.elte.hu/}{\textit{Sect. Comp.}} \textbf{37} (2012), 195--214.

\bibitem{IvanyiL2012AML} A. \href{http://compalg.inf.elte.hu/tanszek/tony/oktato.php?oktato=tony}{Iv\'anyi},
L. \href{http://people.inf.elte.hu/lulsaai}{Lucz}, Degree sequences of multigraphs (Hungarian), \textit{Alkalm. Mat. Lapok} 
\textbf{29} (2012) (to appear). 

\bibitem{IvanyiLMS2011Acta} A. \href{http://compalg.inf.elte.hu/tanszek/tony/oktato.php?oktato=tony}{Iványi},
L. \href{http://people.inf.elte.hu/lulsaai}{Lucz}, T. F. \href{mailto:moritamas@ludens.elte.hu}{Móri}, 
P. \href{http://people.inf.elte.hu/sopsaai}{Sótér}, On the Erd\H os-Gallai and Havel-Hakimi algorithms. 
\textit{Acta Univ. Sapientiae,} \href{http://www.acta.sapientia.ro/acta-info/informatica-main.htm}{\textit{Inform.}} \textbf{3,} 2 (2011) 230--268. 

\bibitem{IvanyiLMS2011A004251} A. \href{http://compalg.inf.elte.hu/tanszek/tony/oktato.php?oktato=tony}{Iványi},
L. \href{http://people.inf.elte.hu/lulsaai}{Lucz}, T. F. \href{mailto:moritamas@ludens.elte.hu}{M\'ori}, 
P. \href{http://people.inf.elte.hu/sopsaai}{Sótér},
The number of degree-vectors for simple graphs, in: ed. by  N. J. A. 
\href{http://www2.research.att.com/~njas/}{Sloane}, \textit{The On-Line Encyclopedia of Integer Sequences,} 
2011. http://oeis.org/A004251

\bibitem{IvanyiP2011} A. \href{http://compalg.inf.elte.hu/tanszek/tony/oktato.php?oktato=tony}{Iványi},     
S. \href{http://maths.uok.edu.in/Faculty5.aspx}{Pirzada}, Comparison based ranking, in: 
ed. A. \href{http://compalg.inf.elte.hu/tanszek/tony/oktato.php?oktato=tony}{Iványi}, 
\href{http://www.tankonyvtar.hu/}{\textit{Algorithms}} \textit{of Informatics, Vol. 3,}    
\href{http://www.antoncom.hu/books.htm}{AnTonCom}, Budapest 2011, 1209--1258.

\bibitem{KannanTV1999} R. Kannan, P. \href{http://people.math.gatech.edu/~tetali/}{Tetali}, S. Vempala, Simple Markovian-chain algorithms for generating bipartite graphs and tournaments. 
\href{http://onlinelibrary.wiley.com/journal/10.1002/(ISSN)1098-2418;jsessionid=9397758EE5C1A8ED7CD528AB47AADCC4.d02t03}{\textit{Random}} \textit{Struct. Algorithms} \textbf{14,} 4 (1999) 293--308.

\bibitem{KayibiKPI2012} K. Kayibi, M. A. \href{http://faculty.kfupm.edu.sa/PYP/malikhan/}{Khan}, 
S.  \href{http://maths.uok.edu.in/Faculty5.aspx}{Pirzada}, 
A.  \href{http://compalg.inf.elte.hu/tanszek/tony/oktato.php?oktato=tony}{Iv\'anyi}, 
Random sampling of minimally cyclic digraphs with given imbalance sequence. 
\emph{Acta Univ. Sapientiae,}  \href{http://www.acta.sapientia.ro/acta-math/matematica-main.htm}{\textit{Math.}}  
 (submitted).

\bibitem{KemnitzD1997} A. Kemnitz, S. Dolff, Score sequences of multitournaments. 
\textit{Congr. Numer.} \textbf{127} (1997) 85--95. 

\bibitem{Keri2011} G. \href{http://www.oplab.sztaki.hu/cv_kg_hu.htm}{K\'eri}, 
On qualitatively consistent, transitive and contradictory judgment matrices emerging from 
multiattribute decision procedures, \href{http://www.springerlink.com/content/1435-246x/19/2/}{\textit{CEJOR}} 
\textit{Cent. Eur. J. Oper. Res.} \textbf{19,} 2 (2011) 215--224. 

\bibitem{KimTMESz2009} H. \href{http://www.phys.vt.edu/people/hkim.shtml}{Kim}, Z. \href{http://obelix.phys.nd.edu/~toro/}{Toroczkai}, I. 
\href{http://www.renyi.hu/~miklosi/}{Miklós}, P. L. Erdős, L. A. \href{http://www.math.sc.edu/~szekely/}{Székely}, 
Degree-based graph construction, \textit{J. Physics}: \href{http://iopscience.iop.org/1751-8121/42/39/392001/}{\textit{Math. Theor. A}} \textbf{42,} 39 (2009), 392001-1-3920001.10.

\bibitem{Kiraly2012} Z. \href{http://www.cs.elte.hu/~kiraly/}{Király}, Recognizing graphic degree sequences and generating all realizations, Technical Report of  Egerváry Research Group, TR-2011-11, Budapest. 
Last modification 23 April, 2012. \url{http://www.cs.elte.hu/egres/}

\bibitem{Kiraly2012DS} Z. \href{http://www.cs.elte.hu/~kiraly/}{Király}, \textit{Data Structures}  (Lecture notes in Hungarian),  \href{http://www.elte.hu/en}{Eötvös}  Loránd University, Mathematical Institute, Budapest, 2012.  
 \url{http://www.cs.elte.hu/~kiraly/Adatstrukturak.pdf}
 
\bibitem{KleitmanW1981} D. J. \href{http://www-math.mit.edu/~djk/}{Kleitman}, K. J. Winston, 
Forests and score vectors, \href{http://www.combinatorica.hu/kezdolap.html}{\textit{Combinatorica}} \textbf{1,} 1 (1981) 49--54.

\bibitem{KovacsP2002} G. Zs. Kovács, N. Pataki, Analysis of ranking sequences (Hungarian), Scientific student paper, 
 \href{http://www.elte.hu/en}{Eötvös}  Loránd University, Faculty of Sciences, Budapest 2002.
 
\bibitem{LaMar2010} M. D. LaMar, Algorithms for realizing degree sequences of directed graphs. arXiv, (2010). 
\href{http://arxiv.org/abs/0906.0343}{http://arxiv.org/abs/0906.0343}.
 
\bibitem{Landau1953} H. G. Landau, On dominance relations and the structure of 
\href{http://www.springerlink.com/content/765042v152l07721/}{animal societies}. III. 
The condition for a score sequence, \textit{Bull. Math.} 
\href{http://www.springerlink.com/content/765042v152l07721/}{\textit{Biophys.}} \textbf{15,} (1953) 143--148. 

\bibitem{LiljerosEASA2001} F. \href{http://people.su.se/~liljeros}{Liljeros}, C. R. Edling, L. \href{http://amaral.northwestern.edu/people/amaral/}{Amaral}, H. E. \href{http://polymer.bu.edu/hes/}{Stanley}, 
Y. Aberg, The web of human sexual contacts. \href{http://www.nature.com/}{\emph{Nature}} \textbf{411} (2001) 907--908.

\bibitem{Lucz2012}  L. \href{http://people.inf.elte.hu/lulsaai}{Lucz}, \textit{Analysis of degree sequences of graphs} (Hungarian), MSc Thesis, 
 \href{http://www.elte.hu/en}{Eötvös}  Loránd University, Faculty of Informatics, 
Budapest, 2012. \href{http://people.inf.elte.hu/lulsaai/diploma}{http://people.inf.elte.hu/lulsaai/diploma}

\bibitem{LuczI2012MaCS} L. \href{http://people.inf.elte.hu/lulsaai}{Lucz}, A. 
\href{http://compalg.inf.elte.hu/tanszek/tony/oktato.php?oktato=tony}{Iványi}, 
Testing and enumeration of football sequences, in: \textit{MaCS'12. 9th Joint Conference in Mathematics and Computer 
Science} (Siófok, Hungary, February 9--12, 2012, ed. by Z. \href{http://people.inf.elte.hu/csz/}{Csörnyei}), ELTE IK, Budapest, 2012, 63--63.

\bibitem{McKayW1996} B. D. McKay, X. Wang, Asymptotic enumeration of tournaments with a given score sequence. 
\textit{J. Comb. Theory A}, \textbf{73,} 1 (1996) 77--90. 

\bibitem{MeierlingV2009} D. Meierling, L. \href{http://www.math2.rwth-aachen.de/de/mitarbeiter/volkm}{Volkmann}, A remark on degree 
sequences of multigraphs. \textit{Math. Methods Oper. Res.} \textbf{69,} 2 (2009) 369--374. 

\bibitem{MetropolisS1980} N. Metropolis, P. R. Stein, The enumeration of graphical partitions, 
\href{http://www.sciencedirect.com/science/journal/01956698}{\textit{European}} \textit{J. Comb.} \textbf{1,} 2 (1980) 139--153.

\bibitem{MiklosES2010} I. \href{http://www.renyi.hu/~miklosi/}{Miklós}, P.  L.  \href{http://www-history.mcs.st-and.ac.uk/Mathematicians/Erdos.html}{Erdős}, 
L. \href{http://www.renyi.hu/~soukup/}{Soukup}, Towards random uniform sampling of bipartite graphs with given degree sequence, 
\textit{arXiv} 1004.2612v3 [math.CO] (14 Sep 2010), \href{http://arxiv.org/pdf/1004.2612v3.pdf}{http://arxiv.org/pdf/1004.2612v3.pdf} 

\bibitem{Miller2012} J. W. \href{http://www.dam.brown.edu/people/jmiller/}{Miller}, Reduced criterion for degree sequences, 
\href{http://arxiv.org/pdf/1205.2686v1.pdf}{\textit{arXiv}}, arXiv:1205.2686v1 [math.CO] 11 May 2012, 18 pages.

\bibitem{Moon1962} J. W. Moon, On the score sequence of an $n$-partite tournament.   
\href{http://cms.math.ca/cmb/}{\textit{Can.}} \textit{ Math. Bull.} 
\textbf{5} (1962) 51--58. 

\bibitem{Moon1963} J. W. Moon, An extension of Landau's theorem 
on tournaments, \href{http://msp.berkeley.edu/pjm/about/journal/cover.html}{\textit{Pacific J. Math.}} \textbf{13} (1963), 1343--1345.

\bibitem{Moon1968} J. W. Moon, \textit{Topics on Tournaments}. Holt, Rinehart and Winston. 
New York, 1968. 
 
\newpage\bibitem{NarayanaB1964} T. V. Narayana, D. H. Bent, Computation of the number of score sequences in 
round-robin \href{http://cms.math.ca/10.4153/CMB-1964-015-1}{tournaments}, 
\href{http://cms.math.ca/cmb/}{\textit{Canad.}} \textit{Math. Bull.} \textbf{7,} 1 (1964) 133--136.

\bibitem{NewmanBW2006} M. \href{http://www-personal.umich.edu/~mejn/}{Newman}, A. L. \href{http://www.barabasi.com/}{Barabási}, 
D. J. Watts, \textit{The Structure and Dynamics of Networks.} Princeton University Press, (2006).

\bibitem{Ozkan2011} S. Özkan, Generalization of the Erdős-Gallai inequality. 
\href{http://bkocay.cs.umanitoba.ca/arscombinatoria/}{\textit{Ars Combin.}} \textbf{98} (2011) 295--302.

\bibitem{Palvolgyi2009} D. \href{http://www.cs.elte.hu/~dom/}{Pálvölgyi}, Deciding soccer scores and partial orientations of graphs. 
\textit{Acta Univ. Sapientiae,} \href{http://www.acta.sapientia.ro/acta-math/matematica-main.htm}{\textit{Math.}} \textbf{1,} 1 (2009) 35--42.

\bibitem{PatrinosH1976} A. N. Patrinos, S. L. Hakimi, Relations between graphs and integer-pair sequences. 
\href{http://www.sciencedirect.com/science/journal/0012365X}{\textit{Discrete}} \textit{Math.} \textbf{15} 4 (1976) 347--358

\bibitem{PecsySz2000} G. Pécsy, L. Szűcs,   
\href{http://www.cs.ubbcluj.ro/~studia-i/2000-2/2-Pecsy.pdf}{Parallel} verification and enumeration of tournaments,  
\textit{Stud. Univ.} \href{http://www.cs.ubbcluj.ro/~studia-i/contents.php}{Babe\c s-Bolyai}, 
\textit{Inform.} \textbf{45,} 2 (2000) 11--26.

\bibitem{Pirzada2012} S. \href{http://maths.uok.edu.in/Faculty5.aspx}{Pirzada}, \textit{An Introduction to Graph Theory}. Orient BlackSwan, Hyderabad, 2012. 

\bibitem{PirzadaZI2011} S. \href{http://maths.uok.edu.in/Faculty5.aspx}{Pirzada}, G. \href{mailto:gfzhou@nju.edu.cn}{Zhou}, 
A. \href{http://compalg.inf.elte.hu/tanszek/tony/oktato.php?oktato=tony}{Iványi},  
Score lists of multipartite hypertournaments, \textit{Acta Univ. Sapientiae, Inform.} \textbf{2,} 2 (2011) 184--193.  

\bibitem{Reid1996} K. B. Reid, Tournaments: Scores, kings, 
generalizations and special topics, \textit{Congr. Numer.} \textbf{115} (1996) 171--211.

\bibitem{ReidZ1998} K. B. Reid, C. Q. Zhang, Score sequences of semicomplete digraphs, 
\textit{Bull. Inst. Combin. Appl.} \textbf{24} (1998) 27--32.

\bibitem{RodsethST2009} {\O}. J. R{\o}dseth, J. A. Sellers, H. Tverberg, Enumeration of the degree sequences 
of non-separable graphs and connected graphs. \textit{European J. Comb.} \textbf{30,} 5 (2009) 1309--1317. 

\bibitem{RuskeyCES1994} F. \href{http://webhome.cs.uvic.ca/~ruskey/}{Ruskey}, F. R. Cohen, P. Eades, A. Scott,  
Alley CATs in search of good homes. \textit{Congr. Numer.} \textbf{102} (1994) 97--110. 

\bibitem{Ryser1957} H. J. Ryser, Combinatorial properties of matrices of zeroas and ones, \textit{Canad. J. Math.} 
\textbf{9} (1957) 371--377.

\bibitem{Schoenfield2008A064626} J. E. \href{mailto:jscho@hiwaay.net}{Schoenfield}, The number of football score sequences, in: ed. by 
N. J. A. \href{http://www2.research.att.com/~njas/}{Sloane}, \textit{The On-Line Encyclopedia of Integer Sequences,} \newline
2012. http://oeis.org/A064626

\bibitem{SierksmaH1991} G. \href{http://www.rug.nl/staff/g.sierksma/index}{Sierksma}, 
H. \href{http://www.cs.uu.nl/staff/slam.html}{Hoogeveen}, Seven criteria for integer sequences being graphic, 
\textit{J.} \href{http://onlinelibrary.wiley.com/journal/10.1002/(ISSN)1097-0118/issues}{\textit{Graph Theory}} 
\textbf{15,} 2 (1991) 223--231.

\bibitem{Siklosi2001} B. Siklósi, \textit{Comparison of sequential and parallel algorithms solving sport problems}.  
Master thesis. \href{http://www.elte.hu/en}{Eötvös}  Loránd University, Faculty of Sciences, Budapest, 2001.
 
\bibitem{Sloane2011A004251} N. J. A. \href{http://www2.research.att.com/~njas/}{Sloane}, 
The number of degree-vectors for simple graphs.
In (ed. N. J. A. Sloane): \textit{The On-Line Encyclopedia of the Integer Sequences}. 2011. http://oeis.org/A004251 

\bibitem{Soroker1990} D. Soroker, \textit{Optimal parallel construction of prescribed tournaments,} 
\href{http://www.sciencedirect.com/science/journal/0166218X}{\textit{Discrete}} \textit{Appl. Math.}  
\textbf{29,} 1 (1990) 113--125.

\bibitem{Stanley1991} R. P. \href{http://www-math.mit.edu/~rstan/}{Stanley}, 
A zonotope associated with graphical degree sequence, in: 
\textit{Applied Geometry and Discrete Mathematics, Festschr. 65th Birthday Victor Klee.}  
DIMACS Series in Discrete Mathematics and Theoretical Computer Science. \textbf{4} (1991) 555--570.

\bibitem{SzekelyCE1992} L. A. \href{http://www.math.sc.edu/~szekely/}{Székely}, L. H. Clark, R. C. Entringer. An inequality for degree sequences. 
\href{http://www.sciencedirect.com/science/journal/0012365X}{\textit{Discrete}} \textit{Math.}
\textbf{103,} 3 (1992) 293--300. 

\bibitem{Takahashi2007} M. Takahashi, \textit{Optimization Methods for Graphical Degree Sequence Problems 
and their Extensions}, PhD thesis, Graduate School of Information, Production and systems, Waseda University, 
Tokyo, 2007. \href{http://hdl.handle.net/2065/28387}{http://hdl.handle.net/2065/28387}    
    

\bibitem{Temesi2011} J. \href{http://portal.uni-corvinus.hu/index.php?id=24294&no_cache=1&tx_efcointranet_pi1$\%$5Bfomenu$\%$5D=elerhetoseg&tx_efcointranet_pi1$\%$
5Bcusman$\%$5D=jtemesi}{Temesi}, Pairwise comparison matrices and the error-free property of the decision maker, 
\href{http://www.springerlink.com/content/1435-246x/19/2/}{\textit{CEJOR}} \textit{Cent. Eur. J. Oper. Res.} 
\textbf{19,} 2 (2011) 239--249.


\bibitem{Tetali1998} P. \href{http://people.math.gatech.edu/~tetali/}{Tetali}, A characterization of unique tournaments. 
 \textit{J.} \href{http://www.sciencedirect.com/science/journal/00958956}{\textit{Combin.}} 
\textit{Theory Ser. B} \ \textbf{72,} 1  (1998) 157--159.  

\bibitem{TripathiT2008} A. \href{http://maths.iitd.ac.in/people/faculty/amitabh_tripathi.php}{Tripathi}, H. Tyagy, 
A simple criterion on degree sequences of graphs.  
\href{http://www.sciencedirect.com/science/journal/0166218X}{\textit{Discrete}} \textit{Appl. Math.}  \textbf{156,} 18 (2008) 3513--3517.

\bibitem{TripathiV2003} A. \href{http://maths.iitd.ac.in/people/faculty/amitabh_tripathi.php}{Tripathi}, 
S.  \href{http://www.math.illinois.edu/~sujith/}{Vijay}, A note on a theorem of Erdős \& Gallai. 
\href{http://www.sciencedirect.com/science/journal/0012365X}{\textit{Discrete}} \textit{Math.} 
\textbf{265,} 1--3 (2003) 417--420.

\bibitem{TripathiVW2010} A. \href{http://maths.iitd.ac.in/people/faculty/amitabh_tripathi.php}{Tripathi}, 
S. Venugopalan, D. B. \href{http://www.math.uiuc.edu/~west/}{West}, A short constructive proof of the Erdős-Gallai characterization of graphic lists. 
\href{http://www.sciencedirect.com/science/journal/0012365X}{\textit{Discrete}} \textit{Math.} 
\textbf{310,} 4 (2010) 833--834.

\bibitem{Tyskevich2000} R. Tyskevich, Decomposition of graphical sequences and unigraphs, 
\href{http://www.sciencedirect.com/science/journal/0012365X}{\textit{Discrete}} \textit{Math.} \textbf{220,} 1--3 (2000) 201--238.

\bibitem{Emde1975} P. \href{http://www.illc.uva.nl/People/show_person.php?Person_id=Emde+Boas+P.+van}{van Emde Boas}, 
Preserving order in a forest in less than logarithmic time, \textit{Proc. 16th Annual Symp. Found Comp. Sci.} \textbf{10} 
(1975) 75--84.    

\bibitem{Weisstein2012DS} E. W. \href{http://mathworld.wolfram.com/about/author.html}{Weisstein}, 
\href{http://mathworld.wolfram.com/DegreeSequence.html}{\textit{Degree Sequence,}} 
From MathWorld---Wolfram Web Resource, 2012. 
 
\bibitem{Weisstein2012GS} E. W. \href{http://mathworld.wolfram.com/about/author.html}{Weisstein}, 
\href{http://mathworld.wolfram.com/GraphicSequence.html}{\textit{Graphic Sequence,}}
From MathWorld---Wolfram Web Resource, 2012.

\bibitem{WinstonK1983} K. J. Winston, D. J. \href{http://www-math.mit.edu/~djk/}{Kleitman},  
On the asymptotic number of tournament score sequences.  
\textit{J.} \href{http://www.sciencedirect.com/science/journal/00973165}{\textit{Combin.}} \textit{Theory Ser. A.} 
\textbf{35} (1983) 208--230. 

\bibitem{ZverovichZ1992} I. E. Zverovich, V. E. \href{http://www.cems.uwe.ac.uk/~vzverovi}{Zverovich}, Contribution to the theory of graphic sequences, 
\href{http://www.sciencedirect.com/science/journal/0012365X}{\textit{Discrete}} \textit{Math.} \textbf{105} (1992) 293--303.

\end{thebibliography}
\end{document}